\makeatletter
\@ifundefined{@parse@version@dash}{%
\def\@parse@version#1{\@parse@version@0#1}
\def\@parse@version@#1/#2/#3#4#5\@nil{%
\@parse@version@dash#1-#2-#3#4\@nil}
\def\@parse@version@dash#1-#2-#3#4#5\@nil{%
  \if\relax#2\relax\else#1\fi#2#3#4 }
}{}
\makeatother
\documentclass[aip,jcp,numerical,reprint]{revtex4-2}
\usepackage{latexsym,eucal,amsfonts,amsmath,amsthm,amssymb} 
\usepackage{scalerel}
\usepackage{blindtext}
\usepackage[table]{xcolor}
\usepackage{graphicx}
\usepackage{psfrag}
\usepackage{amsbsy}      
\usepackage{etoolbox}
\usepackage[final]{changes}

\usepackage{comment}
\usepackage{siunitx}
\usepackage{chemfig}
\usepackage{mathtools}
\usepackage{xfrac}
\usepackage{color}
\usepackage{dsfont}
\usepackage{tikz}
\usepackage{bm}% bold math
\usepackage[utf8]{inputenc}
\usepackage[T1]{fontenc}
\usepackage{etoolbox}
\usepackage{newfloat}

\usepackage{hyperref}
\usepackage{cleveref}

\usepackage{soul}
\usepackage{enumitem} % change enumerate item style

\graphicspath{{fig/}}

\newcommand{\oline}[1]{\mkern 2mu\overline{\mkern-2.5mu#1\mkern-2mu}\mkern 2mu}

\DeclareMathOperator{\Htd}{H2A-H2B}
\DeclareMathOperator{\HtHf}{(H3-H4)_2}

\def\ve{\varepsilon}

\def\kd{k_{\rm d}}
\def\kon{k_{\rm on}}
\def\tkon{k_{\rm on}}
\def\koff{k_{\rm off}}

\def\v{{\bf v}}
\def\x{{\bf x}}

\def\pa{p_{\rm a}}
\def\pd{p_{\rm d}}
\def\qa{q_{\rm a}}
\def\qd{q_{\rm d}}

\def\n{{\bf n}}

\def\D{\Delta}
\def\0{{\bf 0}}
\def\A{{\bf A}}
\def\B{{\bf B}}

\def\C{{\bf C}}

\def\D{{\bf D}}

\def\F{{\bf F}}
\def\H{{\bf H}}
\def\I{{\bf I}}

\def\M{{\bf M}}
\def\P{{\bf P}}
\def\Q{{\bf W}}

\begin{document}

%\title{Stochastic dynamics of passive and active histone remodeling}
\title{Stochastic nucleosome disassembly mediated by remodelers and histone fragmentation}

%dynamics of passive and cofactor-mediated histone disassembly
%and nucleosome remodeling}

\author{Xiangting Li}
\email{xiangting.li@ucla.edu}
\affiliation{Department of Computational Medicine, 
University of California, Los Angeles, CA 90095-1766 USA}
\author{Tom Chou}
\email{tomchou@ucla.edu}
\affiliation{Department of Computational Medicine, 
University of California, Los Angeles, CA 90095-1766 USA}
\affiliation{Department of Mathematics, University of California, Los Angeles, CA
90095-1555 USA}

%\runningpagewiselinenumbers
%\runninglinenumbers

\begin{abstract}
  We construct and analyze monomeric and multimeric models of the
  stochastic disassembly of a single nucleosome. \added{Our monomeric
  model predicts the time needed for a number of histone-DNA contacts
  to spontaneously break, leading to dissociation of a non-fragmented
  histone from DNA. The dissociation process can be facilitated by DNA
  binding proteins or processing molecular motors that compete with
  histones for histone-DNA contact sites.}
%
% Both spontaneous detachment and protein-mediated active processes
%  that rive off histones are modeled.
%
Eigenvalue analysis of the corresponding master equation allows us to
  evaluate histone detachment times under \added{both spontaneous
  detachment and protein-facilitated processes.}  We find that
  competitive \added{DNA} binding of remodeling proteins can
  significantly reduce the typical detachment time but only if these
  remodelers have \added{DNA-}binding affinities comparable to those
  of histone-DNA contact sites. In the presence of processive motors,
  the histone detachment rate is shown to be proportional to the
  product of the histone single-bond dissociation constant and the
  speed of motor protein procession. Our simple intact-histone model
  is then extended to allow for multimeric nucleosome kinetics that
  reveal additional pathways of disassembly. In addition to a
  dependence of complete disassembly times on subunit-DNA contact
  energies, we show how histone subunit concentrations in bulk
  solution can mediate the disassembly process by rescuing partially
  disassembled nucleosomes. Moreover, our kinetic model predicts that
  remodeler binding can also bias certain pathways of nucleosome
  disassembly, with higher remodeler binding rates favoring
  intact-histone detachment.
\end{abstract}

\begin{comment}
Significance Statement: 
Our work provides a detailed kinetic model and predictions for
nucleosome disassembly that incorporates formation and breaking of
histone-DNA contacts, Brownian ratcheting via binding of remodeler
proteins, and histone subunit fragmentation. We applied perturbation
theory within these stochastic models estimates of the expected times
to nucleosome disassembly.  These estimates compared favorably with
exact computations of first disassembly times. Finally, we quantify
the probability the histone falls off as a single unit or sequentially
as subunits break off.  The relative fluxes of the monomeric and
multimeric disassembly pathways where found to depend on histone
subunit binding strength, remodeler affinity, and bulk histone
concentration.
\end{comment}

\maketitle
%\pacs{05.60.-k,87.16.Ac,05.10.Ln}

%%%%===========introduce basic notations===================
\section{Introduction}\label{SEC:INTRO}
%%%%=======================================================

In eukaryotic cells, ~147 base pairs of DNA wrap around each histone
octamer. DNA binds to the histone octamer at approximately 14 sites to
form a nucleosome core particle. Nucleosomes, in turn, help compact
meters of DNA inside the nucleus \cite{SCHIESSEL2001,LUGER1997},
protecting DNA from other proteins and unwanted enzymatic activity
\cite{LORCH1987,RICHMOND2003}.  On occasion however, nucleosomes have
to partially or completely release the substrate DNA to allow access
by DNA-processing enzymes. Histones thus have to simultaneously
perform two contradictory functions \cite{LORCH1987}.  While there is
consensus that histone modification and chromatin remodeling are
critical in epigenetic regulation \cite{Gibney2010}, the details of
how nucleosomes dynamically perform different tasks are not yet fully
understood \cite{AKEY20036}. Therefore, it is essential to first
understand the molecular mechanics and dynamics of histone-DNA
interactions.

DNA at both nucleosome ends is transiently accessible due to
spontaneous bond breaking. This nucleosome ``breathing'' has been
identified using single-molecular biophysics techniques
\cite{ANDERSON2000,LI2005,TOMSCHIK2005,Bundschuh2011}. Based
on these observations, a rigid base-pair nucleosome Markov model was
proposed and computationally explored to characterize the mechanical
response to external
tensions \cite{Bundschuh2011,BRUIN2016,Tompitak2017}, sequence
dependence and positioning of nucleosomes
\cite{TOMPITAK2017505,Culkin2017}, and salt
dependence \cite{VANDEELEN2020}. Recently, similar discrete stochastic
binding and unbinding models have been used to describe target search
by pioneer transcription factors
\cite{Shvets2018aug,Iwahara2021feb,Mondal2022dec,Mondal2023apr}.

In molecular dynamics studies, coarse-grain models and even all-atom
molecular models of nucleosome unraveling have also been discussed
recently, characterizing the free energy landscapes of nucleosomes and
capturing the finer details during the process of unwrapping
\cite{ZHANG2016,LEQUIEU2016,DAVID2018,Takada2021}. Despite these mechanistic
studies and modeling efforts, quantification of histone unwrapping
using the above approaches is computationally expensive.  In
particular, these simulation approaches make it difficult to study:

\begin{itemize}[itemsep=0pt,parsep=0pt,topsep=0pt,partopsep=0pt]
  \item[(i)] rare but decisive events such as complete spontaneous
  unwrapping; and
  \item[(ii)] indirect interactions with other DNA binding proteins via
  transient nucleosome breathing.
\end{itemize}

\noindent Thus, simple analytic descriptions of the dynamics of
histone-DNA and nucleosome-protein interactions can provide a useful
tool for estimating and efficiently testing molecular hypotheses of
nucleosome-mediated chromatin remodeling.  In this paper, we develop
discrete stochastic Markov models that relate different elements of
histone-DNA interactions to overall rates of nucleosome disassembly.

%By defining both the relevant coarse-grained molecular configurations and
%states and the transitions among them, we construct 
%equilibrium distributions and kinetics of nucleosome disassembly.

% Our approach is to analyze a discrete state Markov model describing
% the time-evolution of a probability vector ${\bf P}$ of molecular
% configurations which obeys $\partial_{t}{\bf P} = \Q{\bf P}$, where
% $\Q$ is a model-dependent transition matrix. From this analysis, we
% will derive results that predict the distribution of configurations
% and the statistics of disassembly times. Different detachment
% scenarios involving remodeling factors will also be considered.

In the next section, we formulate two classes of models, one in which
histones remain as an intact single molecule, and another in which
they are composed of three major subunits that can successively
dissociate from DNA. The first \added{abstraction} describes DNA as
linearly unspooling from a contiguous footprint defined by the histone
particle and extends earlier work \cite{chou_2007}.  The state-space
structure of \added{this simple} model is then nested to describe the state space
of more molecularly realistic models of histone
fragmentation. Finally, catalysis of nucleosome disassembly can be
mediated by remodeling factors such as transcription
factors \cite{SHEU2022,MOROZOVMBOC}.  We will also model such
cofactor-facilitated histone removal by incorporating
competitive \added{DNA-}protein binding within each of these two
classes (intact-histone and fragmenting histone) of models.

Our primary goal is to provide a quantitative characterization of the
first passage time (FPT) from an initial configuration to a totally
dissociated state. \deleted{In the context of maintenance of epigenetic
information and DNA replication, it is also important to evaluate the
relaxation time of the \textit{reversible} process that allows for the
rebinding of histones.} We aim to provide a closed form expression or
numerical procedure for evaluating these timescales under specific
biophysical conditions.

\section{Mathematical Models and Results}

The approach we will take for all of our following models is to analyze a
discrete state Markov model describing the time-evolution of a
probability vector ${\bf P}$ of molecular configurations which obeys
$\partial_{t}{\bf P} = \tilde{\Q}{\bf P}$, where $\tilde{\Q}$ is a
model-dependent transition matrix. The state space and the transition
matrix $\tilde{\Q}$ will be appropriately defined for each type of
model, including variants that incorporate protein-catalyzed
nucleosome disassembly.  By analyzing the specific subsets of the
state space and the eigenvalues of the associated transition matrices
$\tilde{\Q}$, we derive results that predict the distribution of
configurations and the statistics of disassembly times.

% We present two classes of kinetic models to describe histone
% dissociation from DNA. The first class of models assume an intact
% histone that incrementally delaminates from DNA.  This ``linearly
% unbinding'' model can also be extended to include facilitated
% disassembly mechanisms such as competitive protein binding and
% peeling by processive motors. In a second class of models, we
% consider histone subunits that can bind each other as well as to the
% DNA substrate. Here, each subunit can linearly delaminate from DNA
% while also detaching from themselves. We also develop a complete
% subunit model that incorporates cofactor facilitation.

The complete state space in our models, $\Omega\cup\Omega^{*}$,
consists of the set of bound states $\Omega$ and the set of detached
states $\Omega^{*}$. In general, the transition matrix coupling all
states is $\tilde{\Q}$. However, since transitions into $\Omega^{*}$
from $\Omega$ are typically irreversible in our analyses, we can
define $\tilde{\Q}$ to operate on states within and out of $\Omega$.
Henceforth, we describe the eigenvalues of $\tilde{\Q}$,
$\{\tilde{\lambda}_{j}\}_{j \geq 0}$, in descending order of their
real parts. The principal eigenvalue $\tilde{\lambda}_0$ of
$\tilde{\Q}$ will be that with the largest real part. When transitions
to $\Omega^*$ are assumed to be irreversible, $\tilde{\Q}$ defined on
$\Omega$ represents a sub-matrix with all eigenvalues having negative
real parts. Using this nomenclature, the inverse of the eigenvalues
describes the timescales associated with the stochastic dynamics of
sets of configurations (described by eigenvectors) within the state
space.  For example, $-1/\tilde{\lambda}_0$ \deleted{$\sim \tau_{0}$} is the
slowest timescale of decay to $\Omega^*$ in the stochastic dynamics.

\added{Quantities like $\tilde{\Q}$ carry a physical dimension of rate
(1/time). To make our mathematical analysis notationally simpler, we
will normalize $\tilde{\Q}$ by the fastest rate in the model to make
it dimensionless. In the rest of the paper, the dimensionless
transition matrix and its associated dimensionless eigenvalues are
denoted $\Q$ and $\lambda_0$, respectively.  Other conventions for
mathematical symbols and objects are summarized in
Table~\ref{tab:nomenclature}.}

\subsection{Linear peeling, simple histone model}

Here, we first consider the stochastic dynamics of how a single
histone particle peels from the DNA wrapping it. This approach is
similar to that taken in Kim et al. \cite{SHEU2022}, but we track
simultaneous peeling from both ends of the histone particle and assume
uniform binding and unbinding rates along the DNA substrate.
Parameters and variables used in this model are listed in
Table~\ref{tab:parameters}.
%
%\begin{widetext}

% A table to list all the parameters used in the paper
\begin{table*}[ht]
%\centering
  \begin{center}
    \begin{tabular}{lcc}
\hline \textbf{parameter/variable} & \textbf{symbol} & \textbf{typical value}
\\ 
\hline 
total number of DNA-histone contact sites & $N$ & 14 \\ 
no. of open contacts right of the right-most
protein-bound site on the right & $n_{1}$ & -- \\
no. of open contacts left of the left-most protein-bound
site on the right & $n_{2}$ & -- \\
position of right-most protein-bound contact on the left & $m_{1}$ & -- \\
position left-most protein-bound contact on the right & $m_{2}$ & -- \\
DNA-histone contact site attachment rate & $k_{\rm{on}}$  & 20-90s$^{-1}$
 \\
DNA-histone contact site detachment rate & $k_{\rm{off}}$  & $\sim 4$s$^{-1}$ \\
%\added{$\varepsilon = e^{-2} $}   \\
%
detachment rate of the final contact site & $k_{\rm{d}}$  & $\sim \koff$   \\
%\added{$s = e^{-2} $} \\
%
contact site binding free energy & $E_{\rm c}=\log (k_{\rm off}/k_{\rm
  on})$ & $-2$ \\
remodeler protein\added{-DNA} binding rate & $p_{\rm{a}}$  & -- \\
remodeler protein\added{-DNA} unbinding rate & $p_{\rm{d}}$  & -- \\
remodeler protein\added{-DNA} binding free energy & $E_{\rm p}=\log (p_{\rm
  d}/p_{\rm a})$ & -- \\ 
\hline
\end{tabular}
  \caption{Parameters and variables used in linear peeling,
intact-histone models.  The distances between the inner-most bound
contact and the inner-most remodeler-bound sites on the left and right
are defined as $n_{1}$ and $n_{2}$, respectively.  The distances from
the inner-most remodeler-bound sites to the left and right ends of the
$N$-total length contact segment are denoted $m_{1}$ and $m_{2}$,
respectively, as shown in Fig.~\ref{FIG1}.  In all subsequent
analyses, we will measure all energies in units of $k_{\rm
B}T$. \added{Since $\kon$ is the fastest rate in this system, our
models and analyses will typically be presented in dimensionless form
with rates measured in units of $\kon$ and dimensionless parameters
$\varepsilon \equiv \koff/\kon \ll 1$ and $s=\kd/\kon \ll 1$.}}
\label{tab:parameters}
\end{center}
\end{table*}

%\end{widetext}
%
\begin{figure}[h!]
\centering
\includegraphics[width=3.1in]{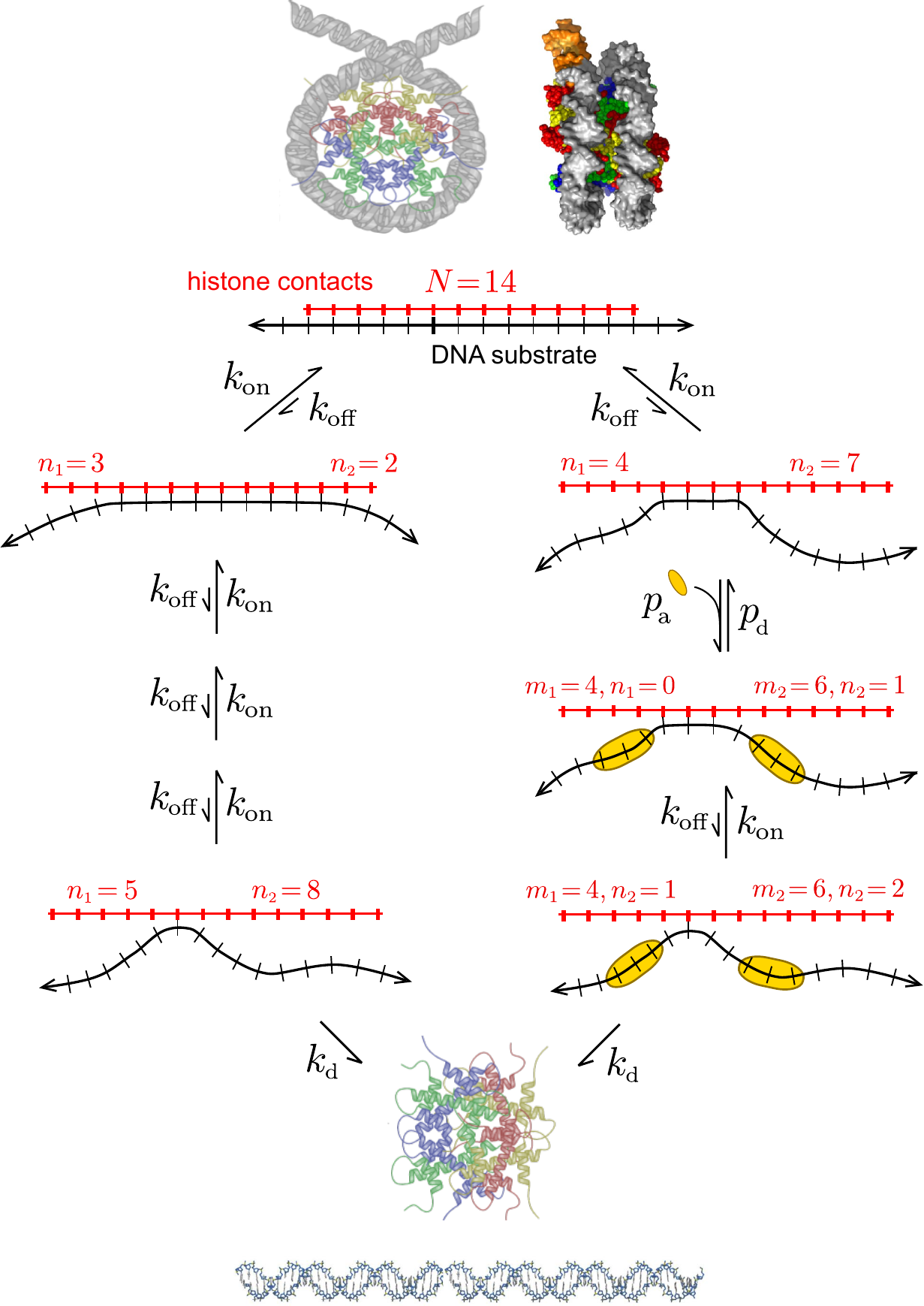}
\caption{A schematic of simple, intact-histone detachment. The 
unfacilitated and remodeler-facilitated pathways are shown on the left
and right, respectively. Top image shows en face and sagittal views of
a histone-DNA complex. Histone-DNA attachment points are described by
discrete sites on a one-dimensional lattice. In this example, we
illustrate $N=14$ contact sites, \added{evenly spaced by $\sim 10$ DNA
base pairs,} that each unbind and rebind with rates $\koff$ and
$\kon$, respectively. \added{Protein or ``remodelers'' (yellow) can
bind the DNA, occluding certain contacts sites and preventing them
from rebinding DNA. Thus remodelers generate a ratchet mechanism
accelerating nucleosome dissociation.} In the remodeler-assisted
model, $m_{1}$ and $m_{2}$ represent the number of cofactor-occluded
contact sites on the left and right, respectively, and $n_{1}$,
$n_{2}$ now represent the number of open contacts further to the right
and left of $m_{1}$ and $m_{2}$, respectively. Detachment of the final
contact occurs at rate $\kd$, which may be comparable to
$\koff$. \deleted{After complete dissociation, histones can rebind
from bulk solution at a rate $k_{\rm a}$ proportional to their bulk
concentration $C_{\rm b}$.}}
\label{FIG1}
\end{figure}
\subsubsection{Spontaneous histone-DNA detachment} Histone-DNA
interactions typically consist of $N \approx 14$ possible contact
sites. Each contact site on the DNA lattice may be in a bound (1) or
unbound (0) configuration.  If all contact sites are unbound at a
specific time, the histone can be considered to be completely
dissociated from the DNA at that time.  Due to steric constraints,
unbinding of the contacts will be assumed to occur sequentially from
either end, as depicted in Fig.~\ref{FIG1}. Thus, the only way an
interior site can be open is if all sites to the left or right of it
are in an unbound state.  In other words, histones can be peeled off
only from the ends of their contact footprint. Under this assumption,
the full configuration space $\{0,1\}^N$ can be reduced to a
bound-histone state space $\Omega=\{(n_1,n_2): n_1+n_2 < N\}$ and a
detached state $\Omega^{*}=\{(n_1,n_2): n_1+n_2 =N\}$, where $n_1$ and
$n_2$ denote the number of detached histone-DNA bonds at the two ends
of the histone-DNA contact footprint. In order to characterize the
timescale associated with complete disassembly, we assume that the
histone leaves the system once all contacts break.  This defines a
FPT problem to an ``absorbing'' detached state
$\Omega^*$.

%Eventually, rebinding of a histone from the bulk solution
%$(\Omega^{*} \to \Omega)$ can be treated as a separate process
%involving spatial diffusion, target search, and reformation of bound
%contacts, as schematized in Fig.~\ref{FIG2}(a).

\deleted{The detached state $\Omega^{*}$ might correspond to a histone with all
contact sites detached.  The master equation for this process is
determined by the transition matrix $\tilde{\Q}$ on the bound states
$\Omega$.}

\begin{figure}[h!]
\centering
\includegraphics[width=3.35in]{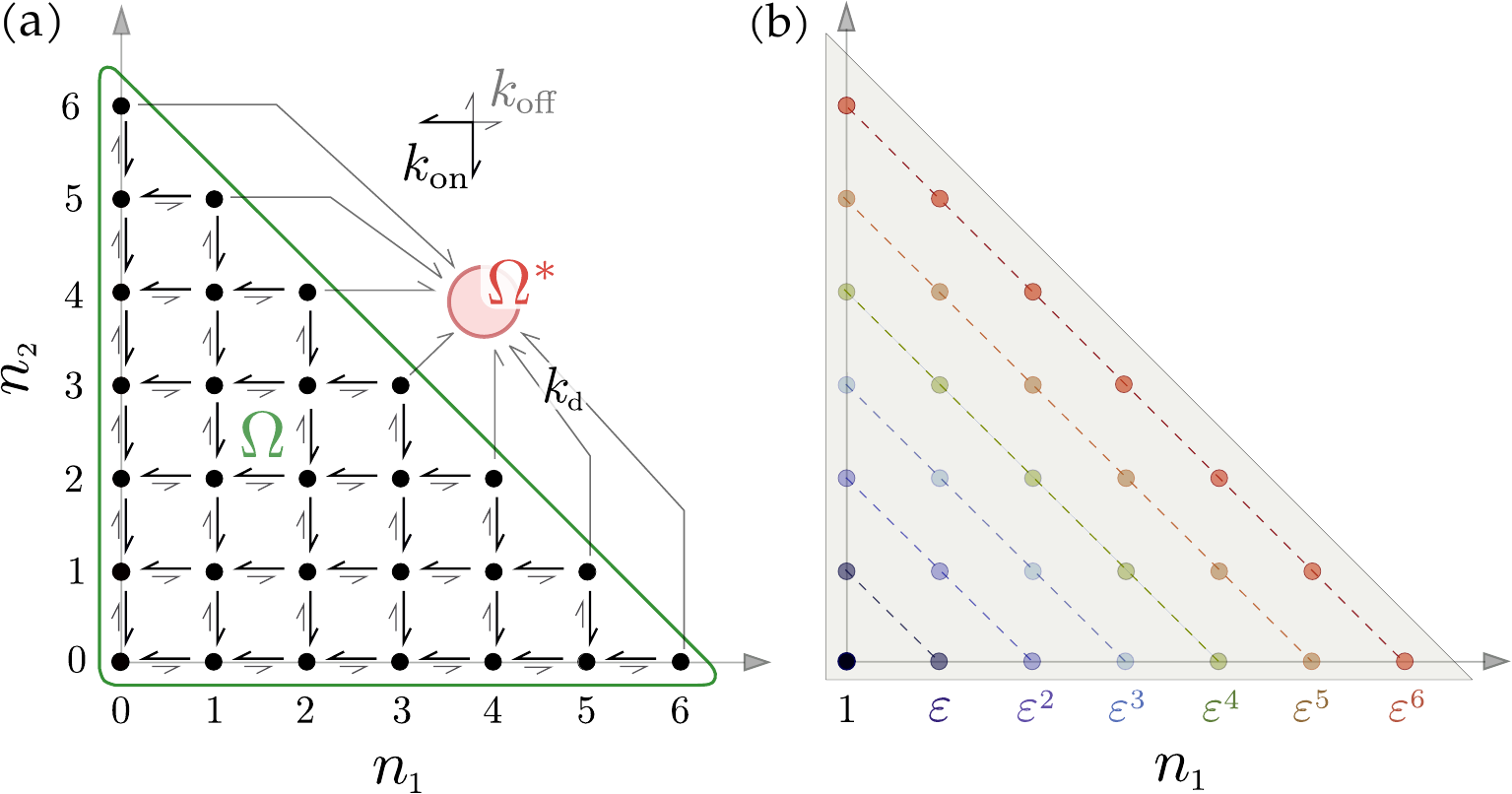}
\caption{\added{(a) Schematic of a hypothetical 
attached-histone} state space $\Omega$ for $N=7$ (seven contact
  sites). Since there are $N(N+1)/2=28$ bound states, the transition
  matrix $\tilde{\Q}$ is $28\times 28$.  Histone-DNA contacts increase
  and decrease by one with rate $\kon$ and $\koff$, respectively,
  except the last contact which breaks with rate $\kd$.  The
  completely detached absorbing state is indicated by
  $\Omega^{*}$.  \deleted{Rebinding from the bulk solution can occur
  at any of the $N$ contact sites at a rate $k_{\rm a}$, which will
  depend on histone bulk concentration. Details of histone rebinding
  from bulk solution are discussed in
  Appendix~\ref{Appendix:linear-relaxation-time}.} (b) For a strongly
  binding system confined to $\Omega$, $\varepsilon \equiv k_{\rm
  off}/k_{\rm on}\ll 1$, and a quasi-steady state distribution arises
  in which state probabilities $\sim \varepsilon^{n_{1}+n_{2}}$.  The
  most probable states are those with small $n_{1}+n_{2}$,
  corresponding to a tightly wrapped histone.}
%
% (d) the relaxation rate $\lambda_1$ as a function of rebinding rate
%  $k_{\rm on}$, numerical results compared with estimates.}
  \label{FIG2}
  \end{figure}

The state space and the transitions within it can be visualized by
random walks along the points in the triangular array along the $n_1$
and $n_2$ axes shown in Fig.~\ref{FIG2}(a).  The transitions are
driven by spontaneous detachment and attachment of single histone-DNA
bonds with possibly sequence- and position-dependent rates $k_{\rm
off}$ and $k_{\rm on}$, respectively.  We allow the dissociation rate
$k_{\rm d}$ of the final contact to be different from $k_{\rm off}$,
since no other DNA-histone contact holds the histone in place.  We
expect this final-contact detachment rate to have magnitude $k_{\rm
d} \sim k_{\rm off}$.
In bulk genomic DNA, most sequences have similar binding energies with
the histone octamer \cite{Lowary1997,LOWARY1998}. Thus, we first assume
homogeneity in histone-DNA contact site binding energies and uniform
association and dissociation rates $\kon$ and $\koff$.

\deleted{As detailed in
Appendix~\ref{Appendix:spontaneous-intact-histone},} We \deleted{can}
define a dimensionless transition matrix by dividing the master
equation by $k_{\rm on}$, which we assume to be the fastest
kinetic rate in our model. \added{As detailed in
Appendix~\ref{sec:matrices-si}}, the dimensionless transition matrix
$\Q = \tilde{\Q}/k_{\rm on}$ can be further decomposed as
\begin{equation}
  \Q(s) \coloneqq  \A + \varepsilon \B + s \C,
\label{QDEFN}
\end{equation}
where $\A$ represents the transitions in which one extra bond is
formed ($n_{1}+n_{2}$ decreases by one), $\B$ describes the
transitions of one bond being broken without leading to the detached
state, and $\C$ indicates the transitions involving the breaking of
the last contact, leading to the detached state.  Matrices involving
detachment, $\B$ and $\C$, are multiplied by the \deleted{Arrhenius}
\added{Boltzmann} factor $\varepsilon \equiv \koff/\kon 
\equiv e^{E_{\rm c}}$ and $s \equiv \kd/\kon$, respectively. 
Here, $E_{\rm c}$ represents the change in free energy of forming
contact site bond. For \added{strong-binding} contacts, $E_{\rm c} \ll
-1$, and $\varepsilon, s\ll 1$. \added{Physicochemical considerations
suggest $s \sim \ve$, but in our subsequent analysis, we allow $s$ to
vary independently of $\ve$.}
%
% \frac{k_{{\rm off}}}{k_{\rm on}}$.  $\A$ represents the transitions
% after which one extra bond is formed.  Elements in the matrix $\B$
% represent one bond at either boundary being broken \added{but there
% are still DNA-histone bonds in the system}.  Finally, the matrix
% $\C$ refers to transition in which the last bond is broken and the
% histone leaves the DNA segment of interest.
%
% The dimensionless transition matrix $\Q$ in Eq.~\eqref{QDEFN} allows
% for the detachment rate $\kd$ and final contact free energy $E_{\rm
% c}'$ of the last contact to be possibly different from those when
% more that one contact is intact.

\begin{figure}[htbp]
  \centering
\includegraphics[width=2.7in]{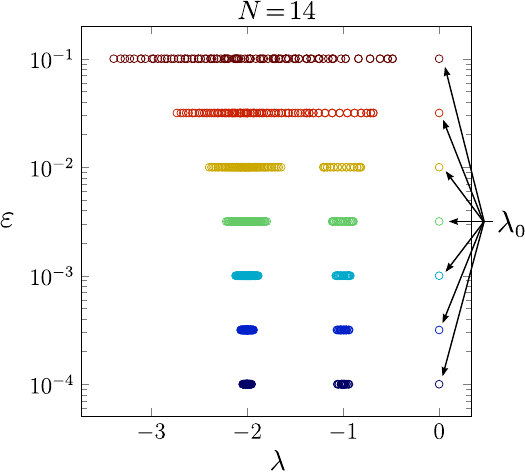}
\caption{Eigenvalues $\lambda$ of \added{the dimensionless} 
transition matrix $\Q = \tilde{\Q}/\kon$ associated with $\Omega$ for
    $N=14$ and $\varepsilon=s=0.1, 0.03, 0.01, 0.003,0.001, 0.0003$,
    and $0.0001$. The principal dimensionless eigenvalues are
    $\lambda_{0}\lesssim 0$, while two other groups cluster near $-1$
    and $-2$ as $\varepsilon \to 0$.}
\label{EIGEN0}
\end{figure}

We separate different detachment processes by $\B$ and $\C$ because
$\A + \varepsilon \B$ is the transition matrix of a reversible Markov
process, while the $s \C$ process describes full detachment into an
absorbing state, disrupts reversibility.
%
% First, we define all rates relative to $\kon$ and consider the
% unperturbed matrix $\Q_0=\A + \varepsilon \B$ whose eigenvalues can
% be characterized analytically.
%
$\A$ represents the binding reactions and is upper triangular with
eigenvalues $\{0,-1, \cdots, -1,-2, \cdots, -2\}$; hence, the
dimensionless eigenvalues of $\Q(0) \equiv \A + \varepsilon \B$
fall into three groups:
\begin{itemize}
  \item $\{\lambda: \lambda \sim O(\varepsilon) \lesssim 0  \}$, unique;
  \item $\{\lambda: \lambda \sim -1+ O(\varepsilon) \}$, degeneracy $2(N-1)$;
  \item $\{\lambda: \lambda \sim -2 + O(\varepsilon) \}$, degeneracy $\frac{(N-2)(N-1)}{2}$.
\end{itemize}
These groups of values are mainly controlled by the ``on-rate''
transition matrix $\A$ and control the pattern of the eigenvalues of
the full matrix $\Q(s) = \A + \varepsilon \B + s \C$.  Fig.~\ref{EIGEN0}
shows numerically computed eigenvalues of $\Q$ for different values of
$s=\varepsilon$. For sufficiently small $\varepsilon$, they fall into
the three clusters governed by $\A$.

The principal eigenvalue of $\Q$, $\lambda_{0}$, can be computed using
a two-step perturbation analysis. Adding $\varepsilon \B$ to $\A$
yields the matrix $\Q(0) \equiv \A+\varepsilon\B$, which represents
the internal transitions of the bound states $\Omega$, and as such,
has a unique eigenvalue $0$ and an associated equilibrium distribution
${\bf v}_0$ as its eigenvector.  Such internal transitions make the
system an irreducible and reversible Markov process. Therefore, the
equilibrium distribution \added{${\bf v}_0$} can be found
as \replaced{$v_0(n_1, n_2)$}{${\bf v}_0(n_1,n_2)$}$ \propto
\varepsilon^{n_1+n_2}$, \added{with $v_0(n_1,n_2)$ indicating
the component of ${\bf v}_0$ on the element
$(n_1,n_2) \in \Omega$. This scaling relation} indicates that for
small $\varepsilon$, the most probable states are those with small
$n_{1}+n_{2}$ (fully wrapped histones).

Applying perturbation theory to calculate \added{the principal
eigenvector ${\bf v}_0$ as a function of $s$,} ${\bf v}_0(s)$\added{,}
under the small change $\Q(0) \to \Q(0) + s \C$, one can see that each
component of ${\bf v}_0(s)$ is
\replaced{approximately}{shifted}
\replaced{$v_0(n_1,n_2;s)=\big(1+O(s)\big)v_0(n_1,n_2;0)$}{$x^{(i)}_0(s)
= (1+ O(s))x_{0}^{(i)}(0)$}, \added{as shown by
Eq.~\eqref{eq:x0_approx} in Appendix~\ref{sec:matrices-si-perturb}}.
Consequently, the eigenvalue structure of the perturbed matrix $\Q(0)
+s\C$ is preserved not only for $s \ll \varepsilon$, but also for
$s \sim \varepsilon$. Hence, we can use the
\replaced{principal}{equilibrium} eigenvector ${\bf v}_0(0)$ \added{at
equilibrium} to approximate the
principal \replaced{eigenvector}{eigenvalue} under the perturbation
$s\C$. This procedure of switching on an absorbing boundary on an
otherwise equilibrium system is commonly used to evaluate
FPTs of rare events, usually known as the
\textit{absorbing boundary method} or generalized Fermi's Golden rule
\cite{Nordholm1979}.  In the $\ve, s \to 0^{+}$ limit, we find (see
\added{Eq.~\eqref{eq:lambda_0_approx} in}
Appendix~\ref{sec:matrices-si-perturb}) the dimensionless
principal (largest) eigenvalue of the perturbed matrix $\Q(s) = \Q(0)
+ s \C$ to be approximately

\begin{equation}
\lambda_0(s) = -Ns\varepsilon^{N-1}\big[1+O(s)\big].
\label{EQUATION:PERTURBED}
\end{equation}
\added{After reintroducing the physical rate $\kon$, the 
eigenvalue $\tilde{\lambda}_{0} = \kon\lambda_{0}$ associated with
$\tilde{\Q}$ sets the slowest physical timescale representing the
effective rate of detachment from an equilibrium state.}
Eq.~\eqref{EQUATION:PERTURBED} can be motivated by
considering the barrier-crossing rate or probability flux
\added{, i.e. the transition rate multiplied by corresponding equilibrium
probability,} from an equilibrium state to the detached state
$\Omega^{*}$. The energy barrier confining the equilibrium state is
$(N-1)E_{\rm c}$ while there are $N$ transition states. Therefore, the
probability flux of disassembly is $\sim N \kd e^{(N-1)E_{\rm c}}$,
which corresponds to a dimensionless principal eigenvalue of
$Ns \varepsilon^{N-1}$.

The other eigenvalues $\lambda_{i>0}$ are ordered as $\lambda_0 \gg
\lambda_{1} \geq \cdots \geq \lambda_{\|\Omega\|}$. These other
eigenvalues reflect the faster timescales associated with other states
(eigenvectors).  The difference between the principal eigenvalue and
other eigenvalues, the \added{\textit{spectral gap}}, is an
important indicator of the dynamics of the system. If the system
starts in any initial configuration in $\Omega$, \added{and the
spectral gap is very large,} it will quickly (with rate $\sim \vert
\lambda_{i>0}\vert$) reach the near-equilibrium state ${\bf v}_0$
before ultimately dissociating with rate $\vert\lambda_{0}\vert$. As a
result, the mean first passage time (MFPT) from any initial bound
state ${\bf x}$ (such as ${\bf v}_0$) to the fully detached state
$\Omega^*$ can be approximated by finding the MFPT that is dominated
by the time from ${\bf v}_0$ to $\Omega^*$. \added{We find \textit{the
mean dimensionless nucleosome disassembly time} (the MFPT)}
\begin{equation}
  \mathbb{E}\big[T({\bf x})\big]
% \approx \tau_0
%
 \approx  \frac{1}{\vert
    \lambda_0\vert} \simeq \frac{1}{N s  e^{(N-1)E_{\rm
        c}}} \approx \frac{1}{N  e^{NE_{\rm
        c}}},
        \label{eq:3}
\end{equation}
\noindent \added{where the last approximation assumes $s \approx \ve$. Theoretical
justification and further discussion of this approximation are
provided in Appendix~\ref{sec:eigenvalues}, where Eq.~\eqref{eq:3} is
proved as Eq.~\eqref{eq:mean-first-passage-time-simple-case}.}
%

%
%The structure of the eigenvector ${\bf v}_0$ associated
%with the eigenvalue $0$ is shown in Fig.~\ref{FIG2}(c).
%
In the context of the histone problem, according to Li {\textit{et
al}.\cite{LI2005}, single histone-DNA binding sites are highly
dynamic, with an opening rate $\koff \sim 4 s^{-1}$ and a closing rate
$\kon \sim 20-90 s^{-1}$. According to Eq.~\eqref{EQUATION:PERTURBED},
this leads to an effective mean overall disassembly rate of
$\vert\tilde{\lambda}_0\vert \approx \tkon\vert \lambda_0\vert  \simeq N \tkon e^{-N E_{\rm c}}\approx  4.6 \times 10^{-8}{\rm s}^{-1}$,
corresponding to a \added{mean nucleosome disassembly time
$\mathbb{E}[T({\bf x})]/\kon \sim 15$ years. Typically, the
disassembly rate is defined by the inverse of the MFPT from the bound
state to the detached state.  In the case of multiple bound states, it
is not easy to define a simple measure of disassembly rate given the
complexity of the dynamics. A reasonable choice is to consider the
weighted average of MFPTs from all bound states, with weights given by
the (quasi-)equilibrium distribution of the bound states, which leads
to the strict identity between the disassembly rate and $1/{|
\lambda_0|}$. For a proof of this identity, see
Eq.~\eqref{eq:Ptot_eigenvalue}. Fortunately, in the histone
disassembly model, as we have argued above, the MFPT from all bound
states to the detached state are similar and thus $1/{|
\lambda_0|}$ is a reasonable measure of the overall disassembly rate.}

\added{In light of the above estimate for 
$\mathbb{E}[T({\bf x})]/\kon$, cells need to dynamically remodel their
histone binding patterns during DNA replication and changes in gene
expression, processes that occur on a much shorter timescale.}
Fortunately, a variety of intracellular remodeling factors, such as
SWI/SNF-type ATPases \cite{Fyodorov2002aug,
VargaWeisz2006apr,Chen2023jan,MOROZOVMBOC}, can catalyze this
remodeling process. Next, we will extend our model to incorporate
mechanisms of remodeling cofactors that can compete for DNA or histone
contacts.

\deleted{Finally, rebinding of histones from bulk solution is also an important
process, particularly when investigating nucleosome reassembly, after,
\textit{e.g}, disassembly or DNA replication. When rebinding is
allowed, as indicated in Figs.~\ref{FIG1} and \ref{FIG2}(a) by the
rebinding steps with rate $k_{\rm a}$, there is no longer an absorbing
state and the states within the system $\Omega \cup \Omega^*$ are
\textit{reversible}. The largest eigenvalue is then $\lambda_{0} = 0$
(associated with an equilibrium configuration distribution) and the
spectral gap allows us to determine the dynamics of the system's
relaxation to equilibrium, particularly the relaxation time $T_{\rm
r}$, as we consider in detail in Appendix
\ref{Appendix:linear-relaxation-time}.
}

% This approximation facilitates the derivation of simple estimates
% for the relaxation time of the \textit{reversible} system $\Omega
% \cup \Omega^*$, as illustrated in Fig.~\ref{FIG2}(a), using a
% coarse-grained model shown in
% Fig.~\ref{fig:coarse-grained-model}(a).

\subsubsection{Remodeler-facilitated linear detachment}

Regulation of histone-DNA binding and acceleration of disassembly by
other proteins/cofactors can be achieved in two ways: (i) competitive
binding of proteins may block reattachment of histone contact sites to
DNA and (ii) cofactors may allosterically inhibit histone-DNA binding.
Recent studies suggest that a number of DNA-binding proteins interact
with the histone-DNA complex by competing for open contact
sites \cite{Bundschuh2011,Clement2022feb,Makasheva2022feb,McCauley2022feb}.
Here, we model such \replaced{a}{as} mechanism via ratcheted blocking mechanism
whereby nucleosome remodeler proteins block rebinding of DNA, thereby
facilitating disassembly.  The second, allosteric mechanism can be
modeled directly by modification of site binding and unbinding rates
$k_{\rm on}$ and $k_{\rm off}$. Therefore, allostery can be subsumed
under the spontaneous disassembly model. In the following discussion,
we will focus on the blocking mechanism and refer to the intervening
cofactor as a nucleosome remodeler. We develop a model that can be
applied both to proteins that slide along DNA and to those that directly
bind and occlude DNA-histone contact sites. While most known
nucleosome remodelers are ATPases that slide along DNA
\cite{Fyodorov2002aug,VargaWeisz2006apr,Chen2023jan}, our model is
also intended to describe the general interaction between DNA-binding
proteins and the nucleosome and to better understand why other
proteins cannot effectively evict histones from DNA.

Assume nucleosome remodeler proteins compete with histones on the same
DNA binding sites and have binding rates $\pa$ and dissociation rates
$\pd$, as illustrated in Fig.~\ref{FIG1}.  Bound contact sites must
detach before cofactors such as remodeling protein can bind. However,
if a remodeler first binds to and occludes a DNA or histone contact
site, this site is unavailable for histone reattachment or binding,
promoting histone detachment. We describe the state of
DNA-histone-remodelers by a four-integer tuple $(m_1,m_2,n_1,n_2)$. In
this enumeration, $m_{1}$ and $m_{2}$ are the rightmost and leftmost
contact sites occluded by a remodeler protein measured from the left
and right ends of the contact footprint. These remodelers can bind
to either the histone or the DNA substrate as shown in Fig.~\ref{FIG1}. In
the presence of bound remodeler proteins ($m_{1}>0$ and/or $m_{2} >0$),
the remaining available sites for direct DNA-histone interactions will
be reduced to $N-m_1-m_2$. The associated state space of $(n_1,n_2)$
is then reduced correspondingly. In the presence of bound remodelers,
$n_{1}$ and $n_{2}$ now measure the unbinding progress of the histone
octamer and represent the additional numbers of opened binding sites
inward from the rightmost and leftmost remodeler binding site. In this
notation, the fully detached state is visited only when
$m_{1}+n_{1}+m_{2}+n_{2} = N$.

Since $(m_1,m_2)$ accounts only for the most inwardly occluded contact
sites, the information about remaining remodelers is not delineated in
this state space $\Omega_{\rm p} \coloneqq \{(m_1,m_2,n_1,n_2)\in
\mathbb{N}^4:m_1+m_2+n_1+n_2 < N\} \cup \Omega^*_{\rm p} \coloneqq
\{(m_1,m_2,n_1,n_2)\in \mathbb{N}^4:m_1+m_2+n_1+n_2 = N\}$. \added{In
the following, we will use the subscript ``p'' to indicate quantities
associated with the remodeler-facilitated disassembly model.}
Consequently, the full remodeler adsorption pattern is not fully
captured by $m_1$ and $m_2$. Multiple cofactors could cooperatively
bind \added{(where a DNA-bound cofactor accelerates binding of another
cofactor near it)} and compete for open sites amongst themselves,
leading to complex dynamics of assisted histone displacement.  We can
simplify the model by considering, \textit{e.g.},
\replaced{stepwise}{incremental} increases of $(m_{1}, m_{2})$,
\added{in which case $m_1,m_2$ can only change by 1 at a time}.
This restriction is appropriate for remodelers that are motor proteins
processing along DNA \cite{chou_2007} and yields an overall upper
bound to remodeler-facilitated disassembly rates.  Since molecular
motors such as SWI/SNF complexes typically attack nucleosomes from one
side, we explicitly modify our formulae in Appendix
\ref{Appendix:one_sided} to account for one-sided peeling.

Within the undissociated state space $\Omega_{\rm p}$, the transition
matrix $\H$ can be constructed from matrices defined in the previous
section. Because of occlusion by remodelers, histone detachment can
now occur after spontaneous separation of $n \leq N$ binding sites.
We denote the spontaneous transition matrix with $n$ binding sites as
$\Q_n$ and define $\Q_{n:m}$ to be block diagonal with $m$ $\Q_n$s on
the diagonal.  By arranging the states $(m_1,m_2,n_1,n_2)$ as
described in Appendix \ref{Appendix:linear_facilitated_Q}, the
transition matrix $\Q_{\rm p} = \Q_{N,{\rm p}}$ can be written as

\begin{widetext}
\begin{equation}
%\begin{aligned}
\Q_{N,{\rm p}} = \left[\begin{array}{ccccc}
      \Q_N & \0  & \cdots & \0 \\
      \0 & \Q_{N-1 :2}  & \ddots & \vdots \\
      \vdots  & \ddots & \ddots & \0 \\
      \0  & \cdots & \0 & \Q_{1 : N}
      \end{array}\right]  + \frac{\pa}{\kon} \left[\begin{array}{ccccc}
      {\bf M}_N & \0 & \cdots  & \0 \\
      {\bf M}_{N-1, N} & {\bf M}_{N-1} & \ddots & \vdots \\
      \vdots & \ddots & \ddots & \0 \\
      {\bf M}_{1, N}  & \cdots & {\bf M}_{1,2} & \0
      \end{array}\right] + \frac{\pd}{\kon}
      \left[\begin{array}{ccccc}
        \0 & {\bf G}_{N, N-1} & \cdots  & {\bf G}_{N,1} \\
        \0 & {\bf G}_{N-1}  & \ddots & \vdots \\
        \vdots  & \ddots & \ddots & {\bf G}_{2,1} \\
        \0  & \cdots & \0 & {\bf G}_1
        \end{array}\right]. \label{HN}
%  \end{aligned}
\end{equation}
\end{widetext}
\noindent \added{In Eq.~\eqref{HN}, the states are grouped by the sum of
$m_1+m_2$ in ascending order. The first block entry represents the
states with no remodeler bound, the second block entry represents the
states with one remodeler bound, and so on.} The transition matrices
${\bf M}_{m,n}$ and ${\bf G}_{m,n}$ describe changes in state
associated with remodeler binding and unbinding, respectively, and
\added{the explicit construction rules of $\Q_{N,{\rm p}}$ are given
by Eqs.~(\ref{eq:MG_Motor}-\ref{eq:MG_Binding})} in
Appendix \ref{Appendix:linear_facilitated_Q}. Specifically, column
sums of ${\bf M}_{m,n}$ and ${\bf G}_{m,n}$ are zero, reflecting
conservation of probability.
We will again employ perturbation theory to find approximations for
the principal eigenvalue. The unperturbed process corresponds to $\pa
= \pd = 0$. Even though there are multiple eigenvectors associated
with the eigenvalue $0$ of the matrix $\Q_{N,{\rm p}}$ with $\pa = \pd = 0$, we
are interested only in the eigenvector that embeds the previous
eigenvector ${\bf v}_0$ of $\Q_N$. The embedding is implemented by
appending zeros to the end of the original ${\bf v}_0$.  This new
${\bf v}_0$ serves as the starting point of our subsequent
perturbation analysis.

We will classify scenarios based on the ability of remodelers to
occlude a binding site, defined by the remodeler\added{-DNA} binding
energy $E_{\rm p} = \log(\pd/\pa)$.  An $E_{\rm p} > 0$ indicates $\pd
> \pa$ and a weakly binding remodeler; negative $E_{\rm p} < 0$ means
an attractive remodeler-DNA interaction and strong remodeler
binding. Remodeler proteins compete directly with histones for DNA
contact sites; this competition is quantified by comparing $E_{\rm p}$
to $E_{\rm c}$. If $E_{\rm p} > E_{\rm c}$, histone-DNA binding
is \textit{stronger} than remodeler binding; if $E_{\rm p} < E_{\rm
c}$, histone-DNA binding is weaker than remodeler binding. The complex
state space and parameters of this problem, however, do not allow for
simple analytical solutions.

\deleted{We briefly present the main results of
this model in limiting cases and leave the technical details to
Appendix \ref{Appendix:linear_facilitated_Q}.}

\vspace{2mm}
\noindent \textit{Weak remodelers - }In the the weak remodeler binding
limit ($E_{\rm p} > E_{\rm c}$), the eigenvector corresponding to the
largest eigenvalue is only weakly perturbed by the presence of
remodelers but we can still use the total binding energy
\replaced{$E(m_1,m_2,n_1,n_2)=\left[\left(m_1+m_2\right)
\left(E_{\mathrm{p}}-E_{\mathrm{c}}\right)-\left(n_1+n_2\right)
E_{\mathrm{c}}\right]$}{$E(n_1,n_2,m_1,m_2)$} associated with each
state $(m_1,m_2,n_1,n_2)$ to approximate the steady state distribution
\added{${\bf v}_0$ via the Boltzmann relation $v_0(m_1,m_2,n_1,n_2)\propto
e^{-E(m_1,m_2,n_1,n_2)}$ \cite{gibbsmeasure}}. \added{The principal
eigenvalue can be found by via relation}
%
% $\lambda_0={{\bf 1}^{\intercal}
% \Q_{N,{\rm p}} {\bf v}_0}/{{\bf 1}^{\intercal} {\bf v}_0}$.} 
%
\deleted{To estimate
the principal eigenvector ${\bf v}_0$, we employ the quasi-steady
state assumption by considering the probability distribution on
$\Omega_{\rm p}$ given by ${\bf v}_0(\omega), ~\forall \omega\in
\Omega_{\rm p}$, the \textit{steady-state} distribution.  In this
case, the probability of the system being in state $(m_1,m_2,n_1,n_2)$
is proportional to $\exp[(m_1+m_2)(E_{\rm p}-E_{\rm
c})-(n_1+n_2)E_{\rm c}]$.}
\begin{equation}
  \begin{aligned}
   \lambda_0 &= \frac{1^{\intercal}  \Q_{N,{\rm p}}
     {\bf v}_0}{1^{\intercal} {\bf v}_0}, \\
    & \approx \frac{\sum_{({\bf m,n})} \sum_{({\bf m',n'})}
    W_{N,{\rm p}}({\bf m'},{\bf n}', {\bf m}, {\bf n} )e^{-E({\bf m},{\bf n})}}
    {\sum_{({\bf m,n})} e^{-E({\bf m},{\bf n})}},
  \end{aligned}
  \label{eq:lambda0_ss_long_exp}
\end{equation}
\noindent \added{where $({\bf m}, {\bf n})$ represents
the tuple $(m_1,m_2,n_1,n_2)$ and $({\bf m',n'})$ represents the tuple
$(m'_1,m'_2,n'_1,n'_2)$. ${W}({\bf m',n',m,n})$ represents the
transition rate from state $({\bf m,n})$ to state $({\bf m',n'})$.}

\added{We proceed to simplify the expression in
Eq.~\eqref{eq:lambda0_ss_long_exp}.} At steady state, the most
probable configuration is fully bound: $(m_1,m_2,n_1,n_2) =
(0,0,0,0)$, \added{and other states are much less likely.} The
boundary states with positive \replaced{transition rate}{flux} to full
disassembly are characterized by the condition $m_1+m_2+n_1+n_2 =
N-1$.  \added{States $({\bf m,n})$ away from the boundary satisfies
$\sum_{({\bf m',n'})}W({\bf m',n',m,n})=0$ because of conservation of
probability. States $({\bf m,n})$ on the boundary satisfies
$\sum_{({\bf m',n'})}W({\bf m',n',m,n})=-s$.} When $E_{\rm p} > 0$,
the most probable boundary states are still those with $m_1 = m_2 = 0$
and $n_1 + n_2 = N-1$, whose probability is proportional to
$e^{(N-1)E_{\rm c}}$.  When $E_{\rm p} < 0$, the most probable
boundary states become those with $m_1 + m_2 =N-1$ and $n_1 =n_2 =0$,
whose probability is proportional to $e^{(N-1)(E_{\rm c}-E_{\rm p})}$.
In both cases, there are $N$ identical most-probable boundary states
because the attack comes from both ends, forming a triangular state
space. \added{Instead of investigating every state $({\bf m,n})$, we
simplify the expression in Eq.~\eqref{eq:lambda0_ss_long_exp} by
considering only the state $({0,0,0,0})$ with energy $0$ and relative
weight $1$, and $N$ boundary state\added{s} with energy $(N-1)(E_{\rm
p}^- - E_{\rm c})$ and weight $e^{(N-1)(E_{\rm p}^- -E_{\rm
c})}$. Here, $E_{\rm p}^{-} \coloneqq \min \left\{E_{\rm p},
0 \right\}$ to account for different most-probable boundary states
under different conditions.} \replaced{With this approximation, we
derive a physical estimate of the principal eigenvalue by summing
Eq.~\eqref{eq:lambda0_ss_long_exp} over the $N+1$ most probable states
in the interior and on the boundary}{An approximation of the principal
eigenvalue is given by} 
\begin{equation}
  \lambda_0 \approx \hat{\lambda}_{0}(E_{\rm p}  > E_{\rm c}) \coloneqq  
  -\frac{s N e^{(N-1) (E_{\rm c}- E_{\rm p}^{-})}}{1+Ne^{(N-1) (E_{\rm c}- E_{\rm p}^{-})}}.
  \label{eq:intact_weak_facilitation}
\end{equation}
\noindent \added{If} $E_{\rm p}>0$, Eq.~\eqref{eq:intact_weak_facilitation}
reduces to the spontaneous disassembly rate \added{given} in
Eq.~\eqref{EQUATION:PERTURBED} (since $E_{\rm c}\ll -1$).
\added{A more refined approximation of Eq.~\eqref{eq:lambda0_ss_long_exp}
that sums over more states is provided in
Eq.~\eqref{eq:weak_perturbation2}.} \deleted{where $E_{\rm
p}^{-}\coloneqq \min\{E_{\rm p},0\}$}

\vspace{2mm}
\noindent \textit{Strong facilitation limit -- }In the 
$E_{\rm p} \rightarrow -\infty$ limit, corresponding to irreversible
remodeler binding ($p_\textrm{d} \to 0$), the structure of the
principal eigenvector ${\bf v}_0$ embedded in $\Omega_{\rm p}$ is
preserved under small a perturbation ($p_{\rm a} \ll
\kon$) \added{as shown in  Eq.~\eqref{EXPRESSION:IRR_APP}},
\begin{equation}
\label{EXPRESSION:IRR}
{\bf v}_0(p_{\rm a}) =\left[1+O(\ve)+
O\big(\tfrac{\pa}{\kon}\big)\right]{\bf v}_{0}(0).
\end{equation}
We can then use ${\bf v}_{0}(p_{\rm a})$ in the relation $ {\bf
v}_0^{\intercal} \Q_{N,{\rm p}} {\bf v}_0 = \lambda_0 \|{\bf
v}_0\|^2_2$ to extract \added{an estimate of} dimensionless principal
eigenvalue \added{(see Eq.\eqref{EQUATION:IRR_APP})}
\begin{equation}
\begin{aligned}
% \lambda_0(p_{\rm a}) \approx
& \hat{\lambda}_{0}(E_{\rm p}\! \rightarrow -\infty, p_{\rm a}\! \ll\! \kon) \\
& \hspace{1.2cm} \coloneqq -\Big[Nse^{(N-1) E_{\rm c}}\!
+\frac{p_{\rm a}}{\kon}\!\sum_{j=1}^{N-1}(j+1)e^{j E_{\rm c}}\Big],
%
%\Big[1+O\big(\frac{k_{\rm off}+p_{\rm a}}{k_{\rm on}}\big)\Big]
\label{EQUATION:IRR}
\end{aligned}
\end{equation}
valid when $p_{\rm a} \ll k_{\rm on}$.

When $p_{\rm a} \sim k_{\rm on}$, the most probable state moves to the
boundary since one may consider $p_{\rm d}=0$ as a limit of $E_{\rm
p} \rightarrow -\infty$, in which case the boundary states carry the
lowest energy. Although the probability distribution is no longer
proportional to $e^{-E}$, it provides intuition for the behavior of
the system in this limit. The rate-limiting step is the one-step
unbinding with rate $\kd$.  Therefore, the perturbed principal
eigenvalue $\lambda_{0}(p_{\rm a})$ is given by
\begin{equation}
\begin{aligned}
\hat{\lambda}_{0}(E_{\rm p}\! \rightarrow -\infty)
\coloneqq \max\left\{\hat{\lambda}_{0}(E_{\rm p}\!\rightarrow -\infty, p_{\rm a}\! \ll\! \kon), -s\right\}\!.
  \label{eq:intact_strong_facilitation}
\end{aligned}
\end{equation}
Since the most probable state is shifted from those in the interior to
those at the boundary, the principal eigenvalue approximates the
inverse of the MFPT to $\Omega^*$ starting near the
boundary.  On the other hand, starting from the fully bound state, the
system will first take an average time $(N-1)/\koff$ to reach the
boundary \added{in the $p_{\rm a} \gg \kon$ limit}. Although
MFPTs to the disassembled absorbing state differ for
different initial configurations, for \replaced{$E_{\rm c} \ll -1~
(\varepsilon \ll 1)$}{$E_{\rm c} \ll 0$ ($\varepsilon \ll 1$)}, they are all
on the same order of magnitude determined by the unbinding rate
$\koff$ and $\kd$.

% In terms of reversible binding, we use the equilibrium flow intensity
% $j(\Omega^*|\Omega)$ as an (upper bound) estimate of
% $\lambda_0^{\Omega}$ for the detachment process.

\vspace{2mm}
\noindent \textit{Effective facilitation -- }We have characterized the
principal eigenvalue in the case of weak facilitation $E_{\rm p} >
E_{\rm c}$ and strong facilitation limit $E_{\rm p} \rightarrow -
\infty$. Of interest is the very typical intermediate case $E_{\rm p} < E_{\rm c}$
as it can effectively contribute to remodeling.  However, in this
limit, simple analytic approximations cannot be found, and we must
compute the eigenvalues numerically. \added{By using established
numerical methods for evaluating the eigenvalues
in \texttt{JuliaLang}\cite{Bezanson2017jan},} we find that the
principal eigenvalue under intermediate $E_{\rm p} < E_{\rm c}$ is
larger (smaller magnitude) than that of the strong facilitation limit
$E_{\rm p} \to -\infty$ given by
Eq.~\eqref{eq:intact_strong_facilitation}. The strong facilitation
limit leads to shorter nucleosome disassembly times.  Moreover, the
right-hand side of Eq.~\eqref{eq:intact_weak_facilitation} can be
identified as approximately the probability flux intensity
\added{$j(\Omega^*_{\rm p}\,\vert\,\Omega_{\rm p}) \coloneqq
\sum_{{\bf x}\in \Omega_{\rm p}, {\bf y}\in \Omega_{\rm p}^{*}}
{W}_{\bf y,x} {v}_0({\bf x})/\sum_{\bf x \in \Omega} v_0({\bf x})$
into the detached state $\Omega^*_{\rm p}$} from a quasiequilibrium
configuration \added{${\bf v}_0$} in $\Omega_{\rm p}$. It is
well-known that the principal eigenvalue is always dominated by the
flux intensity \cite{aldous2002}. Consequently, we can obtain an
overall upper bound on the facilitation effect as the maximum of the
two analytic approximations given by
Eqs.~\eqref{eq:intact_weak_facilitation} and
\eqref{eq:intact_strong_facilitation}:
\begin{equation}
  \hat{\lambda}_{0,\textrm{p}}
 \added{\coloneqq} \max\left\{\hat{\lambda}_{0}(E_{\rm p}\! >\!
  E_{\rm c}),\, \hat{\lambda}_{0}(E_{\rm p}\! \rightarrow
  -\infty)\right\}.  
\label{eq:intact_effective_facilitation}
\end{equation}

Further mechanistic insight can be gained via a coarse-grained
model \added{shown in Fig.~\ref{fig:coarse-grained-model},} that
ignores the fine structure of histone-DNA interaction by projecting
the original undissociated state space $\Omega_{\rm
p}=\{(m_1,m_2,n_1,n_2):m_1+m_2+n_1+n_2 < N\}$ onto
$\widetilde{\Omega}_{\rm p}\coloneqq\{(m_1,m_2): m_1+m_2 < N\}$.
\added{Justification of this approximation is provided in
Appendix~\ref{sec:eigenvalues} while
Appendix~\ref{sec:remodeler_coarse-graining} provides some physical
intuition and discussion.}  Since we now track the transition of the
states of only the nucleosome remodelers, the structure of this
coarse-grained model resembles the original spontaneous linear
detachment model, as shown in Fig.~\ref{fig:coarse-grained-model}(a),
where the effective \deleted{binding and unbinding} rates \added{$\pd$
and $\pa e^{E_{\rm c}}$ can be intuitively explained by considering
the fine structure within a coarse-grained state shown in
Fig.~\ref{fig:coarse-grained-model}(b).}

\begin{figure}[htbp]
  \centering
\includegraphics[width=3.25in]{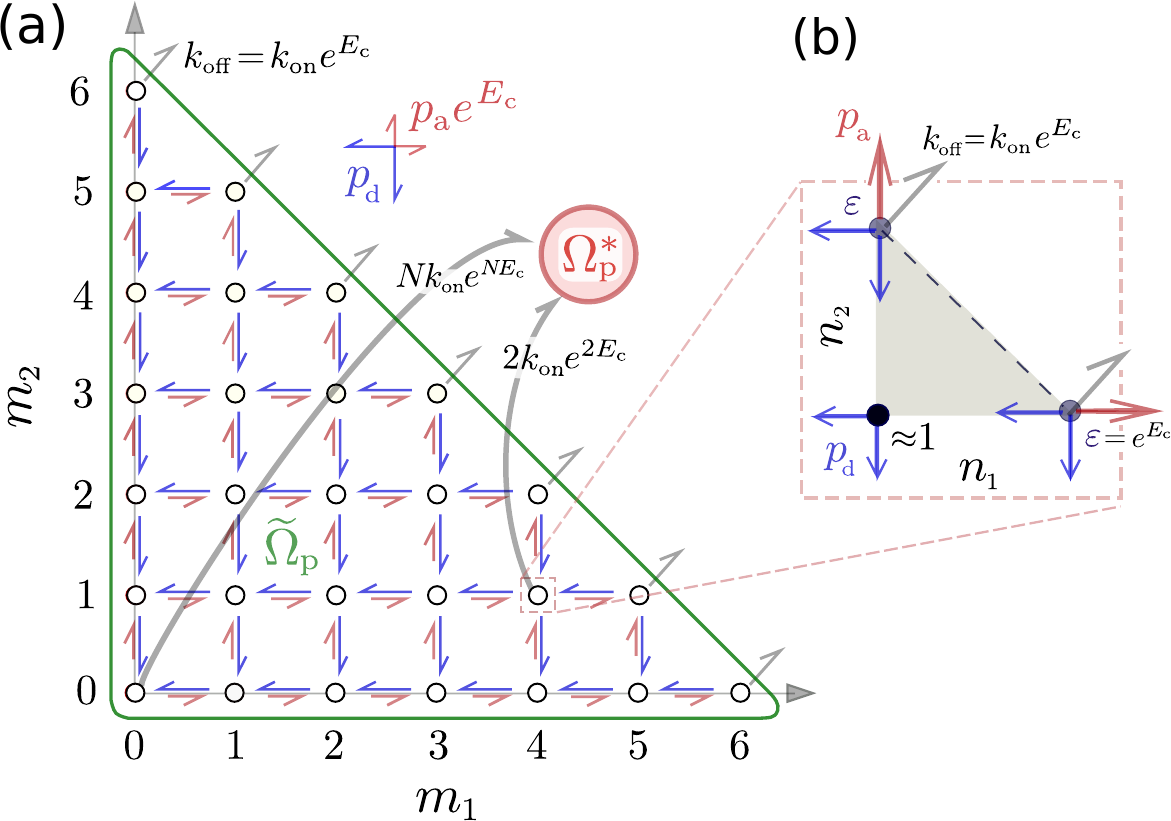}
\caption{A simple coarse-grained approximation of the
\emph{facilitated} \deleted{and \emph{unfacilitated}}
intact-histone model.  (a) The state space $\widetilde\Omega_{\rm p}$
and $\Omega^*_{\rm p}$ for the coarse-grained model for the linear
facilitated detachment. Each node represents the DNA occupancy $(m_1,
m_2)$ of remodeling factors.  Red and blue arrows represent effective
transitions corresponding to invading and retreating leading
remodelers, respectively.  The gray arrows (not all shown) represent
the transition from the coarse-grained state $(m_1,m_2)$ to the fully
dissociated state $\Omega^*_{\rm p}$ with rate $(N-m_1-m_2){k_{\rm
on}}e^{(N-m_1-m_2)E_{\rm c}}$, where we have assumed $k_{\rm
d}=\koff$. (b) \added{An illustration of the ``internal structure''
$\{(n_1,n_2):n_1 +n_2 \leq 1\}$ within a coarse-grained state
$(m_1,m_2)$ in which $m_1 + m_2 = N-2$. The internal dynamics are much
faster than the transitions to external states indicated by different
arrows in the schematic.  The fast internal state is well
characterized by a quasi-steady state distribution
$v_{0}(n_{1}=0,n_{2}=0)\approx 1$, $v_{0}(1,0)\sim
v_{0}(0,1)\approx \varepsilon=e^{E_{\rm c}}$, effectively lumping the
state space shown in Fig.~\ref{FIG2}(b) into one with two binding
sites. The $\varepsilon$-probability states are allowed to transit to
$\Omega^*_{\rm p}$ with rate $k_{\rm off}$, and remodelers may bind to
the exposed DNA with rate $p_{\rm a}$ in this case.  If the internal
states are in $(0,0)$, the remodeler cannot bind to DNA and no direct
transition to $\Omega^*_{\rm p}$ is allowed. In all these states, the
bound remodeler may detach with rate $p_{\rm d}$. Multiplying the
steady state probability of the internal structure and the
corresponding transition rate yields the effective transition rates
shown in (a). For example, binding of additional remodelers to DNA
requires exposed DNA. Therefore, transition to higher $(m_1,m_2)$ is
not allowed when the internal state is $(0,0)$. The probability of at
least one site being exposed is $\sim e^{E_{\rm c}}$, resulting in an
effective remodeler-DNA binding rate $p_{\rm a}e^{E_{\rm c}}$. For
remodeler unbinding, there is no restriction on the internal state,
and the effective unbinding rate is $p_{\rm d}$.} \deleted{The
coarse-grained model for the linear unfacilitated case can be obtained
as a limit of $\pa \rightarrow 0$ in the linear facilitated case. In
this limit, the undissociated state space is composed of a single
state $\widetilde{\Omega}=(0,0)$. We incorporate rebinding of the
histone from solution with rate $N k_{\rm a}$ into the model.}}
\label{fig:coarse-grained-model}
\end{figure}

Finally, to capture a crucial structural feature of histone-DNA
interactions, we incorporate into the coarse-grained model an
additional hopping rate from state $(m_1,m_2)$ to the fully detached
state $\Omega^*_{\rm p}$, given by $(N-m_1-m_2)k_{\rm
on}e^{(N-m_1-m_2)E_{\rm c}}$, assuming $k_{\rm d}=k_{\rm off}$.  These
hopping transitions to $\Omega^*_{\rm p}$ are inconvenient to
visualize completely in Fig.~\ref{fig:coarse-grained-model}(a), so we
indicate only three effective transitions. 

As shown in Fig.~\ref{FIG:RESULTS}, we compare the numerical
eigenvalues predicted by the coarse-grained model to those of the full
model. The coarse-grained model approximates the original model well
in all regimes of $E_{\rm p}$ provided $\pd \ll \kon$; however,
analytic approximations to the principal eigenvalue are still
inaccessible.
\begin{figure}[htbp]
\begin{center}
  \includegraphics[width=3.4in]{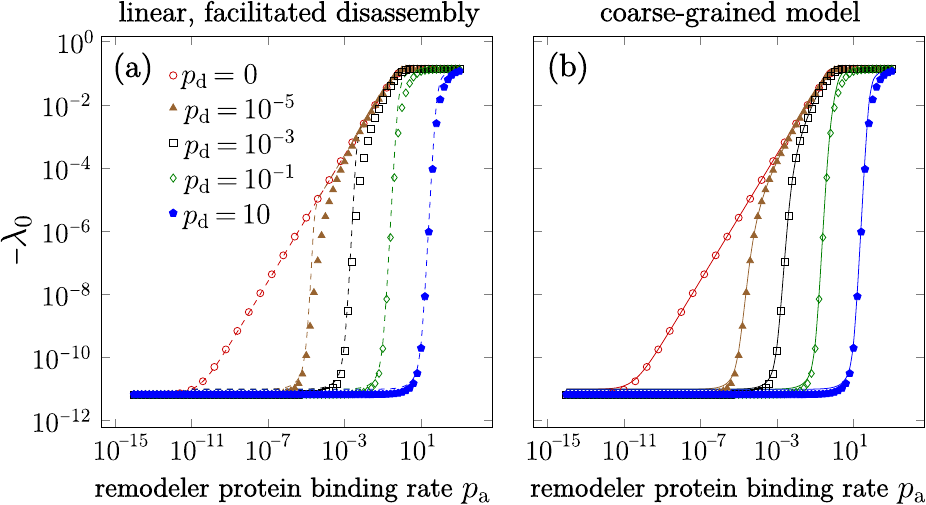}
\end{center}
\caption{Values of $-\lambda_{0}$ (principle eigenvalue of the
transition matrix ${\bf H}_{N}$), \added{a surrogate for the
disassembly rate of nucleosome under remodeler facilitation,} were
numerically computed (symbols, both panels). (a) Numerically computed
eigenvalues $-\lambda_{0}$ are compared to the approximation in
Eq.~\eqref{eq:intact_effective_facilitation} (dashed lines) using
$E_{\rm c}=-2$. \textit{Here, and in all subsequent plots, all rates
are normalized with respect to $\kon$}. (b) The same numerically
computed values of $-\lambda_{0}$ are compared to the numerical
predictions of the coarse-grained model (solid lines) indicating the
accuracy of our coarse-graining.}
\label{FIG:RESULTS}
\end{figure}

Summarizing, our simplified model describing processive motors and
nucleosome remodelers that bind strongly and cooperatively assumed
stepwise transitions of $(m_1, m_2)$. For remodelers that bind
independently, the values of $(m_{1}, m_{2})$ can undergo
longer-ranged jumps as multiple cofactors bind. \deleted{However, the
probability of increasing $m_{1}$ or $m_{2}$ by $\Delta m$ due to
remodeler binding is at most proportional to $e^{\Delta m E_{\rm
c}}$,} \added{Under quasi-steady state conditions, the probability of
exposing $\Delta m$ DNA-binding sites for remodeler binding is
proportional to $e^{\Delta m E_{\rm c}}$; thus, the probability of
increasing $m_{1}$ or $m_{2}$ by $\Delta m$ due to remodeler binding
is at most proportional to $e^{\Delta m E_{\rm c}}$. The probability
of decreasing a certain number of sites depends on the position of the
trailing remodelers and hence on the bulk remodeler concentration.}

When remodeler binding is strong ($E_{\rm p}$ is very negative) or
cooperative, $m_{1}$ and $m_{2}$ will \replaced{seldom}{not often}
make large jumps so their dynamics can be treated as stepwise. On the
other hand, \replaced{when $p_{\rm a} \leq p_{\rm
d}$}{when remodeler binding is weak}, facilitation is minimal since
the rate-limiting step is spontaneous peeling. \replaced{Even for
independent remodelers with weak binding energy}{In this
scenario}, \replaced{the stepwise model predicts the numerically
computed full-model disassembly rate reasonably well despite the
possibility of large jumps to lower $(m_1,m_2)$ states.}{the stepwise
model is again quantitatively reasonable.}  Variances in our
predictions under randomly distributed histone-DNA contact energies is
considered in Appendix
\ref{Appendix:random_landscapes}. Remodelers that slide along DNA,
such as DNA replication machinery, typically \added{attack the
nucleosome from outside the contact footprint.}  \deleted{In
Appendix~\ref{Appendix:one_sided}, we restrict the model to one-sided
peeling.}

\subsection{Multimeric nucleosome disassembly model}
% Based on recent biochemical studies on the structure of histone and
% DNA, we propose a more realistic multimeric nucleosome disassembly model.
% This model may explain some experimental findings that modifying
% interactions between histone modules can significantly accelerate the
% detachment process.

In this section, we construct models of multistep disassembly
nucleosomes composed of multicomponent histones. In solution, free
histones exist in the form of $\HtHf$ tetramers and $\Htd$
dimers\cite{Thomas1975}. The tetramer is located at the center of the
nucleosome and binds to around 60 base-pairs of nucleosomal DNA. Two
identical H2A-H2B dimers align almost symmetrically at the two ends of
the $\HtHf$ tetramer, each taking up around 30 base-pairs of
nucleosomal DNA. The termini of H3 also attach to the DNA on both
ends, further stabilizing the nucleosome complex \cite{Elbahnsi2018}.

Due to the multicomponent nature of the histone octamer, interesting
questions arise as to whether: (i) octamer breakdown precedes
histone-DNA detachment and (ii) whether the former process facilitates
the latter. Studies have consistently shown that salt-induced
disassembly of nucleosomes occurs stepwise, with H2A-H2B dimers
disassembling first, followed by disruption of the $\HtHf$ tetramer as
a whole
\cite{Khrapunov1997,Hoch2007,Gansen2009,Bohm2010}.  However,
nucleosome disassembly under physiological salt concentrations has yet
to be observed due to the long timescales required.

Here, we propose a kinetic model that captures the multimeric feature
of histone octamers and derive mean times of disassembly. We also
consider the interaction between multimeric histone and nucleosome
remodelers and show that by disrupting the interaction between $\HtHf$
and $\Htd$, the detachment process can be significantly
accelerated \added{compared to the spontaneous, intact histone
model}. This observation is consistent with previous experimental
results \cite{Tachiwana2010}.  Interestingly, we also observe that the
acceleration provided by octamer disassembly and nucleosome remodelers
is sub-additive.  By comparing the multimeric nucleosome disassembly
model to the linear sequential disassembly model, we can predict
disassembly pathways under various conditions.
\begin{figure}[h!]
  \centering
  \includegraphics[width=3.4in]{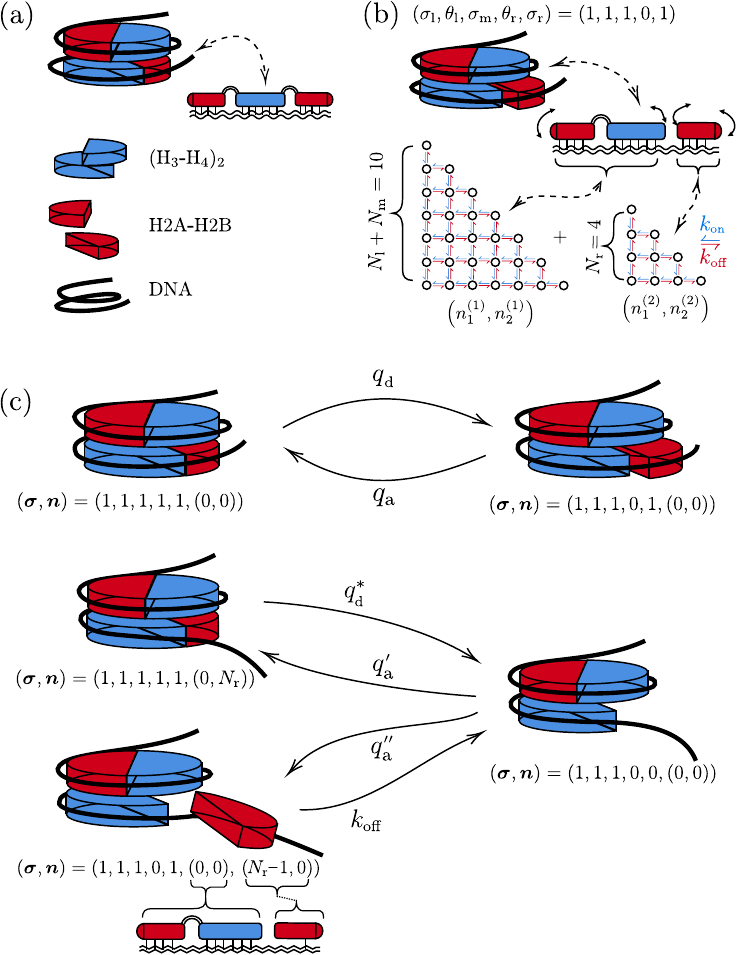}
\caption{Schematic of the multimeric nucleosome disassembly model. (a)
  A histone octamer is composed of one $\HtHf$ tetramer surrounded by
  two $\Htd$ dimers. The presence or absence of the three subunits is
  described by $(\sigma_{\rm l}, \sigma_{\rm m},\sigma_{\rm
  r}) \in \{0,1\}^{3}$, while links between them are described by
  $(\theta_{\rm l}, \theta_{\rm r})\in \{0,1\}^{2}$.  These are
  combined into the string $\boldsymbol{\sigma}=(\sigma_{\rm
  l}, \theta_{\rm l}, \sigma_{\rm m}, \theta_{\rm r}, \sigma_{\rm
  r})\in \{0,1\}^5$. Binding between the subunits and DNA is denoted
  by the vector ${\bf n}$ describing the peeling of contact footprints
  for each linked subunit cluster.  (b) An example of the
  parameterization $(\boldsymbol{\sigma}, {\bf n})$ of state
  space. Here $\boldsymbol{\sigma}=(1,1,1,0,1)$ represents the
  macrostate where all subdomains of the histone bind to DNA but only
  one link exists among them. This leads to two independent linear
  detachment processes denoted by the microstate ${\bf n}
  = \big((n_{1}^{(1)}, n^{(1)}_{2}), (n^{(2)}_{1},
  n^{(2)}_2)\big)$. \added{In this particular case, breaking of the
  DNA-histone contacts can be initiated inside the total nucleosome
  footprint, at the interface between the right dimer and the
  tetramer, as indicated by the small solid-curve arrows.} (c)
  Schematic representation of transitions associated with changes in
  $\theta_{\rm r}$, $\sigma_{\rm r}$, and ${\bf n}$.  For example,
  $(1,1,1,1,1,(0,0))$ represents the state where all subdomains are
  docked and fully bound to DNA and where the links between the
  tetramer and the dimers are intact.}  \label{FIG:MULTIMERIC_MODEL}
\end{figure}
\added{The multimeric is visualized in \ref{FIG:MULTIMERIC_MODEL}
and detailed below.} \deleted{In the multimeric model, we first need
to describe the configuration of histones.  For each histone
configuration, we find a specific state space that describes the state
of each available histone-DNA contact site.}

As discussed in the beginning of this subsection, we simplify the
structure of the histone octamers as a concatenation of two (H2A-H2B)
dimers on both ends of one $\HtHf$ tetramer in the center, as shown in
Fig.~\ref{FIG:MULTIMERIC_MODEL}(a).  To enumerate the presence of the
three different subunits and the links among them, we use the string
$(\sigma_{\rm l}, \theta_{\rm l}, \sigma_{\rm m}, \theta_{\rm
r}, \sigma_{\rm r})\in \{0,1\}^5$ to represent the state of the
histone complex. Here, $\sigma_{j} \in \{0,1\},\, j \in \{{\rm l, m,
r}\}$ represents whether the left, middle, or right part of the
histone modules are present in the complex, while
$\theta_{i} \in\{0,1\},\, i\in \{{\rm l,m}\}$ indicates existence of
links between the left and middle subunits and between the middle and
right subunits, respectively. For any $\theta_{i}=1$, both subunits
that are linked together must be present.

Associated with each state of the histone $(\sigma_{\rm l},
\theta_{\rm l}, \sigma_{\rm m}, \theta_{\rm r}, \sigma_{\rm r})$ is a
state space of ``microstates'' that delineates the underlying states
of DNA-histone bonds.  The representation of microstates depends on
the number $f = \sum_{i = {\rm l, m, r}}\sigma_{i} -\sum_{j={\rm l,
r}}\theta_{j}$ of independent histone modules (a single subunit or a
bound cluster of subunits) that are \textit{not} associated by a
linkage. \added{When the linkage is not present, unbinding of the
DNA-histone contact sites can be initiated at the interface between a
dimer and a tetramer. Each connected module then binds and unbinds
independently in the same way as in the previous intact histone model,
but with fewer contact sites on each module.}  Hence, each state is
represented by a $2f$-tuple $(n_{1}^{(k)}, n_{2}^{(k)})_{k=1}^{f}$,
where $n_{1}^{(k)}$ and $n_{2}^{(k)}$ are analogous to that defined in
Fig.~\ref{FIG2} and are the number of left- and right-detached contact
sites of the $k^{\rm th}$ histone module. For each $k$, $0 \leq
n_{1}^{(k)} + n_{2}^{(k)} < {\textrm{number of DNA binding sites in
the $k^{\rm th}$ module}}$. An example of macrostate $(\sigma_{\rm
l}, \theta_{\rm l}, \sigma_{\rm m}, \theta_{\rm r}, \sigma_{\rm r}) =
(1,1,1,0,1)$ and associated microstates with $f=2$ is shown in
Fig.~\ref{FIG:MULTIMERIC_MODEL}(b).

% A table to list all the parameters used in the paper
\begin{table}[htbp]
\centering
\begin{tabular}{lcc}
\hline \textbf{parameter/variable} & \textbf{symbol} & \textbf{value}
\\ 
%\hline no. of DNA-histone contact sites & $N$ & 14 \\ 
%
\hline 
left, middle, right subunit occupation & $\sigma_{\rm l}, \sigma_{\rm m}, \sigma_{\rm r}$ & $\{0,1\}$ \\
no. of DNA-$(\Htd)$ binding sites & $N_{\rm l}=N_{\rm r}$ & 4 \\
no. of DNA-$\HtHf$ binding sites & $N_{\rm m}$ & 6 \\
l-m and m-r subunit bonds & $\theta_{\rm l}, \theta_{\rm r}$ & $\{0,1\}$ \\
$\HtHf$-$(\Htd)$ association rate & $q_{\rm{a}}$ & - \\
$\HtHf$-$(\Htd)$ dissociation rate  & $q_{\rm{d}}, q_{\rm d}^*$ & {0.01$\kon$} \\
$\HtHf$-$(\Htd)$ binding energy & $E_{\rm q}\!=\!\log\big(\frac{q_{\rm d}}{q_{\rm a}}\big)$ & {$-1$} \\
subunit chemical potential in solution  & $\Delta E_{\rm s}$ & 2 \\
\hline
\end{tabular}
\caption{Parameters and notation used the multimeric histone
  disassembly model. Three histone subunits can occupy the DNA
  substrate and are arranged as left (l), middle (m), and right
  (r). Their presence or absence is enumerated by $\sigma_{\rm
  l}, \sigma_{\rm m}, \sigma_{\rm r} \in \{0,1\}$. The presence of the
  two possible subunit-subunit bonds are indicated by $\theta_{\rm
  l}, \theta_{\rm r}\in \{0,1\}$.}
  \label{tab:parameters2}
\end{table}

% Different from the sequential model, this time
% the origins of peeling off can be located inside the string. A partial loss of
% histone modules leads to temporary inability to recover the the most stable
% state where $y_k = 1, \forall k\in [1,N]\cap \mathbb{Z}$. This feature resembles
% the mechanism for facilitated detachment discussed in the previous section, and
% we expect a significant acceleration compared with sequential model in the
% absence of nucleosome remodelers.

In the multimeric model, we assume that the bonds between H2A-H2B and
(H3-H4)$_2$ can spontaneously break with rate $\qd$ and rebind with
rate $\qa$, provided at least one DNA-histone contacts is
intact. However, once one of the domains loses all its bonds with DNA,
the subunit is no longer held in place and its link with the
neighboring histone domain may break at a somewhat different rate
$\qd^*$.  We also assume that each H2A-H2B carries $N_{\rm l}=N_{\rm
r}=4$ DNA binding sites and the central (H3-H4)$_2$ tetramer carries
the remaining $N_{\rm m}=6$ contact sites. We do not distinguish
different DNA binding sites and let them all have the same $\kon$ and
$\koff$ as in our toy linear delamination model. The DNA-histone
contact energy $E_{\rm c}$ is defined by $E_{\rm c}
= \log(\frac{\koff}{\kon})<0$ as before.  An example of the macrostate
transitions is shown in Fig.~\ref{FIG:MULTIMERIC_MODEL}(c). The
notation and parameters used in the multimeric model is given in
Table \ref{tab:parameters2}.

While it may be reasonable to assume $\qd^* > \qd$ (faster subunit
dissociation if a subunit makes no DNA contacts), for the sake of
simplicity, we will assume $\qd^* = \qd$ in the following discussion.
We will also define the subunit binding affinity $E_{\rm q} =
\log\left(\qd/\qa \right)$ conditioned on the presence of at least one
DNA-histone contact for each of the subunits. Similarly, we also let
$\kd = \koff$ in our subsequent calculations.  

\added{As discussed in the previous section, the disassembly rate
depends on the choice of the initial state. If the initial state is
chosen to be the quasi-steady state ${\bf v}_0$, then the disassembly
rate is given by the principal eigenvalue $-\lambda_0$ of the relevant
transition matrix.} To numerically compute $-\lambda_{0}$, we
constructed a computational algorithm to enumerate all the possible
states and the associated transition matrix of the multimeric histone
disassembly process.  The principal eigenvalue was computed using the
Arnoldi method \cite{Arnoldi1951}. The program is written in
\texttt{JuliaLang} \cite{Bezanson2017jan} and is available through
GitHub at \url{github.com/hsianktin/histone}. We will also numerically
compute the mean first disassembly times $\mathbb{E}[T({\bf 1})]$ of
the multimeric model starting from the fully bound state ${\bf 1}$.
For a stochastic transition matrix $\Q$, the ${\bf
T}\equiv \mathbb{E}[T({\bf x})]$ for all states ${\bf x}\in \Omega$ is
found from solving $\Q^{\intercal}_{\Omega}{\bf T} = - (1, \ldots,
1)^{\intercal}$ and then selecting $\mathbb{E}[T({\bf x}={\bf
1})]$ \cite{FPT}. \added{Our subsequent results show that values of
$-\lambda_0$ and $1/\mathbb{E}\big[T({\bf 1})\big]$ are close to each
other because the most probable state ${\bf v}_0$ in the quasi-steady
state coincides with the fully bound state ${\bf 1}$.}

If the concentration of free $\HtHf$ and $\Htd$ subunits in solution
is negligible, we can treat the detachment of each subunit as
irreversible. If there are appreciable concentrations of histone
dimers or tetramers in solution, their rebinding to a partially
disassembled nucleosome must be considered. An additional parameter
$\Delta E_{\rm s}^{\rm (subunit)}\equiv \log \big(\kon/\qa''^{\rm
(subunit)} \big)$ describing the free energy (or chemical potential)
difference between specific subunits in solution and within a
nucleosome is thus required; due to entropy, the higher the histone
concentration, the lower this difference. The irreversible detachment
of subunit corresponds to the $ \Delta E_{\rm s}= \infty$ limit. A
detailed analysis of reversible multimeric disassembly is given in
Appendix \ref{Appendix:multimeric} \added{where dimers and tetramers
in solution may rebind to a partially disassembled nucleosome, but the
fully detached state is still absorbing in the first passage time
setting.  We will use subscript ``q'' to denote quantities relevant to
the multimeric model.}

\subsubsection{Spontaneous detachment}
We first consider the unfacilitated disassembly of a multimeric
nucleosome and anticipate that subunit-subunit binding and unbinding
rates, $\qa$ and $\qd$, are much faster than their unbinding from DNA,
\added{the rate of which can be estimated by considering
the disassembly rate of a simple intact-histone peeling model
(Eq.~\ref{EQUATION:PERTURBED}) but with $N_{\rm l}$ binding sites:
$\kon N_{\rm l} e^{N_{\rm l} E_{\rm c}}$.}  \added{Additionally, we
assume that a fully linked histone is sufficiently stable so that our
initial condition is an intact octamer. This assumption implicitly
requires $\qa> \qd$ for self consistency and allows us to simply track
unbinding from both ends of the octamer while ignoring the unbinding
from the middle.} It takes an average \added{dimensionless}
time \replaced{$\tau_{\Htd} \approx e^{-N_{\rm l} E_{\rm
c}}$}{$e^{-N_{\rm l} E_{\rm c}}/\kon$} for the $\Htd$ on the left to
unbind from the DNA\added{, whether or not it is attached to the
tetramer.} This estimate is derived in
Appendix \ref{Appendix:one_sided} and comes from
Eq.~\eqref{eq:one_sided_unfacilitated} for the one-sided spontaneous
linear nucleosome disassembly model with $N_{\rm l}$ binding sites. As
with the two-sided spontaneous detachment model in Fig.~\ref{EIGEN0},
there is a \added{large} spectral gap between the first and second
eigenvalues of the transition matrix. Therefore, the expected
unbinding time starting from any bound configuration is given by the
inverse of the principal eigenvalue and similar to that of the
two-sided model. \added{See the Appendix \ref{sec:eigenvalues} for
details.}
\begin{figure}[htb]
  \centering
\includegraphics[width=3.1in]{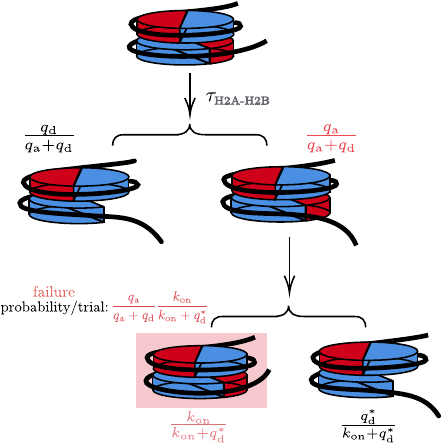}
\caption{\added{Illustration of the geometric trial process.  
Dimer-DNA contacts break after about $\tau_{\text{H2A-H2B}}$. When
this happens the probability that the dimer-tetramer link is also
broken is $\qd/\left(\qa+\qd\right)$. If this is realized, the dimer
breaks free from the system. However, with probability
$1-\qd/\left(\qa+\qd\right) =\qa/\left(\qa+\qd\right)$, the
dimer-tetramer link is intact. From this configuration, there are two
reactions competing with each other: the dimer rebinding to DNA with
rate $k_{\rm on}$ and the dimer-tetramer link breaking with rate
$\qd^{*}$, leading to the dimer breaking free from the system. Thus,
the dimer rebinds to DNA with probability $\kon/(\kon+\qd^{*})$
(failing to disassociate) and breaks free with probability
$\qd^{*}/(\kon+\qd^{*})$.}}
\label{fig:geometric_trial}
\end{figure}

First, consider the expected time $\mathbb{E}[T]$ for the histone
octamer to break down and its subunits to sequentially leave the
system (the multimeric disassembly pathway) when rebinding 
\added{does not occur (when the solution contains no free
histone subunits and $\Delta E_{\rm s} = \infty$).}  Upon unbinding of
$\Htd$ dimer from DNA, the chance that it is linked to the $\HtHf$
tetramer is $\qa/(\qa+\qd)$.  If the $\Htd$ dimer is not bound to the
tetramer $\HtHf$, it will immediately leave the system. Otherwise,
there is a probability $\kon/(\kon+\qd^*)$ that DNA and the dimer will
rebind before the $\Htd$ dimer unlinks from $\HtHf$ and leaves the
system. The expected time for the $\Htd$ dimer to leave the system
from a fully bound configuration ${\bf 1} \equiv (1,1,1,1,1,(0, 0))$
is thus given by

\begin{equation}
\mathbb{E}\big[T_{\Htd}({\bf 1})\big]
\approx \frac{e^{-N_{\rm l}E_{\rm c}}}{1 - \frac{\qa}{\qa + \qd} 
  \frac{\kon}{\kon + \qd^*}}.
  \label{T1_multimeric}
\end{equation}
Here, $1/\left(1 - \tfrac{\qa}{\qa + \qd}
\tfrac{\kon}{\kon + \qd^*} \right)$ measures the expected number of 
trials until the $\Htd$ dimer leaves the system
successfully, \added{as illustrated in
Fig.~\ref{fig:geometric_trial}}.  When an attempt fails, the dimer-DNA
contacts can quickly approach equilibrium because of the spectral gap
for the simple peeling model. Thus, the next dimer removal trial
occurs independently of the last one and the number of trials should
follow a geometric distribution with the probability of failure given
by $\tfrac{\qa}{\qa + \qd}\tfrac{\kon}{\kon + \qd^*}$. The expected
time for both $\Htd$s to leave the system is on the same order of
magnitude as the expected time for one $\Htd$ to leave the system.

After dissociation of \added{the two equivalent dimers, the $\HtHf$
tetramer will unbind from the DNA at a rate of $\kon N_{\rm
m}e^{N_{\rm m} E_{\rm c}}$ according to Eq.~\eqref{EQUATION:PERTURBED}
with $N_{\rm m}$ contact sites. We can then define the approximate
expected time it takes for the entire nucleosome to detach through the
multimeric breakdown pathway as}
\begin{equation}
\widehat{\mathbb{E}\big[T({\bf 1})\big]} \coloneqq 
%
%  \frac{e^{-N_{\rm l}E_{\rm c}}}{1 - \frac{\qa}{\qa
%  + \qd} \frac{\kon}{\kon + \qd^*} } 
%
\alpha \mathbb{E}\big[T_{\Htd}({\bf 1})\big]
+ \frac{e^{-N_{\rm m}E_{\rm  c}}}{N_{\rm m}},
\label{eq:tau1_irr}
\end{equation}
\noindent \added{where $1 < \alpha < 2$ is an additional factor determining 
the expected MFPT for two independent dimers to detach.  For
independent, exponentially distributed waiting times, $\alpha = 3/2$.
In our model, the dynamics of the dimers on opposite sides of the
tetramer are independent, but their detachment times 
are modeled via a multistate geometric attempt processes, and are not 
exponentially distributed. Nonetheless, at this level of approximation 
$\alpha \sim 1 $ suffices to to provide reasonable estimates.}

\begin{figure*}[htb]
  \centering
\includegraphics[width=0.88\textwidth]{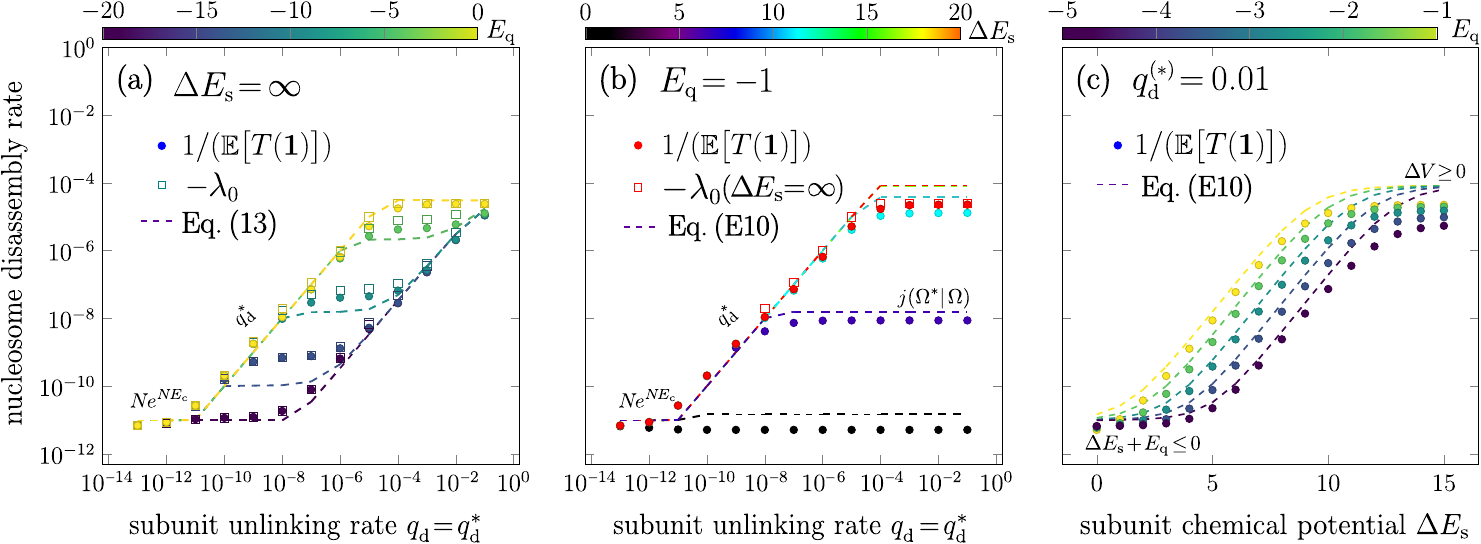}
\caption{Dimensionless rates of remodeler-free, multimeric
nucleosome disassembly measured by the principal eigenvalue and the
inverse of the dimensionless mean disassembly time
$1/\mathbb{E}[T({\bf 1})]$. \added{$-\lambda_0$ and
$1/\mathbb{E}[T({\bf 1})]$ provide similar measures of the disassembly
rate and agree well with each other as indicated in the plots.}  In
all cases, we set $E_{\rm c}=-2$, $\kd = \koff$,
$\qd=\qd^*$, \added{and all the rates are normalized by $\kon$}.  (a)
Rates as a function of the dimensionless rate of subunit unlinking
$q_{\rm d}^{*}/\kon$ in the zero bulk histone concentration ($\Delta
E_{\rm s}=\infty$) limit.
%
%the principal eigenvalue $-\lambda_{0}$ and MFPT
%${1}/{\big(\kon \mathbb{E} [T({\bf 1})]\big)}$ from the fully bound
%state ${\bf 1}$ to $\Omega^*$ as a function of the subunit
%bond-breaking rate $\qd=\qd^{*}$.
%
Numerical results of the principle eigenvalue $-\lambda_{0}$
(\added{open squares}) \added{closely match} those of
$1/\mathbb{E}[T({\bf 1})]$ (\added{filled circles}),
indicating \added{that starting from the \added{fully DNA-bounded
state or from the quasi-steady state yields similar mean dissociation
times.}}  All results are well approximated by the
approximation \added{$-\hat{\lambda}_{0,{\rm q}}(\Delta E_{\rm
s}= \infty)$} for $-\lambda_{0}$ given in
Eq.~\eqref{eq:multimeric_unfacilitated_disassembly} (dashed curves).
%
%\deleted{Eq.~\eqref{eq:tau1_irr} provides a good estimate when $\qd^{*}$ is
%large compared to the inverse of the $\Htd$ detachment timescale. When
%$\qd^*$ is smaller than $1/\mathbb{E}[T]$ given in
%Eq.~\eqref{eq:tau1_irr}, the detachment rate is given by $\qd^* + \kon
%N e^{N E_{\rm c}}$.} 
%
Other parameters are assigned typical values given
in Tables~\ref{tab:parameters} and ~\ref{tab:parameters2}.
(b) Comparison of \added{$1/\mathbb{E}\big[T({\bf 1})\big]$} to the
estimate \added{$- \hat{\lambda}_{0, {\rm q}}$} given in
Eq.~\eqref{eq:lambda_d_mul}. Here, we set $E_{\rm q} = -1$, $\Delta
E_{\rm s}^{(\Htd)} = \Delta E_{\rm s}^{(\HtHf)}$ and vary $\Delta
E_{\rm s}$ and $\qd=\qd^{*}$, which is the rate-limiting step as in
(a).  When $\qd < N e^{N E_{\rm c}}$, the disassembly rate is
approximately $N e^{N E_{\rm c}}$. When $\kon N e^{N E_{\rm c}} < \qd
< \kon j(\Omega^*\,\vert\,\Omega)$ (given by Eq.~\eqref{eq:j_mul}),
the disassembly rate is controlled by $\qd$. When $\qd/\kon$ is large,
the dimensionless disassembly rate is approximated by
$j(\Omega^*\,\vert\,\Omega)$.
(c) Disassembly rates as a function of $\Delta E_{\rm s}$ at different
values of $E_{\rm q}$ for large (not rate-limiting)
$\qd^{*}/\kon=0.01$.  Since $j(\Omega^*\,\vert\,\Omega) \sim N_{\rm m}
e^{N E_{\rm c} + 2(\Delta E_{\rm s} + E_{\rm q})}$, larger $\Delta
E_{\rm s}+E_{\rm q}$ leads to faster disassembly. The value of $\Delta
E_{\rm s}$ at which the disassembly rate saturates can be estimated by
the root to $\Delta V \equiv 2 (E_{\rm q} + \Delta E_{\rm s} + N_{\rm
l} E_{\rm c})=0$. In this example, $N_{\rm l} = 6, E_{\rm c} = -2$, so
when $E_{\rm q}=-1$, the disassembly rate saturates at $\Delta E_{\rm
s}\approx 9$.}
\label{fig:Fig8}
\end{figure*}

\added{The derivation of Eq.~\eqref{eq:tau1_irr} implicitly assumes a
sequential disassembly pathway where the $\Htd$ dimer disassemble
first.  Thus, Eq.~\eqref{eq:tau1_irr} is valid only in the regime
$q_{\rm a}, q_{\rm d}\gg \kon e^{N_{\rm l}E_{\rm c}}$.}

Significant acceleration can be achieved if the links between the
subunits are weak, \added{thereby decreasing the
$\mathbb{E}\big[T_{\Htd}({\bf 1})\big]$ term in
Eq.~\eqref{eq:tau1_irr}.  When subunit links are weak ($E_{\rm
q} \not\ll -1 \Leftrightarrow
\qd/(\qa + \qd) \not\sim 0$) and/or if unlinking is
fast $(\qd^* \gg \kon \Leftrightarrow \kon/(\kon + \qd^*)\sim 0)$,
the factor $1/\left(1 - \tfrac{\qa}{\qa + \qd} \tfrac{\kon}{\kon +
\qd^*} \right) \sim 1$ and thus $\widehat{\mathbb{E}[T({\bf 1})]}
\approx {e^{-N_{\rm m}E_{\rm c}}}/{N_{\rm m}}$.}

\added{Next, we relax the assumption that $q_{\rm d}$ and $q_{\rm a}$
are fast and introduce corrections to Eq.~\eqref{eq:tau1_irr},
obtaining a more general expression for the expected time for
multimeric nucleosome dissociation.}  When the affinity between
subunits is high ($E_{\rm q} \ll -1$) and link breaking is slow
$(\qd, \qd^* \ll \kon)$, the \replaced{mean nucleosome disassembly
time}{stability} of the system depends on the order of the term
$1/\left(1 - \frac{\qa}{\qa + \qd} \frac{\kon}{\kon
+ \qd^*} \right){e^{-N_{\rm l}E_{\rm c}}}$.
If $E_{\rm q} \rightarrow - \infty$, to achieve \replaced{mean
nucleosome disassembly times}{stability} comparable to the linear
peeling model, we need $\qd^* < \kon e^{2 N_{\rm l} E_{\rm c}}$, as
shown by the purple symbols in Fig.~\ref{fig:Fig8}(a).  We assumed
fast $\qd^*$ in the derivation of Eq.~\eqref{eq:tau1_irr}.  When
$\qd^*$ is not fast, the dissociation between histone subdomains can
be a rate-limiting step in the multimeric pathway. In this case the
total time required for the dimers to detach from the system is given
by $\mathbb{E}[T_{\Htd}]+ \kon/\qd^*$.
As indicated by the yellow and cyan symbols in Fig.~\ref{fig:Fig8}(a),
when $\qd^* < \kon/\widehat{\mathbb{E}[T({\bf 1})]}$, the rate of disassembly is
proportional to $\qd^*$. When the dimer-tetramer unbinding rate
further decreases to $\qd^* < \kon  N e^{N E_{\rm c}}$, the monomeric
disassembly (simple histone peeling) is faster than multimeric
breakdown and the dimensionless disassembly rate is $\approx N  e^{N
E_{\rm c}}$ (for $k_{\rm d} = \koff$).

Combining the above results, we obtain the following refined estimate
for the dimensionless disassembly rate:
\begin{equation}
  -\hat{\lambda}_{0,{\rm q}}(\Delta E_{\rm s}=\infty) \coloneqq N e^{N E_{\rm c}}+
  \frac{1}{\widehat{\mathbb{E}\big[T({\bf 1})\big]}
%
%_{q_{\rm a}, q_{\rm d}\gg \kon e^{N_{\rm l}E_{\rm c}}}
%
+ \frac{\kon}{\qd^*}}
\label{eq:multimeric_unfacilitated_disassembly}
\end{equation}
where \added{$\widehat{\mathbb{E}\big[T({\bf 1})\big]}$ is the
expected disassembly time in the $q_{\rm a}, q_{\rm d}\gg \kon
e^{N_{\rm l}E_{\rm c}}$ limit given in Eq.~\eqref{eq:tau1_irr}.}  This
formula provides a good qualitative description of \replaced{both
$1/\mathbb{E}[T({\bf 1})]$ and $-\lambda_0$}{the disassembly rate} in
the $\Delta E_{\rm s} \to \infty$ limit (no \added{subunit} rebinding
from bulk solution) as shown in Fig.~\ref{fig:Fig8}(a). Additionally,
we show close agreement between the numerically obtained principal
eigenvalue $\lambda_0$ and the inverse of the mean dimensionless
disassembly time starting from the fully bound state ${\bf
1}=(\boldsymbol{\sigma}=(1,1,1,1,1), {\bf n}=(0,0))$.

%They agree with each other after transformation to the unit of rates,
%validating our approach of using the principal eigenvalue to
%approximate the disassembly rate.

\deleted{The estimates above hold under finite histone concentrations
($\Delta E_{\rm s} < \infty$) in which partial rebinding of histone
subunits occurs.}  \added{When dimers and tetramers can rebind to a
partially unwrapped nucleosome, with the fully detached state still
absorbing in the first passage time problem, (finite histone subunit
concentration in bulk, $\Delta E_{\rm s} < \infty$),} $\qd^*$ still
serves as a rate-limiting step for the multimeric breakdown pathway
and the disassembly rate for small $\qd^*$ can again be well
approximated by the maximum of $N e^{N E_{\rm c}}$ and $\qd^*$, as
shown in Fig.~\ref{fig:Fig8}(b).  If $\qd^*$ becomes moderately large,
the problem can again be effectively be represented by an irreversible
process \added{that can be analyzed using the absorbing boundary
method.} Since the acceleration factor is approximately $e^{2(\Delta
E_{\rm s} + E_{\rm q})}$, disassembly is sped up only if $(\Delta
E_{\rm s} + E_{\rm q}) \geq 0$.  When $\Delta V/2 \equiv (\Delta
E_{\rm s} + E_{\rm q}) + N_{\rm l} E_{\rm c} > 0$, the acceleration is
limited by the rate of $\HtHf$ detachment, as shown in
Fig.~\ref{fig:Fig8}(c).

Appendix~\ref{section:F1} (Eq.~\eqref{eq:lambda_d_mul}) summarizes the
above discussion and provides an estimate for the principal
eigenvalue, and thus the mean disassembly time of the multimeric
reversible ($\Delta E_{\rm s} < \infty$) detachment model. Both
multimeric and monomeric disassembly pathways are allowed in the full
``multimeric'' model, as illustrated in Fig.~\ref{fig:2paths}.
\begin{figure}[htb]
  \centering
\includegraphics[width=3.25in]{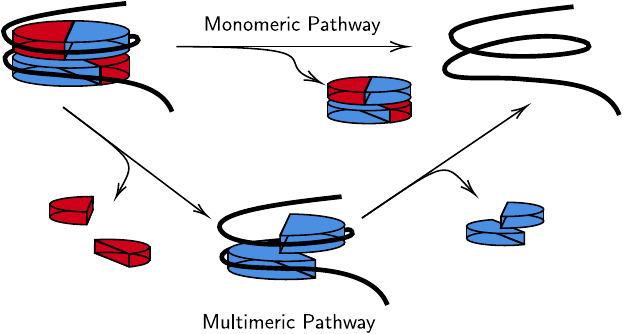}
\caption{Schematic of the two general pathways of nucleosome disassembly
when the histone can break up into its subunits and detach
separately.}
\label{fig:2paths}
\end{figure}
The monomeric disassembly pathway usually occurs at a rate that is a
lower bound to that of the multimeric disassembly pathway. Multimeric
disassembly mediated by two dimer-tetramer links allows for stagewise
dissociation of subunit-DNA contacts, accelerating the overall
process \added{compared to the intact histone disassembly model}.
Factors that limit the rate of multimeric pathways include the
dissociation rate $\qd^*$ and number of trials of dimer disassembly
$1/\left(1 - \frac{\qa}{\qa+\qd} \frac{\kon}{\kon + \qd^*} \right)$ as
described in Eq.~\eqref{eq:multimeric_unfacilitated_disassembly}.  As
detailed in \added{Appendix~\ref{Appendix:multimeric_pathway}, the
disassembly rate is approximated by a weighted sum of the rates
associated with the monomeric and multimeric pathways, as depicted in
Fig.~\ref{fig:2paths}.} \deleted{and thus} This means that disassembly
in the full multimeric model will always be faster than in simple
intact-histone model \added{in which only the monomeric pathway is
present.} \replaced{By contrast}{However}, the multimeric disassembly
pathway (conditioned on histone fragmentation) \added{refers to the
process where one histone module, dimer or tetramer, leaves the DNA
before the whole histone complex dissociates from the
DNA}. \added{This conditioned pathway} can exhibit \textit{slower}
dissociation than that of the monomeric pathway, particularly when
$q_{\mathrm{d}}^*$ is small.

% As
% detailed in Appendix~\ref{section:F2}, the observed disassembly rate
% is a weighted sum of the monomeric and multimeric pathways and thus
% will always be faster than the simple intact-histone disassembly rate.
% However, \deleted{the} the multimeric disassembly pathway (conditioned on
% histone fragmentation) can be slower than monomeric pathway,
% particularly when $\qd^*$ is small.

% Given the expected time for disassembly in Eq.~\eqref{eq:tau1_irr}, we
% estimate the principal eigenvalue of the transition matrix for the
% multimeric model as $\vert \lambda_0\vert \approx 1/\mathbb{E}
% \big[T\big]$. Comparisons of this estimate with numerical results are
% shown in Fig.~\ref{fig:Fig7}(a). Our estimate is in good agreement
% with the numerical results when $1/\qd$ is small compared to the
% timescale of $\Htd$ detachment since Eq.~\eqref{eq:tau1_irr} was
% derived under this parameter regime.

% When $\qd$ is small, subunit dissociation/undocking becomes
% rate-limiting and dominates the principal eigenvalue. When $E_{\rm q}
% \rightarrow - \infty$ and $\qd \rightarrow 0$, the principal
% eigenvalue reduces to that of the linear-binding intact-histone model
% (peeling model). On the other hand, when $\qd$ and $E_{\rm q}$ are
% moderately large, the principal eigenvalue is dominated by the rate of
% $\HtHf$ detachment. This can be several orders of magnitude faster
% than detachment in the linear peeling model and can be estimated by
% $e^{-(N-N_{\rm m})E_{\rm c}}$.
% }
%

\subsubsection{Facilitated multimeric disassembly.}
We now evaluate the interplay between remodelers and multimeric
histones in the disassembly process.  Even though interior octamer-DNA
contacts can be transiently exposed for remodeler binding, for
simplicity and tractability, we assume remodelers can only attack from
the ends of the octamer-DNA contact footprint.  This assumption
changes the underlying geometry of the state space and is valid for
describing the attack from motor proteins such as helicases and RNA
polymerases.
%
%Under this assumption, the attack on $\Htd$ from the remodelers occur
%only to one end.
%
Since remodelers can attack only from the one exterior side of each
$\Htd$ dimer, previous calculations of facilitated detachment in the
linear peeling model can be readily adapted to the one-sided peeling
model (see Appendix~\ref{Appendix:one_sided}).  As in our analysis of
the unfacilitated, irreversible multimeric model, the analysis in this
section begins with the irreversible scenario ($\Delta E_{\rm s}
= \infty$). We first consider a dimer detaching from the system via a
sequence of independent trials, each taking time $\tau_{\Htd}$,
followed by detachment of the remaining tetramer.
%
%This estimate is rate-limited by the dimer-tetramer unlinking rate
%$\qd^*$ and only pertains to the multimeric pathway.  {\color{red}
%Thus, linearly peeling, facilitated disassembly rate derived in
%Eq.~\eqref{eq:intact_effective_facilitation} serves as a lower bound
%for the overall disassembly rate.}
%
Assuming the steady state approximation for each subdomain given by
Eq.~\eqref{eq:one_sided_weak_facilitation} in Appendix
\ref{Appendix:one_sided}, we can estimate the \added{dimensionless}
typical $\Htd$ dimer\added{-DNA} detachment time \added{with
possibility of tethering to the tetramer}
\begin{equation}
\tau_{\Htd} \coloneqq \frac{1+e^{(N_{\rm
      l}-1)(E_{\rm c} -E_{\rm p}^{-})}}{\varepsilon e^{(N_{\rm
      l}-1)(E_{\rm c}-E_{\rm p}^{-})}}
\label{tau_dimer}
\end{equation}
in the weak remodeler regime.  Here,
Eq.~\eqref{eq:one_sided_weak_facilitation} is the one-sided version of
Eq.~\eqref{eq:intact_weak_facilitation} and its inverse results in the
estimate for $\tau_{\Htd}$ and illustrates how one can apply previous
results from the simple, intact-histone peeling model to the peeling
of each of the histone subunits by modifying the number of contact
sites from $N$ to $N_{\rm l}, N_{\rm r}$.

To obtain an estimate for both strong and weak remodeler regimes, we
shall use the one-sided version of
Eq.~\eqref{eq:intact_effective_facilitation} by taking the maximum of
Eq.~\eqref{eq:one_sided_weak_facilitation} and
Eq.~\eqref{eq:one_sided_strong_facilitation}.
%
% Because full detachment of the dimer requires detachment from both
% DNA and the $\HtHf$ tetramer, we use $\tau_{\Htd}$, the detachment
% time only from DNA, to estimate the time for the total removal of
% the dimer.  To fully detach from the system, the $\Htd$ dimer must
% also unlink from the $\HtHf$ tetramer.
%
Following the arguments made for non-facilitated multimeric
disassembly that led to Eq.~\ref{T1_multimeric}, the probability
$q_{\rm a}/(q_{\rm a} + q_{\rm d})$ that the dimer is still attached
to the $\HtHf$ tetramer after its DNA contacts are broken is now
modified by the probability that contacts reform before dimer-tetramer
bond breaking \textit{in the presence of remodeler competition.}  When
there is a strong facilitation by remodelers, they will block DNA
contact sites quickly after the histone dimer unbinds from these
contact sites.  Remodeler binding at rate $\pa$ and dimer dissociation
from the $\HtHf$ tetramer occurring at rate $\qd$ thus compete with
$\Htd$-DNA contact rebinding.  Consequently, in the $q_{\rm a}, q_{\rm
d}\gg \kon e^{N_{\rm l}E_{\rm c}}$ limit, the expected time for
complete $\Htd$ dimer detachment from the nucleosome can be
approximated by
\begin{equation}
\added{
  \widehat{\mathbb{E}\big[T_{\Htd}({\bf 1})\big]}
%_{q_{\rm a}, q_{\rm d}\gg \kon e^{N_{\rm l}E_{\rm c}}}
  \coloneqq \frac{\tau_{\Htd}}{\Big(1
  -\frac{\qa}{\qa + \qd}\frac{\kon}{\kon+\pa+\qd}\Big)}}.
\label{eq:expected_time_detachment_facilitated_H2A-H2B}
\end{equation}
The expected \added{dimensionless} time for detachment of the
remaining $\HtHf$ tetramer is given by \added{$\tau_{\HtHf}
\approx \big(1+ N_{\rm m}e^{(N_{\rm m}-1)(E_{\rm c} -E_{\rm
p}^{-})}\big)/\big(\varepsilon N_{\rm m} e^{(N_{\rm m}-1)(E_{\rm c}
-E_{\rm p}^{-})}\big)$}, analogous to $\tau_{\Htd}$ given in
Eq.~\eqref{tau_dimer} and the inverse of $\hat{\lambda}_{0}(E_{\rm p}>
E_{\rm c})$ given in Eq.~\eqref{eq:intact_weak_facilitation} for the
weak facilitation limit, but with $N_{\rm m}$ tetramer-DNA contact
sites. In analogy to the expected time for detachment of a multimeric
nucleosome in the absence of remodelers (Eq.~\eqref{eq:tau1_irr}), the
expected dissociation time in the presence of remodelers can be
estimated as the sum of the expected time for detachment of both
$\Htd$ dimers and the $\HtHf$ tetramer:
\begin{equation}
 \widehat{\mathbb{E}\big[T({\bf 1})\big]}
\coloneqq \alpha \widehat{\mathbb{E}\big[T_{\Htd}({\bf 1})\big]} + \tau_{\HtHf},
  \label{eq:tau2_irr}
\end{equation}
\noindent valid in the $q_{\rm a}, q_{\rm d}\gg \kon e^{N_{\rm l}E_{\rm c}}$
limit.

In the strong facilitation limit, we simply replace the estimate of the
DNA-detachment times $\tau_{\Htd}$ and $\tau_{\HtHf}$ given in
Eq.~\eqref{eq:tau2_irr} by
Eq.~\eqref{eq:one_sided_strong_facilitation} and
Eq.~\eqref{eq:intact_effective_facilitation}, with $N_{\rm l}$ and
$N_{\rm m}$ number of contact sites.
\begin{figure}[htb]
  \centering
  \includegraphics[width=3.4in]{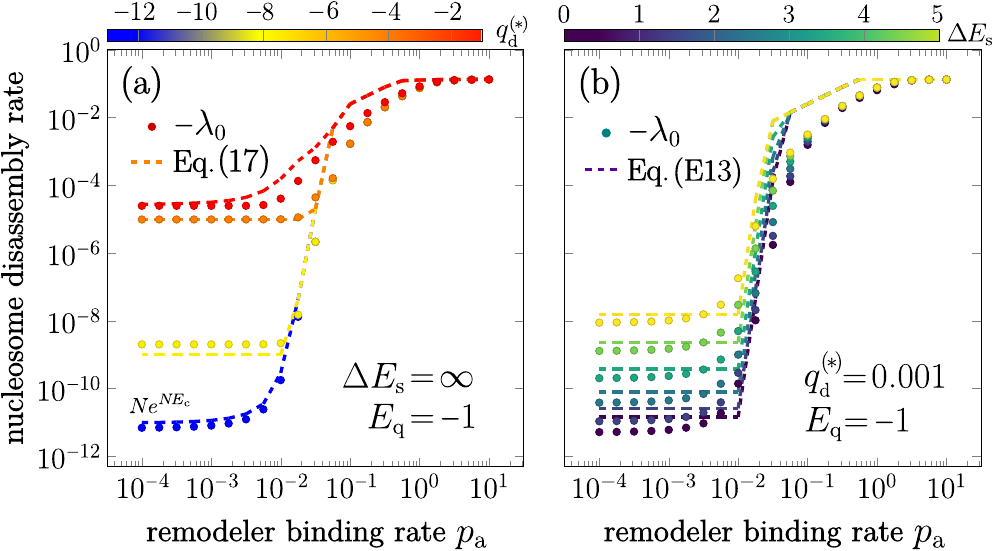}
\caption{Principal eigenvalues -- an estimate of
\added{$1/\mathbb{E}[T({\bf 1})]$} -- from the remodeler-facilitated
disassembly model. (a) For the irreversible model (no subunits in
solution), the dimensionless disassembly rate $-\lambda_0$ is plotted
as a function of remodeler binding rate $\pa$, for fixed $E_{\rm
c}=-2, ~ E_{\rm q} = -1, ~\pd = 0.01$, \added{with estimates
given by Eq.~\eqref{eq:multimeric_facilitated_irreversible}} (b) The
disassembly rate $-\lambda_0$ for different subunit chemical potential
differences $\Delta E_{\rm s}^{(\Htd)}= \Delta E_{\rm s}^{\HtHf} =
\Delta E_{\rm s}$. Estimates given in Eq.~\eqref{eq:lambda_rev} are
plotted as the dashed curves which agree well with numerical results.
\added{In (a) and (b), $q_{\rm d}=q_{\rm d}^*\equiv q_{\rm d}^{(*)}$ 
and all rates are normalized 
with respect to $\kon$}.}
\label{fig:Fig7}
\end{figure}

Likewise, we can estimate the principal eigenvalue of the facilitated
multimeric detachment process by considering the contributions from
the monomeric, simple histone disassembly pathway and other
rate-limiting steps:
\begin{equation}
\added{-\hat{\lambda}_{0,{\rm p,q}}(\Delta E_{\rm s}= \infty) = -\hat{\lambda}_{0,\textrm{p}}
+ \frac{1}{\widehat{\mathbb{E}\big[T({\bf 1})\big]}
%_{q_{\rm a}, q_{\rm d}\gg \kon e^{N_{\rm l}E_{\rm c}}}
+ \frac{\kon}{\qd^*}}}, 
\label{eq:multimeric_facilitated_irreversible}
\end{equation}
\noindent \added{where $\hat{\lambda}_{0,\textrm{p}}$ is given by
Eq.~\eqref{eq:intact_effective_facilitation} and provides an estimate
of the rate of monomeric histone dissociation, while
$\widehat{\mathbb{E}\big[T({\bf 1})\big]}$
%_{q_{\rm a}, q_{\rm d}\gg \kon e^{N_{\rm l}E_{\rm c}}}$ 
%
is given by Eq.~\eqref{eq:tau2_irr}.}  Comparison between this estimate
and numerical results are shown in Fig.~\ref{fig:Fig7}(a). Note that
the requirement $E_{\rm c} - E_{\rm p}^{-} > 0$ for effective
facilitation remains the same as in the simple intact-histone model.
When the remodelers bind weakly, histone fragmentation provides a
strong facilitation to the detachment process. However, when
remodelers bind strongly, histone fragmentation does not significantly
accelerate disassembly.

\added{The case of} finite subunit concentrations in solution (finite
$\Delta E_{\rm s}$) is discussed in more detail in
Appendix~\ref{sec:multimeric_reversible_facilitated}.  An estimate of
the disassembly rate is given in Eq.~\eqref{eq:lambda_rev}.  Analytic
approximations and numerical results are compared in
Fig.~\ref{fig:Fig7}(b) and show qualitative agreement. Nucleosome
remodelers facilitate the disassembly by reducing the energy barrier
for each contact site. This facilitation acts somewhat independently
from histone fragmentation so the threshold $E_{\rm p} < E_{\rm c}$
for effective facilitation is the same as that in the linearly peeling
model, regardless of different values of $\Delta E_{\rm s}$ and
$E_{\rm q}$.
\begin{figure}[htb]
\centering
\includegraphics[width=3.4in]{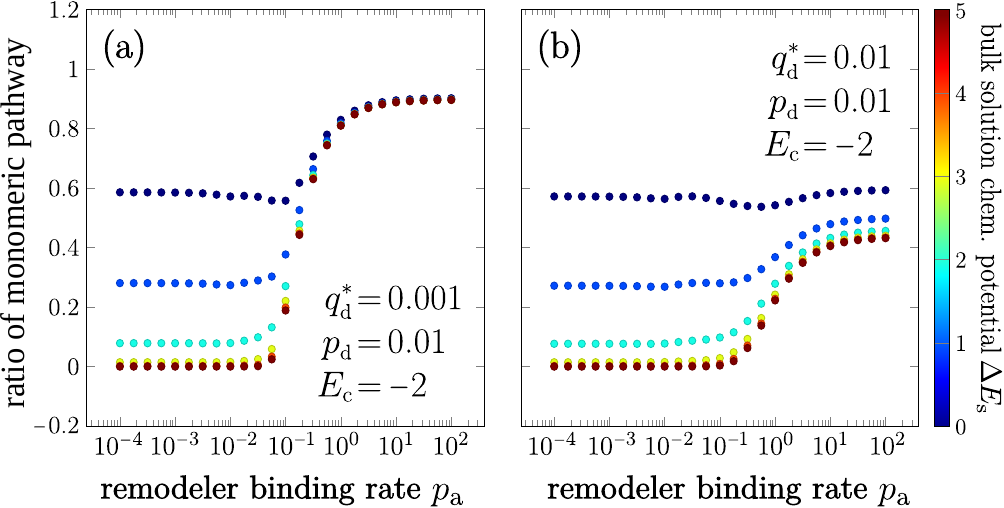}
\caption{The fraction of disassembly pathways that lead to
dissociation of an intact histone octamer. This quantity is defined as
the probability that the histone is an intact octamer at the moment of
full nucleosome disassembly. (a) The probability of monomeric
nucleosome disassembly is plotted as a function of remodeler binding
$\pa$ for different chemical potentials $\Delta E_{\rm s}$. Here a
small subunit unbinding rate $\qd^{*}= 0.001$ and a large
chemical potential (low subunit concentration in solution) allows for
a sharp transition to a monomeric disassembly pathway as facilitation
is increased through $\pa$. (b) The probability of monomeric
disassembly plotted against $\pa$, but with $q_{\rm d}^*=
0.01$. The larger unbinding rate increases the probability of a
fragmented-histone disassembly. \added{In (a) and (b) all rates 
are normalized by $\kon$.}}  
\label{fig:Fig9}
\end{figure}

As detailed in Appendix~\ref{Appendix:multimeric_pathway}, we also
found that remodeler binding and histone subunit concentration in
solution can conspire to bias the disassembly from a more fragmented
dissociation pathway to one in which the histone complex dissociates
intact. Fig.~\ref{fig:Fig9} shows the probability of the histone in
an intact octamer state at the moment of full nucleosome disassembly.
In the weak remodeler facilitation regime, low histone subunit
concentration typically allows a faster multimeric disassembly
pathway, while high histone subunit concentration makes the rates of
monomeric and multimeric disassembly comparable. In the former case,
the histone is more likely to dissociate after fragmentation. On the
other hand, when the remodeler binding is strong, both pathways have
similar rates. In this case, the probability of the histone
dissociating as an intact octamer depends on the dimer-tetramer
unbinding rate $\qd^*$ and \added{the dimensionless mean disassembly
time under the multimeric fragmentation pathway
$\widehat{\mathbb{E}[T({\bf 1})]}$.}
%
%_{q_{\rm a}, q_{\rm d}\gg \kon e^{N_{\rm l}E_{\rm c}}}. 
%
\added{When $\qd^* < \kon/\widehat{\mathbb{E}[T({\bf 1})]}$ (but $q_{\rm a},
q_{\rm d}\gg \kon e^{N_{\rm l}E_{\rm c}}$), the histone is more likely
to dissociate as an intact octamer, as shown in
Fig.~\ref{fig:Fig9}(a).  When
$\qd^* \geq \kon/\widehat{\mathbb{E}[T({\bf 1})]}$ (and $q_{\rm a},
q_{\rm d}\gg \kon e^{N_{\rm l}E_{\rm c}}$),} it is more likely to
\added{fragment before complete dissociation}, as shown in
Fig.~\ref{fig:Fig9}(b).

\section{Summary and Conclusions}

In this study, a suite of Markov chain models was developed to analyze
nucleosome stability.  We delineated a number of mechanisms that
probably contribute to nucleosome stability, including a model of
multimeric histone disassembly.

%When interactions between histones and nucleosome remodelers are
%absent, the system is near equilibrium, and the influence of DNA
%sequence on nucleosome stability arises mainly through the total
%binding energy between DNA and histones.
\vspace{2mm}
\noindent \textit{Linear detachment model.} 
Our first proposed mechanism maintains both high accessibility for few
energy consuming proteins and low accessibility for generic DNA
binding proteins. For the spontaneous detachment problem, the model
can be described by a single parameter, the contact free energy
$E_{\rm c}$, which we assume $\color{blue}{E_{\rm c} \ll -1}$ to reflect strong
histone-DNA binding. The simple-histone linear peeling mechanism
described by our first model applies not only to the histone
detachment problem, but also to a family of nucleic acid-binding
proteins that both protect the nucleic acid from attack and respond to
regulation signals quickly. Examples include \textit{E. coli}
single-stranded DNA binding proteins (\textit{E. coli} SSB) and
replication protein A (RPA) that exists in eukaryotic cells.

In an extended model that incorporates remodeler-facilitated
disassembly, we analyzed the enhancement of dissociation provided by
processive motors moving along DNA, which also serves as a good
estimate of facilitation by generic remodelers binding from
solution. We introduced additional parameters that quantify the
remodeler binding rate $\pa$ and binding energy $E_{\rm p}$.  When the
dissociation rate $\pd = \pa e^{E_{\rm p}}$ is not too slow, a
quasi-steady state approximation provides a tight upper bound on the
effective unbinding rate, revealing a high degree of cooperativity and
a gating mechanism sensitive to the energy cost of the processive
motors or remodelers. Efficient acceleration is possible only if
$E_{\rm c}-E_{\rm p} > 0$; this energy difference controls a ``gate''
that allows certain proteins like polymerases to access DNA while
preventing generic DNA binding proteins from penetrating the
nucleosome.  This simple analysis helps resolve the paradox that
histones must simultaneously bind tightly to DNA yet rapidly release
DNA when accessibility is required, for example, during transcription
or DNA replication. Our prediction is consistent with observations
from previous single-molecule experiments and data-driven modeling
that fast-diffusing remodelers in the absence of ATP consumption do
not significantly affect the nucleosome disassembly
rate\cite{Bundschuh2011}.

Besides nucleosomes, are many other biologically important systems in
which protein-DNA binding and unbinding arise \cite{Chen2014aug}. Many
have been studied in single-molecule experiments that interrogate the
collective dynamics of proteins along a single DNA strand, where
facilitated protein detachment was observed under increased protein
concentration in solution \cite{Gibb2014feb,NAUFER2021}. Our
simple-histone models may provide insight into developing models for
these more general protein-DNA systems.

\vspace{2mm}
\noindent \textit{Multimeric detachment model.} We also derived
explicit formulae for mean first dissociation times of nucleosomes in
which the histone is comprised of octamer subunits (two dimers and a
tetramer).  We first considered the irreversible histone detachment
model in which once a histone subunit (dimer or tetramer) detaches
from the nucleosome complex, it does not rebind.  In a spontaneous,
incremental detachment model, both the binding energy $E_{\rm q}$
between histone dimers and tetramers and their dissociation rates
$\qd$ are additional relevant parameters. The dimer-tetramer
dissociation rate $\qd$ can also depend on the state of the
DNA-histone binding sites.  We thus allow an additional parameter
$\qd^*$ to describe the unbinding rate when at least one of the
histone modules is completely unbound from the DNA. Although binding
affinities between histone subunits have not been experimentally
characterized, we found that the detachment rate can be significantly
upregulated by modulating the binding free energy $E_{\rm q}$ between
the $\HtHf$ tetramer and the $\Htd$ dimers.  This effect comports with
the observation that mutations that reduce the binding affinity
between different modules of histones lead to a much shorter
disassembly time of around 20
minutes \cite{Tachiwana2010,Arimura2018}.

The case of reversible binding (exchange of subunits from bulk
solution) is fully discussed in the Appendix, where we introduced
additional parameters $\Delta E_{\rm s}^{({\rm subunit})}$ and $q_{\rm
a}, q'_{\rm a}, q''_{\rm a}$ to describe the free energy difference
between histone particles in solution and those bound to a nucleosome
(not counting the associated subunit-DNA contacts) and rebinding
rates.  Kinetically, if a histone dimer fully detaches from DNA but is
still linked to the DNA-bound tetramer, it is held close to the DNA,
resulting in a locally high effective dimer concentration.  Since
dimers in solution are much more dilute, the binding rate should be
much smaller than $k_{\rm on}$.  When free histones are present in the
solution, the stability of the nucleosome can be modulated by the
concentration of free histones.  For example, a recent experiment
reported that the free histone concentration is a key modulator of
different responses of nucleosomes to the progression of replication
fork \cite{Gruszka2020sep}; our model can potentially be adapted to
provide a mechanistic insight into this observation.

% \deleted{To include protein-facilitated detachment in our multimeric
% model, we assumed the same association and dissociation rates, $\pa$
% and $\pd$, used in the linear facilitated detachment model.}

  \begin{table}[htbp]
    \centering
    \begin{tabular}{llcc}
    \hline
      \textbf{model} & \textbf{detachment method} & $\vert \lambda_0|$  \\
      \hline
      simple histone \hspace{-8mm} & & &\\

      &   spontaneous & Eq.\eqref{EQUATION:PERTURBED} \\
     &  facilitated & Eq.\eqref{eq:intact_effective_facilitation}  \\
     & one-sided, spontaneous & Eq.\eqref{eq:one_sided_unfacilitated} \\
     & one-sided, facilitated & Eq.\eqref{eq:one_sided_effective_facilitation} \\
      \hline 
      multimeric histone \hspace{-8mm} & & & \\
     &  irreversible spontaneous & Eq.\eqref{eq:multimeric_unfacilitated_disassembly}  \\
     &  reversible spontaneous & Eq.\eqref{eq:lambda_d_mul} \\
     &  irreversible facilitated & Eq.\eqref{eq:multimeric_facilitated_irreversible}  \\      
     & reversible facilitated & Eq.\eqref{eq:lambda_irr_fac} \\
    \hline
    \end{tabular}
 \caption{A summary of main analytical approximations developed in this paper.}
    \label{RESULTS}
\end{table}

% \paragraph{Competition between different disassembly pathways}
Histones can disassemble from DNA either as a whole, or in a piecewise
fashion. Our two classes of models represent two parallel pathways of
nucleosome disassembly. The first pathway is defined by linear
intact-histone detachment, while the second pathway reflects
disassembly involving histone fragmentation.  Preference of one
pathway over the other depends on the subunit unlinking and remodeler
binding rates.  Typically, the multimeric detachment model
disassembles faster than the linear detachment model.  However, strong
nucleosome remodelers, high concentrations of free histones in
solution, and strong binding between histone dimers and tetramers can
render the multimeric disassembly pathway less likely.

%In other parameter regimes, it is necessary to discuss multimeric
%disassembly of nucleosomes.

All of our results are derived assuming uniform binding and unbinding
rates between histone and DNA contacts, and are listed in
Table \ref{RESULTS}.  Numerical tests of heterogeneous $\koff$
performed in Appendix~\ref{Appendix:random_landscapes} suggest that
disassembly of nucleosome that have \textit{random} histone-DNA
contact energy profiles (depending on DNA sequence) can be well
characterized by the average binding energy. However, recent analysis
suggest that these rates may exhibit cooperativity depending on the
amount of unwrapped DNA \cite{SHEU2022}. Our model can be extended to
capture such effects by allowing $\koff(n_{1}, n_{2})$ to explicitly
depend on the state of the system or by simply allowing $\koff$ to be
different constants for the $\Htd$-DNA and $\HtHf$-DNA contacts.

Although our predictions focus on the mean time to disassembly of a
single histone, higher moments or distributions of disassembly times
can in principle be numerically extracted from our stochastic
model. Our suite of models provide the building blocks for
constructing higher-level models of rearrangement of interacting
nucleosome assemblies \cite{epj_chou,Marko2013,Zhao2021,MOROZOVPRE}
that occur during important cellular processes such as transcription
and replication \cite{Thiriet2021} and post-translational modification
of histone binding energies \cite{Bundschuh2011,Cai2021}.

% \section*{Supplementary Materials} Mathematical appendices detailing
% assumptions and derivations used to develop our models are provided
% in the Supplementary Materials.

\begin{acknowledgments}
The authors acknowledge support from the Army Research Office through
grant W911NF-18-1-0345 and the National Institutes of Health through
grant R01HL146552.
\end{acknowledgments}

\section*{Author Declarations}
\noindent {\bf Conflict of Interest}

The authors have no conflicts to disclose.

\section*{Data Availability}
The data that support the findings of this study are available within
the article.
%%%%%%%%%%%%%%%%%%%%%%%%%%%%%%%%%%%%%%%%%%%%%%%%%%%%%%%%%%
%%%%%%%%%%%%%%%%%%%%%%%%%%%%%%%%%%%%%%%%%%%%%%%%%%%%%%%%%%
%%%%%%%%%%%%%%%%%%%%%%%%%%%%%%%%%%%%%%%%%%%%%%%%%%%%%%%%%%
%%%%%%%%%%%%%%%%%%%%%%%%%%%%%%%%%%%%%%%%%%%%%%%%%%%%%%%%%%

\appendix
%
% Label prefix for table and figure references
%
%\renewcommand{\thetable}{S\arabic{table}}
%\renewcommand{\thefigure}{S\arabic{figure}}
% set counter
%\setcounter{table}{0}
%\setcounter{figure}{0}

\begin{table*}[ht]
    \begin{center}
\begin{tabular}{lcc}
  \hline \textbf{object} & \textbf{symbol} & \textbf{examples}
  \\ 
  \hline 
  matrices and vectors & bold letters & ${\Q, {\bf P}(t), {\bf x}, {\bf n}...}$ \\ 
  scalars, components of matrices and vectors & normal letters & $W_{ij},P(n_1,n_2,t), x_i, n_1,n_2$  \\
  eigenvalues and eigenvectors sorted by real parts in descending order & $\lambda_i, {\bf v}_i$ & $\lambda_0, {\bf v}_0, \lambda_1,{\bf v}_1, \cdots$ \\
  the state space for the undissociated histone & $\Omega$ & - \\
  a state in the state space & ${\bf x}$ & - \\
  the state of the dissociated histone & $\Omega^*$ & - \\
  vectors with all entries equal to a number & bold numbers & ${\bf 1, 0}$ \\
  fully histone-DNA bound state with all contact sites closed  & ${\bf 1}$ & ${\bf 1}$ \\
  Euclidean inner product of two vectors & $\langle \cdot, \cdot \rangle$ & $\langle {\bf x}, {\bf y} \rangle$ \\
  transpose of a vector or matrix & $\cdot^{\intercal}$ & ${\bf x}^{\intercal}, \Q^{\intercal}$ \\
  \textit{dimensionless} first passage time (FPT) starting from $x\in \Omega$ to detached state $\Omega^*$ & $T({\bf x})$ & $T({\bf 1})$ \\
  estimates for a quantity & hat over the symbol & $\hat{\lambda}_0, \widehat{\mathbb{E}[T({\bf 1})]}$ \\ 
  vectors with first row removed, or matrices with first row and column removed & $\overline{\cdot}$ & $\oline{\bf v}, \oline{\Q}$ \\
  %
  % quantities with a physical dimension & tilde over the symbol & $\tilde{\Q}, \tilde{\lambda}_0$ \\
  quantities relevant to remodeler-facilitated models & subscript ``p'' & $\hat \lambda_{0,\rm p}, E_{\rm p}$ \\
  quantities relevant to multimeric histone models & subscript ``q'' & $ E_{\rm q}, \hat \lambda_{0, \rm q}, \hat \lambda_{0,\rm p,q} $ \\
  \hline
  \end{tabular}
    \caption{General nomenclature for mathematical symbols and 
    objects.}
  \label{tab:nomenclature}
  \end{center}
\end{table*}

\section{Transition matrices, eigenvectors, and eigenvalues for 
 the intact-histone, spontaneous detachment model}
\label{Appendix:spontaneous-intact-histone}

\subsection{Transition matrix for the intact-histone, spontaneous
detachment model}
\label{sec:matrices-si}

To simplify our mathematical analysis, we normalize
all rates by $\kon$ so that $\koff/\kon = \varepsilon, \kd/\kon = s$,
and $\lambda_{i}$ are dimensionless. It is straightforward to
reconstruct physical rates and times by multiplying or dividing by
$\kon$.  We allow the total number of contact sites $N$ to be a
variable and relabel the transition matrix $\Q$ as $\Q_N \equiv {\bf
A}_N + \varepsilon {\bf B}_N + s {\bf C}_N$, which can be generated
recursively.

The exact form of transition matrix depends on how the different
states $(n_1,n_2)$ of $\Omega$ are enumerated. We choose to order
states by first grouping ones with the same $n_1+n_2$ together, then
ordering the others by ascending order in $n_{1}+n_{2}$. Finally,
states in the same group are ordered in ascending $n_{1}$. For
example, the first few states are $(0,0), (0,1), (1,0), (0,2),
(1,1), \dots$. This bookkeeping scheme allows us to construct the
transition matrices via simple recursion.  Setting $\A_1=0$, $\A_n$ is
\begin{equation}
  {\bf A}_{n} = 
  \setlength\arraycolsep{3pt}
  \left[\begin{array}{cccccc}
 \A_1 & \F_1   &  {\bf 0} & \cdots & \cdots       & {\bf 0}   \\
 {\bf 0} & \D_1   &  \F_2    & {\bf 0} &\cdots & {\bf 0} \\
 {\bf 0} & {\bf 0}&   \D_2     & \F_3 & \ddots &\vdots \\
 {\bf 0} &  \vdots&  \ddots     & \ddots & \ddots & {\bf 0} \\
 {\bf 0} &  \vdots& & \ddots     & \ddots & \F_{n-1} \\
 {\bf 0} &  {\bf 0} & \cdots     & \cdots & {\bf 0} & \D_{n-1} \\
  \end{array}\right],
  \label{eq:A_n}
\end{equation}
where $\F_k$ is a $k \times (k+1)$ matrix, with the two longest
diagonals set to 1 (all other entries are zero), representing the
closure of one open contact site. The matrix
$\D_k=\textrm{diag}\{-1,-2,-2...,-2,-1\}$ is a $(k+1) \times (k+1)$
diagonal matrix determined by setting the column sums of $\A_{k}$ to
0. By construction, $\A_N$ is a $\frac{N(N+1)}{2} \times
\frac{N(N+1)}{2}$ upper triangular matrix with the diagonal entries
$\left\{0,-1, \cdots, -1,-2,\cdots,-2 \right\}$. Specifically, there
is one diagonal entry with value $0$, $2(N-1)$ diagonal entries with
value $-1$, and the remaining $(N-1)(N-2)/2$ diagonal entries with
value $-2$.

Elements in
\begin{equation}
  {\bf B}_{n} =  \setlength\arraycolsep{3pt}
  \left[\begin{array}{cccccc}
{\bf D}_{0}' & {\bf 0} & {\bf 0} & \cdots & \cdots & {\bf 0}\\
{\bf F}_{1}^{\intercal} &  {\bf D}_{1}'   & {\bf 0} & \cdots &\cdots & {\bf 0} \\
{\bf 0} &  \F_2^{\intercal} & {\bf D}_{2}' & \ddots &  &\vdots \\
{\bf 0} &  {\bf 0}  & \ddots & \ddots &\ddots & \vdots \\
{\bf 0} &   \vdots   & \ddots & \ddots & {\bf D}_{n-2}' & {\bf 0} \\
{\bf 0} &   {\bf 0} & \cdots & {\bf 0} & \F_{n-1}^{\intercal} & {\bf 0} \\
  \end{array}\right].
  \label{eq:B_n}
\end{equation}
represent rates of transitions to higher $n_1 + n_2$.  A simple way of
defining $\B_n$ is to transpose $\A_n$ and change the diagonal terms
so that each column adds up to 0 to conserve total probability. The
reason why we can do this is that for every transition lowering
$n_1+n_2$, there is exactly one opposing transition raising $n_1+n_2$.
Since $\Q_{ij}$ represents transition rate from state $j$ to state
$i$, we transpose the matrix to invert the direction of transition.
In Eq.~\eqref{eq:B_n}, the matrix $\F_k^{\intercal}$ is the transpose
of $\F_k$ and ${\bf D}_{k}'$ is a $(k+1) \times (k+1)$ diagonal matrix
with all diagonal entries being $-2$. The last diagonal entry ${\bf
0}$ is an $n\times n$ matrix with all entries being zero.

\vspace{3mm}
Finally, the matrix $\C_n$ represents the transitions leaving the
state space into the absorbing states. For $n\in \mathbb{Z}_{+}$,

\begin{equation}
\C_n = \left(\begin{array}{cc} \0 & \0 \\ \0 & -\I_{n} \end{array}\right),
\end{equation}
where $\C_n$ is an $\frac{n(n+1)}{2} \times \frac{n(n+1)}{2}$ matrix
and $\I_n$ is the identity matrix in $\mathbb{R}^{n\times n}$.

To be concrete, the matrices $\A_3$, $\B_3$ and $\C_3$ are explicitly
\begin{widetext}
\begin{equation}
%  \begin{small}
    \A_3 = \left(\begin{array}{cccccc} 0 & 1 & 1 & 0 & 0 & 0\\
    0 & -1 & 0 & 1 & 1 & 0\\
    0 & 0 & -1 & 0 & 1 & 1\\
    0 & 0 & 0 & -1 & 0 & 0\\
    0 & 0 & 0 & 0 & -2 & 0\\
    0 & 0 & 0 & 0 & 0 & -1 \end{array}\right),\quad
    \B_3 = \left(\begin{array}{cccccc} -2 & 0 & 0 & 0 & 0 & 0\\
    1 & -2 & 0 & 0 & 0 & 0\\
    1 & 0 & -2 & 0 & 0 & 0\\
    0 & 1 & 0 & 0 & 0 & 0\\
    0 & 1 & 1 & 0 & 0 & 0\\
    0 & 0 & 1 & 0 & 0 & 0 \end{array}\right),\quad
    \C_3 = \left(\begin{array}{cccccc} 0 & 0 & 0 & 0 & 0 & 0\\
    0 & 0 & 0 & 0 & 0 & 0\\
    0 & 0 & 0 & 0 & 0 & 0\\
    0 & 0 & 0 & -1 & 0 & 0\\
    0 & 0 & 0 & 0 & -1 & 0\\
    0 & 0 & 0 & 0 & 0 & -1 \end{array}\right)
%    \end{small}
\label{ABC3}
\end{equation}
\end{widetext}

\subsection{Perturbation analysis of the intact-histone,
spontaneous detachment model}
\label{sec:matrices-si-perturb}
We will develop a series expansion of the eigenvector ${\bf v}_{0}$
associated with the principal eigenvalue
$\lambda_0 \equiv \lambda_0(s)$ of $\Q(s)= \A + \varepsilon \B + s \C$
and use it to compute the eigenvalue $\lambda_0(s)$ as a function of
$s$.

We begin with a general observation.  Let $\H$ be a matrix with a
simple eigenvalue 0. \added{Define $\oline{\bf H}$ as the submatrix
of $\H$ obtained by deleting the first row and column, and assume in
addition that $|\oline{\bf H}| \neq 0$. Denote the first column of
$\H$ excluding the first-row element by $\oline{\bf h}$. If ${\bf
v}$ is an eigenvector of $\H$ with eigenvalue $0$ and is written in the
form} ${\bf v} = \left[\begin{array}{c} 1 \\ {\oline{\bf v}}
\end{array}\right]$, \added{then $\H {\bf v} = 0{\bf v}= 0$.
This implies $\oline{\bf h} + \oline{{\bf H}} \oline{\bf v} = 0$ and
the general relationship}
\begin{align}
\oline{\bf v} = -\oline{\bf H}^{-1}\oline{\bf h}.
\label{X1}
\end{align}
\vspace{2mm}

\noindent \textit{Principal eigenvector for $\Q(0)$.} Since
$\Q(0)= \A+\ve\B$ is a transition matrix associated with a reversible
Markov chain, the eigenvector associated with the $0$-eigenvalue is
\begin{equation}
{\bf v}_0(s=0) = \left[\begin{array}{c}
  1 \\
  \varepsilon {\bf 1}_2 \\
  \vdots \\
  \varepsilon^{N-1} {\bf 1}_N
  \end{array}\right],
  \label{eq:x0_unperturbed}
\end{equation}
where ${\bf 1}_i \in \mathbb{R}^{i}$ is a vector of all ones.

\vspace{3mm}
\noindent \textit{Series expansion for ${\bf v}_0(s)$.}
Now, we set $\H \equiv \Q(s) - \lambda_0(s) \I$, denote the associated
principal eigenvector by ${\bf v}(s)$, and express it in the form
${\bf v}_0(s) = \left[\begin{array}{c} 1 \\ \oline{\bf v}_0(s)
\end{array}\right]$. Then, using
Eq.~\eqref{X1},
\begin{widetext}
\begin{equation}
  \begin{aligned}
    \added{\oline{\bf v}_0(s)} & \added{= -
    \big[\oline{\bf W}(0) + s \oline{\bf C} - \lambda_0(s) \I \big]^{-1}
    \oline{\bf w}
    = - \big[\I + s \oline{\bf W}^{-1}(0)
    \oline{\bf C} - \lambda_0(s) \oline{\bf W}^{-1}(0)\big]^{-1} 
            \oline{\bf W}^{-1}(0)  \oline{\bf w}, }
  \end{aligned}
\end{equation}
where \added{$\oline{\bf h}$ in Eq.~\eqref{X1} is set to
$\oline{\bf w}$ which} is equivalent to the first column
of $\Q(s)$, minus the first element, and is independent of $s$. All
terms that depend on the perturbation $s$ are explicitly indicated.
Recall the Neumann series expansion for $(\I + {\bf T})^{-1} =
\sum_{k=0}^{\infty} {\bf T}^k$ provided the operator norm $\|T\| < 1$.
In this case, we can write
\begin{equation}
  \begin{aligned}
   \added{ \oline{\bf v}_0(s)} & \added{= -\left[\I + \sum_{k=1}^{ \infty }
   \Big( \lambda_0(s) \oline{\bf W}^{-1}(0)
   - s \oline{\bf W}^{-1}(0) \oline{\bf C} \Big)^k 
    \right] \oline{\bf W}^{-1}(0) \oline{\bf w}.}
  \end{aligned}
  \label{eq:series_expansion}
\end{equation}

\vspace{2mm}
\noindent \textit{Radius of convergence.} We first estimate the values
of $\lambda$ and $s$ for which series expansion
(\ref{eq:series_expansion}) converges. This amounts to evaluating the
operator norm of the term \added{$(\lambda_0(s) \oline{\bf W}(0)^{-1} 
- s \oline{\bf W}(0)^{-1}\oline{\bf C})$}. Since $\C$ is diagonal
with entries $0$ and $-1$, $\|\added{\oline{\bf C}}\| = 1$,
and we find the bound
\begin{equation}
  \begin{aligned}
  \added{  \|\lambda_0(s) \oline{\bf W}^{-1}(0)
    -  \oline{\bf W}^{-1}(0) s \oline{\bf C} \|  } &
   \added{ \leq  |\lambda_0(s)| \|\oline{\bf W}^{-1}(0)\|
   + s \|\oline{\bf W}^{-1}(0)\| \|\oline{\bf C}\|
    \leq  \big(|\lambda_0(s)| + s\big)\, \|\oline{\bf W}^{-1}(0)\|.}
  \end{aligned}
\end{equation}
Estimating the operator norm of \added{$\oline{\bf W}^{-1}(0)$}
is more involved. We note that ${\bf \hat{Q}}_{1,1}(0)$ is an
$N(N+1)/2-1 \times N(N+1)/2-1$ matrix.  An upper bound for the
operator norm is given by
\begin{equation}
  \begin{aligned}
   \added{ \| \oline{\bf W}^{-1}\| \leq \frac{(N+2)(N-1)}{2} \max_{i,j}
    |\oline{\bf W}^{-1}(i,j)|.}
  \end{aligned}
\end{equation}
We now characterize the entries of \added{$\oline{\bf W}^{-1}(0)$}
by applying the same perturbation formula again to
\added{$\oline{\bf W}(0) = \oline{\bf A} + \varepsilon
\oline{\bf B}$.}  Note that $\A$ and $\B$ are block tridiagonal and
$\A$ is upper-triangular:

\begin{equation}
\added{\oline{\bf W}}(0)= \oline{\bf A}+\varepsilon \oline{\bf B}=
\setlength\arraycolsep{3pt}
\left[\begin{array}{ccccc}
\D_1   & \F_2 & {\bf 0} &\cdots & {\bf 0} \\
{\bf 0} & \D_2 & \F_3 & \ddots &\vdots \\
 \vdots  & \ddots & \ddots & \ddots & {\bf 0} \\
 \vdots   & & \ddots & \ddots & \F_{N-1} \\
 {\bf 0} & \cdots & \cdots & {\bf 0} & \D_{N-1} \\
\end{array}\right] + \varepsilon \setlength\arraycolsep{3pt}
\left[\begin{array}{ccccc}
{\bf D}_{1}'   & {\bf 0} & \cdots &\cdots & {\bf 0} \\
\F_2^{\intercal} & {\bf D}_{2}' & \ddots &  &\vdots \\
{\bf 0}  & \ddots & \ddots &\ddots & \vdots \\
 \vdots   & \ddots & \ddots & {\bf D}_{N-2}' & {\bf 0} \\
 {\bf 0} & \cdots & {\bf 0} & \F_{N-1}^{\intercal} & {\bf 0} \\
\end{array}\right],
\end{equation}
in view of the block matrix representations given by
Eqs~\eqref{eq:A_n} and \eqref{eq:B_n}.  Since $\oline{\bf A}$ is
bidiagonal, its inverse is
\begin{equation}
\oline{\bf A}^{-1} = \setlength\arraycolsep{3pt}
\left[\begin{array}{ccccc}
\D_1^{-1}   & \cdots & \cdots &\cdots & * \\
{\bf 0} & \D_2^{-1} & \ddots & \ddots &\vdots \\
 \vdots  & \ddots & \ddots &\ddots & \vdots\\
 \vdots   & & \ddots & \ddots & \vdots \\
 {\bf 0} & \cdots & \cdots & {\bf 0} & \D_{N-1}^{-1} \\
\end{array}\right]
\end{equation}
and we can expand the inverse $\added{\oline{\bf W}}^{-1}(0)$ as
\begin{equation}
\begin{aligned}
\added{\oline{\bf W}}^{-1}(0) & =\oline{\bf A}^{-1}
+ \sum_{i=1}^{\infty} (-\varepsilon \oline{\bf A}^{-1}
\oline{\bf B})^{i}\oline{\bf A}^{-1} \\
&= \setlength\arraycolsep{5pt}
\left[\begin{array}{ccccc}
\D_1^{-1} +o(1) & \cdots & \cdots &\cdots & * \\
O(\varepsilon) & \D_2^{-1}+o(1) & \ddots & \ddots &\vdots \\
 \vdots  & \ddots & \ddots &\ddots & \vdots\\
 \vdots   & & \ddots & \ddots & \vdots \\
 O(\varepsilon^{N-2}) & \cdots & \cdots & O(\varepsilon) & \D_{N-1}^{-1}+o(1) \\
\end{array}\right] \\[4pt]
&=\left[\begin{array}{ccccc}
  O(1) & \cdots & \cdots &\cdots & O(1) \\
  O(\varepsilon) & O(1) & \ddots & \ddots & O(1) \\
   \vdots  & \ddots & \ddots &\ddots & O(1)\\
   \vdots   & & \ddots & \ddots &  O(1) \\
   O(\varepsilon^{N-2}) & \cdots & \cdots & O(\varepsilon) & O(1)\\
  \end{array}\right].
  \label{eq:Q11_inverse_scaling}
\end{aligned}
\end{equation}
\end{widetext}
Here, each $*$ denotes a block matrix with entries of order $O(1)$.

We can show by induction that the maximum entry of $\oline{\bf
A}^{-1}$ is less or equal to $1$. Therefore, the maximum entry of
$\added{\oline{\bf W}}^{-1}$ is bounded by $1+ O(\varepsilon)$
and we conclude that the radius of convergence of the series expansion
in Eq.~\eqref{eq:series_expansion} is
\begin{equation}
  s+ |\lambda_0(s)| \leq \frac{2}{(N+2)(N-1)\big(1+O(\varepsilon)\big)}.   
\end{equation}
In other words, the series expansion can be valid even if $s \geq
\varepsilon$. The radius of convergence is principally determined by
the operator norm of $\oline{\bf A}^{-1}$.

\vspace{2mm}
\noindent {\bf Perturbations to the eigenvector.} We next explicitly 
evaluate how the eigenvector changes under first order perturbation.
Expanding Eq.~\eqref{eq:series_expansion} to first order in $s +
|\lambda|$, we find
\begin{equation}
  \begin{aligned}
    \added{\oline{\bf v}_0}(s) = & \added{\oline{\bf v}_0}(0)
    + \added{\oline{\bf W}}^{-1}(0)
    \big(\lambda_0(s) \I - s \added{\oline{\bf C}}\big)
    \added{\oline{\bf v}_0}(0)\\
   \: & \hspace{3.5cm} + O\big((s+ |\lambda|)^2\big).
    \label{FIRSTORDER}
  \end{aligned}
\end{equation}
Inserting the estimate of $\added{\oline{\bf W}}^{-1}$ from
Eq.~\eqref{eq:Q11_inverse_scaling}, the definition of
$\added{\oline{\bf C}}$, and $\v(0)$ derived from
Eq.~\eqref{eq:x0_unperturbed} into Eq.~\eqref{FIRSTORDER}, we observe
that
\begin{equation}
\begin{aligned}
\oline{\bf W}^{-1}(0) \oline{\bf v}_0(0) &
= \!\left[\!\begin{array}{c}
    O(\varepsilon){{\bf 1}_2} \\
    O(\varepsilon^2){{\bf 1}_3} \\
    \vdots \\
    O(\varepsilon^{N-1}){{\bf 1}_{N}} \end{array}\right], \\[6pt]
    \oline{\bf W}^{-1}(0) \oline{\bf C} \oline{\bf
    v}_0(0) & =\! \left[\begin{array}{c}
    O(\varepsilon^{N-1}){\bf 1}_2 \\
    O(\varepsilon^{N-1}){\bf 1}_3 \\
\vdots \\
O(\varepsilon^{N-1}){\bf 1}_{N} \end{array}\right].
\end{aligned}
\label{eq:Q11_inverse_x1}
\end{equation}

Let ${\bf 1}$ be the vector with all entries equal to $1$. The
 eigenvalue $\lambda_0(s)$ satisfies the equation
\begin{equation}\label{EQUATION:UNPERTURBED_APP}
\begin{aligned}
\lambda_0(s) &= \frac{\textbf{1}^{\intercal} \Q(s) \begin{bmatrix} 1\\
\oline{\bf v}_0(s) \end{bmatrix}}{\textbf{1}^{\intercal} \oline{\bf v}_0}(s) \\
    \: & = \frac{\textbf{1}^{\intercal} s \oline{\bf C} \oline{\bf
    v}_0(s)}{\textbf{1}^{\intercal} \oline{\bf
    v}_0(s)} \\
    \: & = s \textbf{1}^{\intercal} \oline{\bf C} \oline{\bf v}_0(0)
    + s \textbf{1}^{\intercal}\oline{\bf C} \oline{\bf W}^{-1}(0)
    \oline{\bf v}_0(0) \lambda_0(s) \\ \: & \hspace{2.5cm}
    + s^2 \textbf{1}^{\intercal}\oline{\bf C} \oline{\bf W}^{-1}(0)
    \oline{\bf C} \oline{\bf v}_0(0)\\[6pt] 
\: & = \added{-} N
    s \varepsilon^{N-1} + s O(\varepsilon^{N-1}) \lambda_0(s) + s^2
    O(\varepsilon^{N-1}).
\end{aligned}
\end{equation}
Therefore, the lowest order approximation to the eigenvalue is
\begin{equation}
\lambda_0(s) \approx -N s \varepsilon^{N-1} + O(s^2 \varepsilon^{N-1})
= -Ns \varepsilon^{N-1}\big(1+O(s)\big).
\label{eq:lambda_0_approx}
\end{equation}
This approximation holds whenever $s \ll 1$, (even if
$s \gg \varepsilon$), which guarantees the convergence of the series
expansion in Eq.~\eqref{eq:series_expansion}.

\added{Substituting Eqs.~\eqref{eq:Q11_inverse_x1} and
\eqref{eq:lambda_0_approx} back into Eq.~\eqref{FIRSTORDER},
we find the lowest order approximation to the eigenvector}
\begin{equation}
  \mathbf{v}_0(s)=\left[\begin{array}{c}
    1\\[5pt]
   \big( \varepsilon + \lambda_0(s)O(\varepsilon) + s O(\varepsilon^{N-1})\big)\mathbf{1}_2 \\[5pt]
   \big( \varepsilon^2 + \lambda_0 (s) O(\varepsilon^2) +s O(\varepsilon^{N-1})\big) \mathbf{1}_3 \\[5pt]
    \vdots \\[5pt]
    \big(\varepsilon^{N-1} + \lambda_0 (s)O(\varepsilon^{N-1})+sO(\varepsilon^{N-1})\big) \mathbf{1}_N
    \end{array}\right].
    \label{eq:x_approx}
\end{equation}
\noindent \added{Given that $\lambda_0(s) =O(s \varepsilon^{N-1})$,
for each component $v_0(n_1,n_2;s)$ of ${\bf v}_0(s)$, we have}
\begin{equation}
 v_0(n_1, n_2;s) = \big(1+O(s) \big) v_0(n_1,n_s;0).
  \label{eq:x0_approx}
\end{equation}

\vspace{2mm}
\subsection{Eigenvalues and first passage times}
\label{sec:eigenvalues}
Here, we present some general results on the eigenvalues and
eigenvectors of the transition matrix ${\bf W}(s)$ and their relation
to FPTs.

First, let $\lambda, {\bf v}$ be an eigenvalue and eigenvector of
${\bf W}(s)$, respectively, such that $\lambda \leq 0$ and all
components of ${\bf v}$ are nonnegative. Since the probability vector
${\bf P}(t)$ satisfies ${\mathrm{d}{\bf P}}/\mathrm{d} t = {\bf W}{\bf
P}$, if ${\bf P}(0) = {\bf v}$, then ${\bf P}(t) = e^{\lambda t} {\bf
v}$. 

In a FPT problem, we set the target state $\Omega^*$ to be
absorbing. Restriction of the transition matrix on states other than
$\Omega^*$ makes the total probability $P_{\rm tot}(t)= P[X(t) \notin
\Omega^*]=\langle {\bf 1}, {\bf P}(t) \rangle$ nonincreasing with time
$t$, where $X(t)$ is used to denote a random trajectory of the system,
and $\langle \cdot, \cdot \rangle$ is the Euclidean inner product,
i.e., $\langle {\bf x}, {\bf y} \rangle = \sum_{(n_1,n_2) \in \Omega}
x(n_1,n_2) y(n_1,n_2)$.

In other words, $P_{\rm tot}(t)$ indicates the probability that the
system has not reached the target state $\Omega^*$ by time $t$, and is
equivalent to the survival probability in the context of FPT
problems. $-\mathrm{d}P_{\rm tot}/\mathrm{d}t$ is the probability
density function of the FPT to $\Omega^*$, and is denoted by $f(t)$.

When ${\bf P}(0)={\bf v}$, we have ${\bf P}(t) = e^{\lambda t} {\bf
v}$ and ${P}_{\rm tot}(t) = \langle {\bf 1}, {\bf P}(t) \rangle =
e^{\lambda t} \langle {\bf 1}, {\bf P}(0) \rangle$. In view of the
probabilistic interpretation of $P_{\rm tot}(t)$, we may assume that
${\bf v}$ is normalized, i.e., $\langle {\bf 1}, {\bf v} \rangle =
1$. Therefore, we have
\begin{equation}
P_{\rm tot}(t) = e^{\lambda t},\quad f(t) = - \lambda e^{\lambda t}.
\label{eq:Ptot_eigenvalue}
\end{equation}
Here $f(t)$ represents the distribution of first passage times to
$\Omega^*$ from a normalized non-negative eigenvector ${\bf v}$, and
follows an exponential distribution with rate $-\lambda$. The MFPT is
thus given by $1/(-\lambda)$.

Next, consider the case where eigenvalues of ${\bf W}$ satisfies
$0>\lambda_0 \gg \operatorname{Re}(\lambda_i), \forall i \geq 1$, and
the eigenvector ${\bf v}_0$ associated with $\lambda_0$ is
nondegenerate, nonnegative, and normalized. For simplicity, we assume
that $\Q$ is diagonalizable although this can be relaxed by
considering the Jordan canonical form of non-diagonalizable matrices.

\added{Let ${\bf P}(0)={\bf P}_0$ be an arbitrary distribution over the states
other than $\Omega^*$, then ${\bf P}_0$ admits a unique decomposition
${\bf P}_0 = \sum_{i=0}^{N-1} c_i {\bf v}_i$, where ${\bf v}_i$ is the
eigenvector of ${\bf W}$ associated with $\lambda_i$. By linearity of
the equation ${\mathrm{d}{\bf P}}/\mathrm{d} t = {\bf W}{\bf P}$, the
solution is given by}
\begin{equation}
{\bf P}(t) = \sum_{i=0}^{N-1} c_i e^{\lambda_i t} {\bf v}_i 
  =e^{\lambda_0 t} \sum_{i=0}^{N(N+1) / 2} 
  c_i {\bf v}_i e^{\left(\lambda_i-\lambda_0\right) t}.
 \label{eq:P0_solution}
\end{equation}
When $\operatorname{Re}(\lambda_i) \ll \lambda_0 <0$,
$\operatorname{Re}(\lambda_i - \lambda_0) \ll 0, \forall i \geq 0$ and
we have
\begin{equation}
  \begin{aligned}
  \mathbf{P}(t) & = c_0 e^{\lambda_0 t} {\bf v}_0
  +O\left(e^{-\operatorname{Re}(\lambda_1 - \lambda_0)t}\right),\\
  P_{\mathrm{tot}}(t) & =c_0 e^{\lambda_0 t}+
  O\left(e^{-\operatorname{Re}(\lambda_1 - \lambda_0)t}\right).
  \end{aligned}
\end{equation}
That is to say, in the long time limit, the probability distribution
of the system is dominated by the eigenvector ${\bf v}_0$, and the
survival probability $P_{\rm tot}(t)$ decays exponentially with rate
$\lambda_0$. The MFPT is thus given by $c_0/(-\lambda_0)$.

This analysis applies to a general FPT problem.  In our specific case
of nucleosome disassembly and other scenarios where
the \textit{absorbing boundary method} is applicable, the transition
matrix ${\bf W}$ can be considered as a perturbed transition matrix of
an irreducible Markov chain.  In other words, there exists a
decomposition ${\bf W} = {\bf W}_0 + s \Delta {\bf W}$, where $s$ is a
small parameter. We treat the eigenvectors and eigenvalues as
functions of ${s}$, denoted as $\lambda_i(s)$ and ${\bf v}_i(s)$,
respectively.

Since ${\bf W}_0$ is a transition matrix of a continuous time Markov
chain, we have ${\bf 1^{{\intercal}} W}_0={\bf 0}$, i.e., ${\bf 1}$ is
a left eigenvector of ${\bf W}_0$ associated with eigenvalue
$0$. Therefore, we have
\begin{equation}
  0=\langle {\bf 1}, {\bf W}_0 {\bf v}_i(0) \rangle = 
  \lambda_i (0) \langle {\bf 1}, {\bf v}_i(0) \rangle.
\end{equation}
Irreducibility implies $\lambda_i \neq 0$ for all $i \geq 1$ and thus
$\langle {\bf 1}, {\bf v}_i(0) \rangle = 0$ for all $i \geq 1$.
Therefore, under small perturbation, we have $\left\langle\mathbf{1},
\mathbf{v}_i(s)\right\rangle= \langle \mathbf{1}, {\bf v}_{i}(0) \rangle
+ O(s)\text { as } s \rightarrow 0$. We may in addition require that
$\langle {\bf 1}, {\bf v}_0(0) \rangle=1$.  Note that $\langle {\bf
1}, \mathbf{P}_0\rangle=1$ for any probability vector
$\mathbf{P}_0$. Therefore,
\begin{equation}
  \begin{aligned}
1&=\langle {\bf 1}, \mathbf{P}_0\rangle\\
&= \left\langle {\bf 1}, \sum_{i\geq 0} c_i {\bf v}_i(s) \right\rangle
\\&=c_0\big[1+O(s)\big]+\sum_{i \geq 1} c_i O(s)
\\&=c_0\big[1+O(s)\big]+O(s)
  \end{aligned}
\end{equation}
and $c_0 \sim 1+O(s)$. Consequently, in the case of the
intact-histone, unfacilitated disassembly model, the MFPT
$\mathbb{E}[T({\bf x})]$ from any initial state ${\bf x}$ in $\Omega$
to the fully detached state $\Omega^*$ is given by
\begin{equation}
  \mathbb{E}[T({\bf x})] = \frac{1}{-\lambda_0} + O(s).
  \label{eq:mean-first-passage-time-simple-case}
\end{equation}
Moreover, $T({\bf x})$ is approximately exponentially distributed
with rate $-\lambda_0$ for any initial state ${\bf x}$ in $\Omega$ so
that 
\begin{equation}
  \mathbb{P}(T({\bf x}) \leq t) = 1 - e^{\lambda_0 t} + O(s).
  \label{eq:exponential-distribution}
\end{equation}

The asymptotic exponential distribution and fast relaxation to the
steady state properties of this simple system make it possible to
treat the simple model as a single coarse-grained state, with
transition rates $N e^{-N \varepsilon}$ to $\Omega^*$.

When other slower transitions are present, we can separate the fast
internal relaxation to steady state ${\bf v}_0(s)$ and slow dynamics
for transitions to external states. The transition rates to external
states can be calculated by averaging over the steady state
distribution ${\bf v}_0(s)$ and provides a good approximation to the
full dynamics, as long as the external transition rates are slower
than the relaxation rate $-\operatorname{Re}(\lambda_1)\approx 1$
(measured in units of $\kon$). As an example of this fast-slow
variable separation, we apply this approach to the coarse-graining of
the intact-histone, remodeler-facilitated disassembly model in
Fig.~\ref{fig:coarse-grained-model}. This coarse-graining yields
matched principal eigenvalues shown in Fig.~\ref{FIG:RESULTS}(b).

To formalize the separation of timescales, we consider the following 
general form of the perturbed dynamics:
\begin{equation}
  \frac{\mathrm{d}{\bf P}(t)}{\mathrm{d} t}
  = ({\bf W} + \delta {\bf M}) {\bf P}(t) + \delta {\bf m}(t),
  \label{eq:perturbed-dynamics}
\end{equation}
In Eq.~\eqref{eq:perturbed-dynamics}, $\delta \rightarrow 0$ and ${\bf
M}$ is an additional perturbation to the transition matrix ${\bf
W}={\bf W}_0 + s\Delta {\bf W}$. The vector $\delta {\bf m}(t)$ is a
source term. In the context of coarse-graining of the intact-histone,
remodeler-facilitated disassembly model in
Fig.~\ref{fig:coarse-grained-model}, we restrict the transition matrix
to the microstates within a coarse-grained macrostate. The vector
${\bf m}(t)$ represents the transitions from other macrostates to the
given macrostate, while $\delta {\bf M}$ represents the transitions
from the given macrostate to other macrostates, and ${\bf W}$
represents the transitions within the given macrostate.

We can still apply the diagonalization technique ${\bf W}= {\bf V}
{\bf \Lambda}{\bf V}^{-1}$ where ${\bf \Lambda}$ is a diagonal matrix
with diagonal entries $\lambda_i$ and ${\bf V}$ is a matrix whose
columns are the eigenvectors of ${\bf W}$.
\begin{equation}
  \frac{\mathrm{d}{\bf P}(t)}{\mathrm{d} t}={\bf V}{\bf \Lambda}
     + \delta{\bf M} {\bf V}
  [{\bf V}^{-1} {\bf P}(t)] + \delta {\bf m}(t).
  \label{eq:perturbed-dynamics-diagonalized}
\end{equation}
Left multiply by ${\bf 1}^{\intercal}$ and recall that as $s\to 0$,
${\bf 1}^{\intercal} {\bf V}=[1,0,\cdots,0]+O(s)$ and ${\bf V}^{-1}
{\bf P}(t) = [P_{\rm tot} + O(s), O(s), \cdots, O(s)]^{\intercal}$ for
any nonnegative vector ${\bf P}(t)$. This yields
\begin{equation}
  \frac{\mathrm{d}P_{\rm tot}(t)}{\mathrm{d} t}
  = (\lambda_0 + \delta {\bf 1}^{\intercal}
  {\bf M}{\bf v}_0)P_{\rm tot}(t) + \delta{\bf 1}^{\intercal}
  {\bf m}(t) +  O(\delta s)
   \label{eq:perturbed-dynamics-diagonalized-1}
\end{equation}
as $s \rightarrow 0$. Therefore, the survival probability $P_{\rm
tot}(t)$ corresponding to a coarse-grained macrostate can be
approximated by the following processes: the coarse-grained state
moving to the absorbing state with rate $-\lambda_0$, moving to other
coarse-grained states with rate $-\delta {\bf 1}^{\intercal} {\bf
M}{\bf v}_0$, and other states contributing to the coarse-grained
state with rate $\delta {\bf 1}^{\intercal} {\bf m}(t)$. This
approximation holds when $s$ and $\delta$ are small enough, compared
to $1$, i.e. $\kon$ in the context of our models. In other words, $-
\delta {\bf 1}^{\intercal} {\bf M}{\bf v}_0$ is the rate at which the
original steady state ${\bf v}_0$ leaves the coarse-grained state and
goes to other states under perturbation of $\delta {\bf M}$, and
$\delta {\bf 1}^{\intercal} {\bf m}(t)$ is the rate at which other
states contribute to any state inside the coarse-grained state.

As a specific example, in Eq.~\eqref{TRANSITION:ASSISTED} or
Eq.~\eqref{HN} in the main text, we may write the probability vector
${\bf P}$ in block form: $[{\bf p}_N, {\bf p}_{N-1}, ..., {\bf
p}_{1}]^{\intercal}$. Consider ${\bf p}_N$ as the coarse-grained
state, then ${\bf W}={\bf W}_N$, ${\bf M}={\bf M}_N$, ${\bf
m}(t)=\sum_j \tfrac{\pd}{\pa} {\bf G}_{N, j}{\bf p}_j(t)$, and $\delta
= \pa/\kon$.

% 
%%%%%%%%%%%%%%%%%%%%%%%%%%%%%%%%%%%%%%%%%%%%%%%%%%%%%%%
%%%%%%%%%%%%%%%%%%%%%%%%%%%%%%%%%%%%%%%%%%%%%%%%%%%%%%%
%%%%%%%%%%%%%%%%%%%%%%%%%%%%%%%%%%%%%%%%%%%%%%%%%%%%%%%

\section{Processive motor-assisted histone detachment}
\label{Appendix:one_sided}
Processive motors like DNA helicases slide along the DNA, attacking
the nucleosome from only one side of the histone-DNA footprint. In
this case, the histone is peeled off from the DNA in a one-sided
manner. Analogous to the two-sided peeling model, we can also
construct a one-sided peeling model consisting of the attached state
space $\Omega =\left\{ (m,n): m+n \leq N-1 \right\}$. Here $m$ records
the position of motor protein and $n$ records the number of remaining
histones.

When the remodeler is absent, the energy landscape of the one-sided
peeling model is similar to that of the two-sided peeling model, shown
in Fig.~\ref{FIG2}(b). The main difference lies in the degree of
degeneracy of each energy level. The lowest energy level is $N E_{\rm
c}$, corresponding to the unique $n_1=n_2=0$ state in the two-sided
peeling model and $n=0$ state in the one-sided peeling model. For
other energy levels $(N-j) E_{\rm c}$, there are $j+1$ states in the
two-sided peeling model and only 1 state in the one-sided peeling
model.

The contribution of degeneracy to the principal eigenvalue of the
two-sided model is the factor $N$ in Eq.~\eqref{EQUATION:PERTURBED},
which represents $N$ degenerate states at the energy level of $E_{\rm
c}$. In other words, the associated free energy is given by $E_{\rm c}
+ \log N$. By contrast, there is no degeneracy in the one-sided
peeling model, and the principal eigenvalue is simply given by
\begin{equation}
\label{eq:one_sided_unfacilitated}
  \lambda_0(\varepsilon) =
  -s\varepsilon^{N-1}\big[1+O(s)\big].
\end{equation}

Estimates of the principal eigenvalue of the two-sided
remodeler-assisted peeling model given in
Eqs.~\eqref{eq:intact_weak_facilitation}, ~\eqref{EQUATION:IRR}, and
\eqref{eq:intact_strong_facilitation} are built from the simple
estimate Eq.~\eqref{EQUATION:PERTURBED} of the spontaneous
nucleosome disassembly model. The analogous eigenvalues of the
one-sided peeling model are constructed from
Eq.~\eqref{eq:one_sided_unfacilitated} and are
\begin{align}
\hat{\lambda}_{0}(E_{\rm p} > E_{\rm c}) & \coloneqq -\frac{s e ^{(N-1)\left(E_{\mathrm{c}}-E_{\mathrm{p}}^{-}\right)}}
  {1+ e ^{(N-1)\left(E_{\mathrm{c}}-E_{\mathrm{p}}^{-}\right)}},
\label{eq:one_sided_weak_facilitation}
\end{align}

\begin{equation}
\begin{aligned}
\hat{\lambda}_{0}(E_{\rm p} \rightarrow - \infty)
\coloneqq - \min\Big\{se^{(N-1) E_{\rm c}}
+ \frac{\pa}{\kon}\!\sum_{j=1}^{N-1}\! e^{j E_{\rm c}}, s\Big\},
\label{eq:one_sided_strong_facilitation}
  \end{aligned}
  \end{equation}
  
\noindent \added{and}
\begin{equation}
\begin{aligned}
  \hat{\lambda}_{0, \rm p }
  & \coloneqq \max \left\{\hat{\lambda}_{0}(E_{\rm p}\!\rightarrow - \infty),\,
  \hat{\lambda}_{0}(E_{\rm p} > E_{\rm c})\right\}
  \label{eq:one_sided_effective_facilitation}
\end{aligned}
\end{equation}
\noindent These estimates are very close to those of the two-sided peeling model
as the entropic contribution ($\log N$) is negligible compared to the
enthalpic contribution ($N E_{\rm c}$), especially for strong contacts
$E_{\rm c} \ll -1$.

%%%%%%%%%%%%%%%%%%%%%%%%%%%%%%%%%%%%%%%%%%%%%%%%%%%%%%%%
%%%%%%%%%%%%%%%%%%%%%%%%%%%%%%%%%%%%%%%%%%%%%%%%%%%%%%%%
%%%%%%%%%%%%%%%%%%%%%%%%%%%%%%%%%%%%%%%%%%%%%%%%%%%%%%%%

\section{Transition matrix for intact-histone model with remodeling factors}
\label{Appendix:linear_facilitated_Q}

The linear detachment model is generalized to include remodeling
factors that can bind to DNA or contact sites on the partially
delaminated histone particle.  The total transition matrix that
connects states in the space $\Omega_{\rm p} \coloneqq
\{(m_1,m_2,n_1,n_2)\in \mathbb{N}^4:m_1+m_2+n_1+n_2 < N\}$ is
defined by $\Q_{N,{\rm p}}$, which can be expressed in block form:
\begin{widetext}
\begin{equation}\label{TRANSITION:ASSISTED}
\Q_{N,{\rm p}}=
\setlength\arraycolsep{3pt}
\left[\begin{array}{ccccc}
\Q_{N:1}   & \0 & \cdots &\cdots & \0 \\
\0 & \Q_{N-1:2} & \ddots & \ddots &\vdots \\
 \vdots  & \0 & \ddots &\ddots & \vdots \\
 \vdots   & \vdots & \ddots & \ddots & \0 \\
 \0 & \cdots & \cdots & \0 & \Q_{1:N} \\
\end{array}\right]+\frac{\pa}{\kon}\setlength\arraycolsep{3pt}
\left[\begin{array}{ccccc}
{\bf M}_{N}   & \0 & \cdots &\cdots & \0 \\
{\bf M}_{N-1,N} & {\bf M}_{N-1} & \ddots & \ddots &\vdots \\
 \vdots  & \ddots & \ddots &\ddots & \vdots \\
 \vdots   & & \ddots & \ddots & \0 \\
 {\bf M}_{1,N} & \cdots & \cdots & {\bf M}_{1,2} & \0 \\
\end{array}\right]+\frac{\pd}{\kon} {\bf G}
\end{equation}
Here, $\Q_n$ is the $\frac{n(n+1)}{2}\times\frac{n(n+1)}{2}$ matrix as
defined in the last subsection, and $\Q_{n:m}$ is the
$\frac{mn(n+1)}{2}\times\frac{mn(n+1)}{2}$ matrix constructed by
placing $m$ $\Q_{n}$ matrices along the diagonal blocks.  The matrix
${\bf M}_{i,j}$ describes the connectivity of transitions induced by
remodeler binding while ${\bf G}$ describes connectivity of
transitions \deleted{remodeler} induced by remodeler unbinding. ${\bf
M}$ and ${\bf G}$ depend on the specific transition mechanism.  In the
case of processive motor proteins that peel histones from DNA, $m_1$
and $m_2$ only increase or decrease by $1$ as the motor moves forward
or backward by one step.  For proteins that directly bind to DNA,
$m_1$ and $m_2$ can change by larger distances depending on the
numbers and positions of the collection of bound proteins.  For
example, when two DNA-histone contact sites are exposed, the protein
can bind to either site, and binding to the more interior site results
in $m$ increased by $2$. On the other hand, when the protein unbinds,
since $m$ only tracks the position of inward-most proteins, the next
value of $m$ depends on the position of the second most inward
protein.

\added{The construction of ${\bf H}_N$ depends on 
how the states are enumerated. We provide a possible enumeration
scheme below.}

\begin{itemize}

\item \added{For the states $(m_1,m_2,n_1, n_2)$, we first group the
    states by the value of $m_1+m_2$ in an ascending order. The first
    $N(N+1)/2$ entries correspond to the value of $(m_1+m_2)=0$, the
    next block represents entries satisfying $(m_1+m_2)=1$, where
    there are $2\times (N-1)N/2$ of them, and so on.}

\item \added{Within
    each block, we further group the states by the value of $m_1$ in
    ascending order, then by values of $n_1+n_2$, $n_1$ accordingly
    in  ascending order.}

\item \added{For fixed $(m_1,m_2)$, note that
    possible $(n_1,n_2)$ states are grouped in the same order as in the
    previous non-facilitated model. Therefore, the internal transition
    matrix restricted to those states can be described by the same
    ${\bf W}_{N-(m_1+m_2)}$.}

\item \added{To obtain the whole block with
    $m_1 + m_2$ fixed to a certain value, we just collect the
    corresponding submatrices ${\bf W}_{N-(m_1+m_2)}$ and put them in
    the diagonal entries, giving rise to the notation} ${\bf
    W}_{n:2}=\begin{bmatrix} {{\bf W}_{n}} & \bf 0\\ {\bf 0} & {\bf
    W}_n\end{bmatrix}$ and so on.
\end{itemize}

For transitions represented by ${\bf M}$ and ${\bf G}$, we have
detailed their construction in
Eqs.~(\ref{eq:MG_Motor}-\ref{eq:MG_Binding}).

\subsection{Examples of ${\bf M}$ and ${\bf G}$ for motor proteins}
Instead of giving the explicit matrix forms of ${\bf M}$ and ${\bf
G}$, we characterize them by considering the transitions of $m_1, m_2$
allowed in the model, i.e., the positive entries in ${\bf M}$ and
${\bf G}$. For processive motor proteins, transition of the form $(m_1
\rightarrow m_1 + 1)$ is allowed only if $n_1 \geq 1$. The transition
matrices are then given by
\begin{equation}
  \begin{aligned}
    {\bf M}[(m_1+1, m_2, n_1 - 1, n_2), (m_1, m_2, n_1, n_2)] = 1, && \forall n_1 \geq 1;\\
    {\bf M}[ (m_1, m_2+1, n_1, n_2 - 1), (m_1, m_2, n_1, n_2)] = 1, && \forall n_2 \geq 1;\\
    {\bf G}[(m_1-1, m_2, n_1 + 1, n_2), (m_1, m_2, n_1, n_2)] = 1, && \forall m_1 \geq 1;\\
    {\bf G}[(m_1, m_2-1, n_1, n_2 + 1), (m_1, m_2, n_1, n_2)] = 1, && \forall m_2 \geq 1,
  \end{aligned}\label{eq:MG_Motor}
\end{equation}
where ${\bf M}[j,i]$ indicates the $i\to j$ transition.  The remaining
off-diagonal entries in ${\bf M}$ and ${\bf G}$ are $0$. The diagonal
entries are determined by the normalization condition that the column
sum of ${\bf M}$ and ${\bf G}$ vanishes from conservation of
probability.

One special property of ${\bf M}$ and ${\bf G}$ for motor proteins is
that they are block-tridiagonal matrices. In the block matrix
representation shown in Eq.~\eqref{TRANSITION:ASSISTED}, each block of
rows and columns corresponds to a collection of states with the same
sum $m_1+m_2$.  For example, $\Q_{N}$ represent transitions within the
states with $m_1 + m_2 = 0$ while ${\bf M}_{N-1,N}$
represents transitions from the states with $m_1 + m_2 = 0$ to the
states with $m_1 + m_2 =1$.

\subsection{Examples of ${\bf M}$ and ${\bf G}$ for binding proteins}
For proteins that bind to DNA directly, the transitions of the form
$(m_1 \rightarrow m_1 + k)$ are allowed if $n_1 \geq k$. For the
matrix ${\bf M}$, we have
  \begin{equation}
  \begin{aligned}
    {\bf M}[(m_1+k, m_2, n_1 - k, n_2), (m_1, m_2, n_1, n_2)] = 1,
    && \forall n_1 \geq k;\\
    {\bf M}[ (m_1, m_2+k, n_1, n_2 - k), (m_1, m_2, n_1, n_2)] = 1,
    && \forall n_2 \geq k.
    \label{C3}
  \end{aligned}
\end{equation}
This constraint on $n_1$ and $n_2$ arises naturally from the
requirement that the target state $(m_1',m_2',n_1',n_2')$ must fall
into the state space $\Omega_{\rm p}$.

For the matrix ${\bf G}$, in order to incorporate the
different possibilities in the target state when $m$ decreases, we
consider two limiting scenarios. In the ``high remodeler density''
limit, the matrix ${\bf G}_{\rm hi}$ is identical to that of the motor
proteins, where $m \to m-1$ when the inner-most remodeler unbinds.  In
the ``low remodeler density'' limit, the matrix ${\bf G}_{\rm low}$
represents transitions of the form $m \rightarrow 0$ since only
at most one remodeler is bound per end.
\begin{equation}
\begin{aligned} {\bf G}_{\rm low}[(0, m_2, n_1 + m_1, n_2), (m_1, m_2, n_1, n_2)] = 1, && \forall
m_1 \geq 1;\\
{\bf G}_{\rm low}[ (m_1, 0, n_1, n_2 + m_2), (m_1, m_2,
n_1, n_2)] = 1, && \forall m_2 \geq 1;\\
{\bf G}_{\rm hi}[(m_1-1,
m_2, n_1 + 1, n_2), (m_1, m_2, n_1, n_2)] = 1, && \forall m_1 \geq
1;\\
{\bf G}_{\rm hi}[(m_1, m_2-1, n_1, n_2 + 1), (m_1, m_2, n_1,
n_2)] = 1, && \forall m_2 \geq 1.
\end{aligned}\label{eq:MG_Binding}
\end{equation}

The choices of different ${\bf M}$ and ${\bf G}$ will not
significantly affect the overall histone disassembly rate. For ${\bf
M}$ associated with remodeler binding and motor proteins,
respectively, the effective dissociation rates differ only by
$O(\varepsilon)$.  Moreover, ${\bf G}_{\rm hi}$ and ${\bf G}_{\rm
low}$ yield qualitatively similar outcomes.  When $\pa \lesssim \pd$,
the facilitated states are unlikely and do not contribute to the
histone unbinding. When $\pa \gg \pd$, unbinding itself is unlikely
and their differences are negligible.

% The remodeler unbinding matrix ${\bf G}$
% is an upper-triangular matrix;  for molecular motors and highly
% cooperative proteins, every unbinding event reduces $m_1$ or $m_2$ by 
% 1, with the corresponding transition matrix ${\bf G}_{\rm min}$:
% %
% $${\bf G}_{\rm min}=\setlength\arraycolsep{3pt}
% \left[\begin{array}{ccccc}
% 0   & {\bf G}_{N,N-1} & 0 &\cdots & 0 \\
% 0 & {\bf G}_{N-1} & \ddots & \ddots & \vdots \\
%  \vdots  & \ddots & \ddots &\ddots & 0 \\
%  \vdots   & & \ddots & \ddots & {\bf G}_{2,1} \\
%  0 & \cdots & \cdots & 0 & {\bf G}_{1} \\
% \end{array}\right]$$
% %
% Here, ${\bf G}_{k,k-1}$ represents the change in total available binding sites for
% histone-DNA interaction due to one site released by the remodeler.
%All the rest
%configurations remain the same.

%%%%%%%%%%%%%%%%
\subsection{Irreversible remodeler binding}
%%%%%%%%%%%%%%%%
In this subsection, we assume that $\pd =0$
and \replaced{$\pa \ll \kon$}{$\pa\ll 1 $, (relative to $k_{\rm
on}$)}. Then, Eq.~\eqref{TRANSITION:ASSISTED} becomes
\begin{equation}\label{TRANSITION:A_IRR}
\Q_{N,{\rm p}}(\pa)=
\setlength\arraycolsep{6pt}
\left[\begin{array}{ccccc}
\Q_{N:1}+\frac{\pa}{\kon} \M_N   & 0 & \cdots &\cdots & 0 \\
\frac{\pa}{\kon} {\bf M}_{N-1,N} & \Q_{N-1:2}+\frac{\pa}{\kon}{\bf M}_{N-1} & \ddots & \ddots &\vdots \\
 \vdots  & \ddots & \ddots &\ddots & \vdots \\
 \vdots   & & \ddots & \ddots & 0 \\
 \frac{\pa}{\kon} {\bf M}_{1,N} & \cdots & \cdots & \frac{\pa}{\kon} {\bf M}_{1,2} & \Q_{1:N} \\
\end{array}\right]
\end{equation}
Corresponding to the block matrix representation of $\Q_{N,{\rm
p}}(\pa)$ above, we can write the \added{$i$-th} eigenvector $\v_i$ in
the form of $\v_i=(\v_{i\mid N},\ldots,
\v_{i\mid 1})^{\intercal}$. Here, $\v_{i \mid N} \in \mathbb{R}^{N(N+1)/2}$ corresponds 
to the states with $m_1 =m_2 =0$. If $\lambda$ is the eigenvalue of
this eigenvector, then
\begin{equation}
\label{SIMPLIFIED_EIGENVECTOR}
\lambda(s, \pa) \v_{i\mid N}(s, \pa)=
\left(\Q_{N}+ \frac{\pa}{\kon} \M_N\right)
\v_{i\mid N}(s, \pa) = \left(\Q_{N}(0)+ s \C_N
+ \frac{\pa}{\kon} \M_N\right) \v_{i\mid N}(s, \pa),
\end{equation}
where we have explicitly indicated
the \replaced{dependency}{dependences} on $s$ and $\pa$. If $\v$ is an
eigenvector of $\Q_{N,{\rm p}}$ with nonvanishing $\v_{i \mid N}$
terms, then $\v_{i \mid N}$ is an eigenvector of the matrix
$\Q_{N}(s)+\pa/\kon \M_N$. In the following, we will find an estimate
for the eigenvalue by using perturbation theory for the matrix
$\Q_{N}(s)+\pa/\kon \M_N$ based on the initial state $s = \pa =0$.

First, we will find a proper initial eigenvector to start the
perturbation analysis.  When $\pa = 0$, define
$\v(s,\pa=0)=(\v_{i \mid N}^{\intercal}(s,0), 0 \ldots,
0)^{\intercal}$, where $\v_{i \mid N}(s,\pa=0)$ is the principle
eigenvector of the matrix $\Q_{N}(s)$ associated with the eigenvalue
$N s \varepsilon^{(N-1)}[1+ O(\varepsilon)]$. Then, $\v(s,0)$ is an
eigenvector of the whole matrix $\Q_{N,{\rm p}}(p_{\rm a})$ with
eigenvalue $0$, for all $s$.

We next perturb the initial eigenvector $\v_{i \mid N}(0,0)$ by
applying the same analysis used to obtain
Eq.~\eqref{eq:series_expansion}. We find
\begin{equation}
\label{EIGENVECTOR:EXPANSION_IRR}
  \v_{i \mid N}(s,\pa, \lambda) = \v_{i\mid N}(0,0,0) + \left[ \begin{array}{c}
  0 \\[4pt] -\left(\sum_{i=1}^{\infty} \Big[-\oline{\Q}_{N}^{-1}(0)(s
  {\oline{\bf C}_N}+\pa/\kon \M_N
  -\lambda \I)\Big]^{i}\right)\oline{\bf v}_{i \mid N}(0,0,0) \end{array} \right],
\end{equation}
where $\lambda$ here is treated as an independent
variable. $\oline{{\bf v}}_{i \mid N}(0,0,0) $ denotes $\v_{i\mid
N}(0,0,0)$ excluding the first row, and $\oline{\Q}_{N}(0)$ is the
matrix $\Q_{N}(0)=\A_N+\varepsilon \B_N$ with the first row and column
deleted, as defined earlier.

By applying the same estimate over the deviation, we obtain a formula
analogous to Eq.~\eqref{FIRSTORDER},
\begin{equation}
\label{EXPRESSION:IRR_APP}
\v_{i \mid N}(s,\pa,\lambda) = \v_{i\mid N}(0,0,0)
\Big[1+O\big(s+\vert\lambda\vert
+\tfrac{\pa}{\kon}\big)\Big]
\end{equation}
and calculate the corresponding eigenvalue by the relation $\lambda =
\langle {\bf 1}, \Q {\bf v} \rangle/ \langle {\bf 1}, {\bf v} \rangle$
for the eigenpair $(\lambda, {\bf v})$. \added{In particular, we
consider the principal eigenvalue $\lambda_0(s,\pa)$ and the
corresponding eigenvector $\v_0(s,\pa)$ with its first block component
${\bf v}_{0\mid N}$:}
\begin{equation}
\begin{split}
  \lambda(s,\pa) &  \approx \left\langle
  {\bf 1}_{N(N+1)/2}, \Big[\Q_{N}(s) + \tfrac{\pa}{\kon} \M_N \Big] {\bf x}_N(0,0)\right\rangle
      \Big[1+O\big(s+\tfrac{\pa}{\kon} \big)\Big] \\
    \: & \approx -\Big(N s \varepsilon^{N-1}+\frac{\pa}{\kon} \sum_{i=1}^{N-1}(i+1)
        \varepsilon^{i}\Big)\Big[1+O\big(s+\tfrac{\pa}{\kon}\big)\Big].
  \label{EQUATION:IRR_APP}
  \end{split}
\end{equation}
\added{Eq.~\eqref{EXPRESSION:IRR_APP} provides a justification 
for Eq.~\eqref{EXPRESSION:IRR} in the main text, while
Eq.~\eqref{EQUATION:IRR_APP} provides a justification for
Eq.~\eqref{EQUATION:IRR} in the main text.}

Determining the eigenvalue when $\pa \gg \kon$ is beyond the scope of
this perturbation method because the radius of convergence of the
series expansion is around \replaced{$\pa \sim \kon$}{$\pa \sim 1$
(i.e. $\pa \sim \kon$)}.  Nonetheless, the simple interpolation
formula
\begin{equation}\label{INTERPOLATION_APP}
  \lambda(s,\pa)= \textrm{max}\left\{-\Big[Ns\varepsilon^{N-1}+
    \frac{\pa}{\kon} \sum_{i=1}^{N-1}(i+1)\varepsilon^{i}\Big], -s\right\}
\end{equation}
matches numerical calculations quite well when $s = \varepsilon$.

%%%%%%%%%%%%%%%%
\subsection{Reversible attachment of remodelers}
\label{sec:remodeler_coarse-graining}
We have not found a succinct analytic description of the predictions
of this model; therefore, we adopt a physical approximation by
considering the ``stability'' of $\v_{i\mid N}$ in order to reduce the
block matrix $\Q_{N,{\rm p}}$ into a
$\frac{N(N+1)}{2} \times \frac{N(N+1)}{2}$ matrix connected to
$(m_1,m_2)$. \added{The approximation, or coarse-graining, is shown in
Fig.~\ref{fig:coarse-grained-model} and is motivated by a steady state
assumption under a fast-slow timescale separation as demonstrated and
formalized earlier in Appendix~\ref{sec:eigenvalues}.} Assuming that
$\pa, \pd \ll \kon$, we note that the relaxation time of states
$(n_1,n_2)$ given fixed $m_1,m_2$ is on the order of $\kon$. Before
any remodeler binding and unbinding transition occurs, it is very
likely that the probability distribution of $(n_1, n_2)$ conditioned
on $(m_1,m_2)$ has reached a quasi-equilibrium state close to
\replaced{$\v_0$ with $N-m_1-m_2$ number of contact
sites}{$\v_{n;0}$}. In such a quasi-equilibrium state, the mean rate
of remodeler dissociation will be $\pd$ and the mean rate of another
remodeler binding at a distance $k$ position from a free end will be
$\pa \varepsilon^k$ for binding proteins. The overall approximation
approach seeks to ignore the fine details of $(n_1,n_2)$ given
$(m_1,m_2)$ and approximates the transitions $(m_1,m_2) \to
(m_1',m_2')$ as Markovian. 

For convenience, we further ignore transitions with rate
$\pa\varepsilon^k$ for $k \geq 2$. This truncation allows for a simple
solution for the eigenvector corresponding to the greatest eigenvalue.
In the ``high remodeler density'' limit (stepwise remodeler movement),
the simplified transition matrix ${H}_N'$, defined on
${\Omega}'_{\rm p}\coloneqq\left\{ (m_1,m_2): m_1 + m_2 < N \right\}$,
can be expressed as
\begin{equation}\label{MDA:EQNAPPR}
\Q'_{N,{\rm p}}(\pa,\pd)=
\textrm{diag}\{
  -Ns\varepsilon^{N-1},-(N-1)s\varepsilon^{N-2},...,-s
  \} 
+ \frac{\pd}{\kon} \A_N + \frac{\pa}{\kon}  \varepsilon \B_N.
\end{equation}
The first term describes transitions directly to the detached states
$\Omega^*_{\rm p}$ and other terms describes binding and unbinding of
a remodeler. Analogy of this simplified scenario to the unfacilitated
unbinding model is shown in Fig.~\ref{fig:coarse-grained-model}(a).
\added{The approximation showed 
numerical agreement with the full model in the main text.  In the
following, we employ additional approximation techniques to derive an
analytical expression for the principal eigenvalue when $\pa
+ \pd \gg \koff$.}

We can analytically approximate the principal eigenvalue of
$\Q'_{N,{\rm p}}(\pa,\pd)$ defined in Eq.~\eqref{MDA:EQNAPPR}
only when $\varepsilon \ll \pa+\pd$, where detailed balance
approximately holds.  In this case, we still assume that the structure
$\v_0(m_1,m_2) \propto (\frac{\pa}{\pd})^{k}$ is stable under the
small perturbation determined by $\varepsilon$, providing the estimate
by considering the normalized flux from the bound states
$ \Omega_{\rm p}'$ to the fully open states
$ \Omega_{\rm p}'^*$:
\begin{align}
  \begin{split}
 \lambda_0(\pa,\pd, \varepsilon) 
 & \approx
 \frac{\langle {\bf 1}_{N(N+1)/2},({\H}'(\pa,\pd,0) -
        \textrm{diag}\{Ns\varepsilon^{N-1},-(N-1)s\varepsilon^{N-2},...,s \})\v_0 
        \rangle} {\langle {\bf 1}_{N(N+1)/2}, \v_0 \rangle} \\
\:  & = \varepsilon^N \frac{\sum_{k=0}^{N-1}(k+1)(N-k) K_{A}^k}
  {\sum_{k=0}^{N-1}(k+1) ({\varepsilon K_{A}})^k}, 
    \end{split}
    \label{eq:weak_perturbation2}
\end{align}
where, $K_A\equiv \frac{\pa}{\pd} = e^{-E_{\rm
p}}$. \added{Eq.~\eqref{eq:weak_perturbation2} can be further
simplified to Eq.~\eqref{eq:intact_weak_facilitation} by considering
only the first term ($k=0$) in the numerator, and the first and last
term ($k=0,N-1$) in the denominator.}

\added{The coarse-grained approximation
of the right-hand side of Eq.~\eqref{eq:weak_perturbation2} coincides
with the prediction via the flux intensity $j( \Omega'^*_{\rm p}|
\Omega'_{\rm p})$. In general, for a continuous time Markov chain with
transition rate matrix $W$, let $A$ and $B$ be two disjoint sets of
states, and $\pi$ be the stationary distribution of the Markov chain.
Then the flux intensity from $A$ to $B$ is defined as}
\begin{equation}
j(A \mid B)=\frac{\sum_{a \in A} \sum_{b \in B}
W_{a, b} \pi_b}{\sum_{b \in B} \pi_b}.
\end{equation}
\added{The flux intensity $j({\Omega}'^*_{\rm p} | {\Omega}_{\rm p}')$ serves
as an upper bound for the principal eigenvalue
$\lambda_0$, \textit{e.g.}, Eq. (3.69) in Aldous and
Fill \cite{aldous2002}.}

\end{widetext}

The intuition for the relation between flux intensity and the
eigenvalue is as follows: flux intensity is obtained by assuming that
the eigenvector with an absorbing boundary has the same structure as
that with a reflecting boundary. In reality, presence of an absorbing
boundary will decrease the relative weight of states on the boundary,
and thus making the associated eigenvalue smaller than the flux
intensity. \added{Note that the flux intensity analysis is similar in
both the full facilitated model and the coarse-grained model, which 
provides a further justification of the coarse-graining.}
%
%%%%%%%%%%%%%%%%%%%%%%%%%%%%%%%%%%%%%%%%%%%%%%%%%%%%%%%
%%%%%%%%%%%%%%%%%%%%%%%%%%%%%%%%%%%%%%%%%%%%%%%%%%%%%%%
%%%%%%%%%%%%%%%%%%%%%%%%%%%%%%%%%%%%%%%%%%%%%%%%%%%%%%%
%

\section{Histone detachment with random landscapes}
\label{Appendix:random_landscapes}
Previously, we have assumed that all 14 contact bonds between the
histone core and the DNA are identical with the same binding and
unbinding rates $k_{\rm on}$ and $\koff$. In reality, these can rates
vary depending on local DNA base identity, stiffness and/or
spontaneous curvature. It is estimated that the contact free energies
vary between $1.5$ $k_{\rm b}T$ and $2$ $k_{\rm b}T$
\cite{ANDERSON2000,Polach1995nov,KuliSchiessel}.  To account for this
heterogeneity, we conduct numerical experiments that assume
homogeneous binding rates but random unbinding rates that correspond
to iid binding energies $E_{\rm c}$ that are drawn from a uniform
distribution between $1.5 k_{\rm B}T$ and $2 k_{\rm B}T$.  In this
case, as shown in Fig.~\ref{fig:linear_facilitated_random}, the
variation does not alter the qualitative behavior of the system.  Thus
our model is well parameterized by just the mean binding energy
$E_{\rm c}$.
\begin{figure}[htb]
\centering
\includegraphics[width=2.5in]{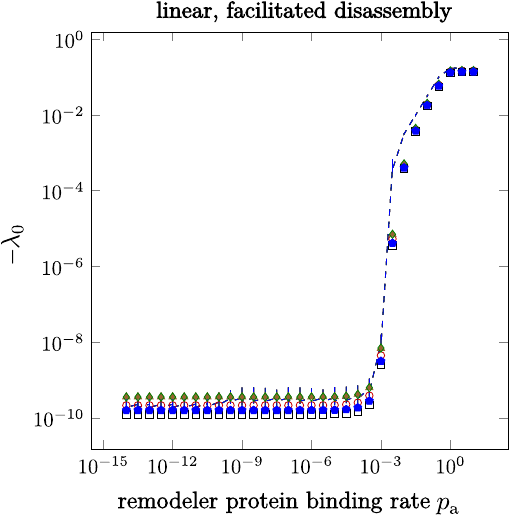}
\caption{Principal eigenvalue of the linear facilitated
detachment model with random binding energy reflected in variations in
  $k_{\rm off}$ that lead to a per-site $E_{\rm c}$ that is uniformly
  distributed between 1.5 and 2 ($k_{\rm B}T$). We set $\pd =10^{-3}$
  (in units of $\kon$) and plot $-\lambda_{0}$ for five randomly
  sampled configurations of $E_{\rm c}$. The dashed line represents
  the prediction based on mean the binding energy and
  Eq.~\eqref{eq:intact_effective_facilitation}.}
\label{fig:linear_facilitated_random}
\end{figure}

\section{Reversible multimeric histone detachment}
\label{Appendix:multimeric}
Histone dimers, tetramers and other transient higher order complexes
in solution may rescue partially disassembled nucleosomes.  They can
initiate rescue of partially disassembles nucleosomes by directly
docking to existing nucleosome subunits (dimers and tetramers) or by
associating with the vacant DNA segments.  We assume that these rates
$\qa'^{(\textrm{subunit})}$ (for docking with another subunit) and
$\qa''^{(\textrm{subunit})}$ (for direct contact with the DNA) are
scaled properly according to their respective equilibrium bulk
concentration to ensure that the overall Markov process considering
these reactions is \textit{reversible}. We have defined $\qa$ as the
docking rate conditioned on both subunits being attached to the DNA.

The primary quantity of interest is the expected time
$\mathbb{E}[T({\bf 1})]$ needed to transition from the fully attached
state ${\bf 1}= \big(\boldsymbol{\sigma}=(1,1,1,1,1), {\bf
n}=(0,0)\big) \equiv (1,1,1,1,1, (0,0))$ to the fully dissociated
state $\Omega^*$. Solving for the mean detachment time requires
inversion of a large matrix over the whole state space, which is
analytically intractable.  We therefore consider the probability flux
intensity $j(\Omega^*\,\vert\,\Omega)$ from the attached states
$\Omega$ to the fully unattached state $\Omega^*$ as a useful
surrogate. \deleted{As previously discussed, the inverse of the flow
intensity $1/j(\Omega^*\,\vert\,\Omega)$ is a lower bound for
$\mathbb{E}[T({\bf 1})]$. The relaxation rate $-\lambda_1$, on the
other hand, characterizes the relaxation of the system to equilibrium
under perturbation.} \added{The general relation between
$j\left(\Omega^*\! \mid\!\Omega\right)$ and $\lambda_0$ is known and
derived in \textit{e.g.}, Eq. (3.69) in Aldous and
Fill \cite{aldous2002}, where the inequality
$j\left(\Omega^* \mid \Omega\right) \geq |\lambda_0|$ is given.}
%
%\begin{equation}\color{blue}
%  j\left(\Omega^* \mid \Omega\right) \geq |\lambda_0|  
%\end{equation}
%\added{was derived.}
%

To obtain a reversible Markov chain, we assume that both bound and
  free histones are in equilibrium. With our definition of
$\qa''^{(\textrm{subunit})}$, the corresponding free energy relative to
bulk solution can be expressed as $\Delta E_{\rm
s}^{(\textrm{subunit})} = \log\big(\kon/\qa''^{(\textrm{subunit})}\big)$.

The scaling relations between $\qa''^{(\textrm{subunit})}$ and
$\qa'^{(\textrm{subunit})}$ must follow the equilibrium conditions
%%%%%%%%%%%%%%%%%%%%%%%
\begin{widetext}
\begin{align}
  \label{EQUATION:EQUILIBRIUM_CONDITION}
  \begin{split}
  \qa'^{({\Htd})} & =  \qd^* \exp(-E_{\rm q} - \Delta E_s^{({\Htd})}),\\
  \qa'^{({\HtHf})} & = \qd^* \exp(-E_{\rm q} - \Delta E_s^{({\HtHf})}),\\
  \qa'^{({\rm Hexamer})}  & = \qd^* \exp(-2E_{\rm q} - \Delta E_s^{({\HtHf})} - \Delta E_s^{({\Htd})}),\\
  \qa''^{({\Htd})} &=  k_{\rm on} \exp(-\Delta E_s^{(\Htd)}),\\
  \qa''^{({\HtHf})} &=  k_{\rm on} \exp(-\Delta E_s^{(\HtHf)}),\\
  \qa''^{({\rm Hexamer})} & = k_{\rm on} \exp(-E_{\rm q} - \Delta E_s^{(\Htd)} - \Delta E_s^{(\HtHf)}),\\
  \qa''^{({\rm Octamer})} & =  k_{\rm on} \exp(-2E_{\rm q} - 2\Delta E_s^{(\Htd)}-\Delta E_s^{(\HtHf)})
  \end{split}
\end{align} 
to satisfy reversibility. The free energy function associated with
each state $(\boldsymbol{\sigma}, \n) \equiv (\sigma_{\rm
l},\theta_{\rm l},\sigma_{\rm m},\theta_{\rm r},\sigma_{\rm r},\n)$
can be expressed as
\begin{align}
  \begin{split}
    E(\sigma_{\rm l},\theta_{\rm l},\sigma_{\rm m},\theta_{\rm r},\sigma_{\rm r},\n) 
    = & (\sigma_{\rm l}+\sigma_{\rm r}) \Delta E_{\rm s}^{(\Htd)} +
    \sigma_{\rm m} \Delta E_{\rm s}^{(\HtHf)} + (\theta_{\rm l}+\theta_{\rm r})E_{\rm q}\\
    \: & \hspace{2cm}+ \Big(N_{\rm l} \sigma_{\rm l} + N_{\rm m} \sigma_{\rm m}
    + N_{\rm r} \sigma_{\rm r}-\sum_{j=1}^f \sum_{k=0}^1  n_k^{(j)}\Big) E_{\rm c}.
  \end{split}
\end{align}
\end{widetext}
In the following, we will further assume $\kd = \koff$, i.e. $s = \varepsilon$,
to reduce the notational complexity.
%%%%%%%%%%%%%%%%%%%%%%%%

\subsection{Estimate of the flux intensity $j(\Omega^*\,\vert\,\Omega)$ for
reversible spontaneous detachment}
\label{section:F1} 
 Let $\Sigma$ denote the collection of \textbf{macrostates}
${\boldsymbol{\sigma}}=(\sigma_{\rm l},\theta_{\rm l},\sigma_{\rm
m},\theta_{\rm r},\sigma_{\rm r})$ that is not equal to $0^{\times
5}$. The microstates on the boundary are characterized by a single
intact DNA-histone contact are defined by $\big(N_{\rm l} \sigma_{\rm
l} + N_{\rm m} \sigma_{\rm m} + N_{\rm r}
\sigma_{\rm r}  - \sum_{k,j}  n_k^{(j)}\big) = 1$ and denoted by $\partial \Omega$.

The equilibrium flow intensity from bound states that can reach the
unbound state in one step can be expressed by enumerating all possible
boundary microstates ${\bf n}$ associated with each macrostate
${\boldsymbol{\sigma}}$ in $\Sigma$ is given by
\begin{align}
  \begin{split}
    j(\Omega^*\,\vert\,\Omega) = \varepsilon \frac{
      \sum_{({\bf {\boldsymbol{\sigma}}},\n)\in \partial \Omega}   e^{-E({\bf {\boldsymbol{\sigma}},n})}
      }{
        \sum_{({\bf {\boldsymbol{\sigma}}},\n)\in \Omega} e^{-E({\bf {\boldsymbol{\sigma}},n})}
      },
  \end{split}\label{EQUATION:Spontaneous_FLUX}
\end{align}
where the free energy can be separated into component energies
$E({\boldsymbol{\sigma}},{\bf n}) \equiv  U({\bf n}) + V({\boldsymbol{\sigma}})$ where

\begin{equation}
\begin{aligned}
  U({\bf n}) = & - E_{\rm c}\sum_{j=1}^f \sum_{k=0}^l n_k^{(j)}, \\
  V({\boldsymbol{\sigma}}) = & (\sigma_{\rm l}+\sigma_{\rm r}) \Delta E_{\rm s}^{(\Htd)}
+ \sigma_{\rm m}\Delta E_{\rm s}^{\HtHf} \\
  \: &  \hspace{1.3cm} + (\theta_{\rm l}+\theta_{\rm r})E_{\rm q}
  + N({\boldsymbol{\sigma}}) E_{\rm c},\\[3pt]
   N({\boldsymbol{\sigma}}) = & N_{\rm l} \sigma_{\rm l} + N_{\rm m} \sigma_{\rm m} + N_{\rm r}
  \sigma_{\rm r}, \\ 
\end{aligned}
\end{equation}
where $U({\bf n})$ describes the peeling energy cost of the
DNA-histone contacts in the microstate ${\bf n}$ and
$V({\boldsymbol{\sigma}})$ is the energy of the most probable
microstate ${\bf n}^*_{\boldsymbol{\sigma}}$ given macrostate
${\boldsymbol{\sigma}}$.  $N({\boldsymbol{\sigma}})$ is the number of
available DNA-histone contacts in macrostate ${\boldsymbol{\sigma}}$.
The denominator in Eq.~\eqref{EQUATION:Spontaneous_FLUX} is
the \textit{partition function} of the equilibrium distribution on
$\Omega$.

We can simplify the expression of
Eq.~\eqref{EQUATION:Spontaneous_FLUX} by grouping degenerate states
$({\boldsymbol{\sigma}}, {\bf n})\in \partial \Omega$ associated with
each macrostate ${\boldsymbol{\sigma}}$ in the numerator and
identifying the most probable microstate ${\bf
n}^*_{\boldsymbol{\sigma}}$ for each macrostate
${\boldsymbol{\sigma}}$ in the denominator.  The most probable
microstate ${\bf n}^*_{\boldsymbol{\sigma}}$ corresponds to the state
with the largest number $N({\boldsymbol{\sigma}})$ of DNA-histone
contacts.  The relative energy of the boundary microstates ${\bf
n}_{\rm b}$ compared to the most probable microstate ${\bf
n}^*_{\boldsymbol{\sigma}}$ for a specified ${\boldsymbol{\sigma}}$ is
$U({\bf n}_{\rm b}) = -\big[N({\boldsymbol{\sigma}})-1\big] E_{\rm
c}$.  For $\varepsilon = \koff/\kon \ll 1$,
$j(\Omega^*\,\vert\,\Omega)$ simplifies to

\begin{equation}
  \begin{aligned}
    j(\Omega^*\,\vert\,\Omega) & \approx j(E_{\rm c}) \\
   \: & =\frac{
                \sum_{{\boldsymbol{\sigma}}\in S} \varepsilon
                N({\boldsymbol{\sigma}}) e^{- V({\boldsymbol{\sigma}}) + E_{\rm c}[N({\boldsymbol{\sigma}}) -1]}
                }
                {
                \sum_{{\boldsymbol{\sigma}}\in S} 
                  e^{-V({\boldsymbol{\sigma}})}
                } \\
  \: &   =  \frac{
                  \sum_{{\boldsymbol{\sigma}}\in S} 
                  N({\boldsymbol{\sigma}}) e^{- V({\boldsymbol{\sigma}}) + E_{\rm c}[N({\boldsymbol{\sigma}})]}
                  }
                  {
                  \sum_{{\boldsymbol{\sigma}}\in S} 
                    e^{-V({\boldsymbol{\sigma}})}
                  }
  \end{aligned}.
  \label{eq:j_mul}
\end{equation}

Note that the exponents in the factor $\sum_{{\boldsymbol{\sigma}}\in
S} e^{-V({\boldsymbol{\sigma}})}$ include all possible macrostates,
with contributions from both histone-histone interactions ($\Delta
E_{\rm s}$ and $E_{\rm q}$) and DNA-histone contacts ($E_{\rm
c}$). Conversely, the exponents in $\sum_{{\boldsymbol{\sigma}}\in S}
N({\boldsymbol{\sigma}}) e^{-V({\boldsymbol{\sigma}}) + E_{\rm
c}[N({\boldsymbol{\sigma}})]}$ take into account only histone-histone
interactions.

%%%%%% Added Part for Further Approximation of J %%%%%%%%%%

\begin{figure}[htbp]
  \centering
  \includegraphics[width=3.3in]{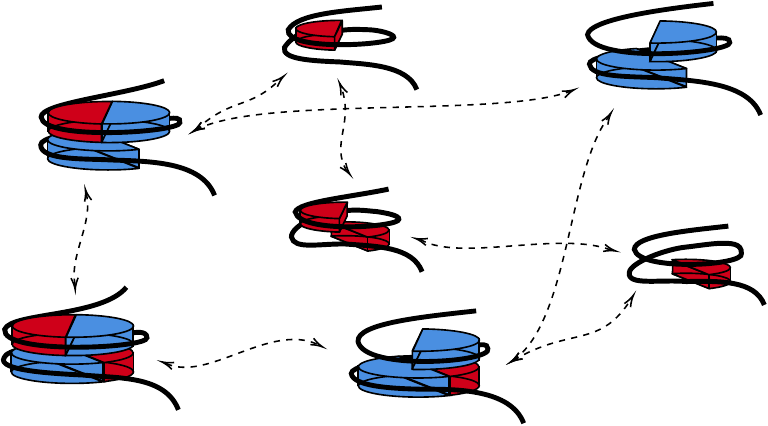}
\caption{A schematic of possible macrostates
of $(\sigma_{\rm l}, \sigma_{\rm m}, \sigma_{\rm r})$. The states of
linkage, $\theta_{\rm l}, \theta_{\rm r}$, are omitted for
simplicity. For different values of $E_{\rm q}$ and $ \Delta E_{\rm
s}$, the most probable states are only chosen from either the fully
bound state, shown in the lower left corner, or the state where only
the $\HtHf$ tetramer is bound, shown in the upper right corner. Other
states are less probable transient states.}
\end{figure}
We further simplify the formula of $j(\Omega^*\,\vert\,\Omega)$ by
considering the relative probability of two main macrostates, the
fully bound state $\boldsymbol{\sigma}_1 =(1,1,1,1,1)$ and the
state where only the $\HtHf$ tetramer is bound
$\boldsymbol{\sigma}_{\rm m} =(0,0,1,0,0)$. Assuming that $\Delta
E_{\rm s}^{\HtHf} = \Delta E_{\rm s}^{\Htd}$, the energies of the two
macrostates are given by $V(\boldsymbol{\sigma})$:
\begin{equation}
  \begin{aligned}
    V(\boldsymbol{\sigma}_1) = & 3 \Delta E_{\rm s} + 2 E_{\rm q} + N E_{\rm c},\\
    V(\boldsymbol{\sigma}_{\rm m}) = & \Delta E_{\rm s} + N_{\rm m} E_{\rm c}.
%
%    \Delta V \coloneqq & V(\boldsymbol{\sigma}_1) - V(\boldsymbol{\sigma}_{\rm m}) \\
%     = & 2 \Delta E_{\rm s} + 2 E_{\rm q} + (N - N_{\rm m}) E_{\rm c}. 
  \end{aligned}
\end{equation}

By tracking only these two macrostates, we approximate $j(\Omega^*\,\vert\,\Omega)$
in Eq.~\eqref{eq:j_mul} by
\begin{equation}
  \begin{aligned}
    j(\Omega^*\,\vert\,\Omega) & \approx  \frac{N \varepsilon^N e^{- V(\boldsymbol{\sigma}_1)} 
   + N_{\rm m} \varepsilon^{N_{\rm m}} e^{- V(\boldsymbol{\sigma}_{\rm m})}}
   {e^{- V(\boldsymbol{\sigma}_1)}+ e^{- V(\boldsymbol{\sigma}_{\rm m})}} \\
    & =  \frac{N \varepsilon^N + N_{\rm m} \varepsilon^{N_{\rm m}} e^{ \Delta V}}
                  {1 + e^{ \Delta V}}\\ 
%    & =  N \varepsilon^N \frac{1 + (N_M/N)
%            e^{\Delta V - (N-N_M)E_{\rm c}}}{1 + e^{ \Delta V}}\\
    & =  N \varepsilon^{N} \frac{1 + (N_{\rm m}/N)e^{2 (E_{\rm q} + \Delta E_{s})}}{1 + e^{ \Delta V}},
  \end{aligned}
  \label{eq:j_mul_approx}
\end{equation}
where
\begin{equation}
\begin{aligned}
\Delta V & \equiv V(\boldsymbol{\sigma}_1) - V(\boldsymbol{\sigma}_{\rm m}) \\
\: & =  2 \Delta E_{\rm s} + 2 E_{\rm q} + (N - N_{\rm m}) E_{\rm c}.
\end{aligned}
\end{equation}
Given the discussion of irreversible nucleosome disassembly in the
main text, here, we focus on understanding the role of $\qd^*$ in
nucleosome disassembly and how $\Delta E_{\rm s}$ affects reversible
histone rebinding.  Eq.~\eqref{eq:j_mul_approx} explicitly shows the
roles of $\Delta E_{\rm s}$ and $E_{\rm q}$ in the reversible
multimeric nucleosome disassembly.

Irreversible subunit unbinding arises when $\Delta E_{\rm
s} \rightarrow \infty$ which is equivalent to $\qa', \qa'' = 0$. In
this limit, the most probable state is the $\HtHf$-bound state, with
$j(\Omega^*\,\vert\,\Omega) \sim N_{\rm m} e^{N_{\rm m}E_{\rm c}}$. The
transition point between an effectively irreversible scenario and a
reversible unbinding scenario is when $\Delta V \approx 0$, above
which the fully bound macrostate $\boldsymbol{\sigma}_1$ is no longer
the most probable state.  This transition point is characterized by
$(E_{\rm q} + \Delta E_{\rm s}) = (N_M - N) E_{\rm c} /2$.

When $\Delta E_{\rm s}$ is small but still positive, and $\Delta
E_{\rm s} + E_{\rm q} > 0$, the most probable state is the fully bound
state with $N$ DNA-histone contacts. However, the boundary states can
be stabilized by absence of one or more histone modules, with
$j(\Omega^*\,\vert\,\Omega) \sim N_{\rm m} e^{N E_{\rm c} + 2(\Delta
E_{\rm s} + E_{\rm q})}$.  However, when $\Delta E_{\rm s}$ is
negative, absence of one or more histone subunits cannot stabilize the
boundary states. In this case,
$j(\Omega^*\,\vert\,\Omega) \sim N e^{N E_{\rm c}}$, which is close
to the linear intact histone model.

In the context of a FPT problem from the
fully attached state ${\bf 1}=(\boldsymbol{\sigma}_1, {\bf n}=(0,0))$
where $N$ histone-DNA contacts must be dissociated to be reach
$\Omega^*$, which we formally treat as an absorbing state while still
allowing for partial rebinding. The flux intensity
$j(\Omega^*\,\vert\,\Omega)$ serves as an estimate of $-\lambda_0$, the
principal eigenvalue of the transition matrix with absorbing state
$\Omega^*$, which in turn is inversely related to the
MFPT $\mathbb{E}[T({\bf 1})]$.  Thus, Eq.~\eqref{eq:j_mul}
captures the dependence of $\mathbb{E}[T({\bf 1})]$ on $\Delta E_{\rm
s}$.

We can provide a better estimate of the principal eigenvalue
$\lambda_0$ of the detachment process under partial histone rebinding
by incorporating the rate-limiting effects of the unlinking step into
the flux intensity $j(\Omega^*\,\vert\, \Omega)$ and the contribution
from the monomeric pathway $N e^{N E_{\rm c}}$:
\begin{equation}
  \hat{\lambda}_{0,\rm q} (E_{\rm c}, \qd^*) = - \min \left\{ 
                (\qd^*/\kon \vee N  e^{N E_{\rm c}}), 
                j(\Omega^* | \Omega) \right\},
    \label{eq:lambda_d_mul}
\end{equation}
\added{where $(\qd^*/\kon \vee N e^{N E_{\rm c}}) \coloneqq
\max \{\qd^*/\kon, N e^{N E_{\rm c}} \}$.}  The
results of numerical calculations of $\mathbb{E}[T({\bf 1})]$ and its
comparison to $-\lambda_0$ in the irreversible case, as well as the
estimates in Eq. \eqref{eq:lambda_d_mul} are shown in
Fig.~\ref{fig:Fig8}(b-c). Good agreement between
Eq.~\eqref{eq:lambda_d_mul} and numerical results is observed.

\subsection{Limits of remodeler facilitation}
\label{sec:multimeric_reversible_facilitated}
We now consider the case where the disassembly of nucleosomes is
facilitated by additional nucleosome remodelers.
%
% In principle, previous analysis based on the flow intensity
% $j(\Omega^*\,\vert\,\Omega)$ and the overall flow parameter $k_c$
% can also be adapted to this case, though the analysis is more
% complicated.
%
We make the following observations in different limits of the
remodeler strength.  These observations parallel the corresponding
limits in the linear peeling intact-histone model.
%

%\subsection{Strong facilitation limit} 

When the binding energy $E_{\rm p}$ of the remodeler to DNA is
strongly negative, and the binding rate $\pa > N_{\rm l} k_{\rm on}
e^{N_{\rm l} E_{\rm c}}$, after the dissociation of the histone
modules, the remodeler will bind to the DNA and prevent the
reassociation of the histone modules. Consequently, the scenario is
equivalent to the irreversible, facilitated disassembly of nucleosomes
as discussed in the main text, where we have the estimate through
Eq.~\eqref{eq:tau2_irr} which defines $\mathbb{E} \big[T\big]$:
\begin{equation}
-\hat{\lambda}_{0,{\rm p,q}}(E_{\rm p}\! \rightarrow - \infty)
= - \hat{\lambda}_{0}(E_{\rm p}\! \rightarrow - \infty)
+ \frac{1}{\mathbb{E}\big[T\big] + \tfrac{\kon}{\qd^*}}
\label{eq:lambda_irr_fac}
\end{equation}
%

%\subsection{Weak facilitation limit}
%\label{sec:multimeric_reversible_facilitated}
When the binding energy $E_{\rm p}$ of the remodeler to DNA is weakly
negative, and the binding rate $\pa$ is fast enough, the remodeler
effects are limited to modifying the effective contact energy between
the histone and DNA in Eq.~\eqref{eq:lambda_d_mul}.
\begin{equation}
  \hat{\lambda}_{0,{\rm p,q}}(E_{\rm p} > E_{\rm c})=  
\hat{\lambda}_{0, {\rm q}}(E_{\rm c} - E_{\rm p}^{-}, \qd^*),
  \label{eq:lambda_rev_ss}
\end{equation}
where $E_{\rm p}^{-} \coloneqq \min \{E_{\rm p}, 0\}$.  A general
estimate of the disassembly rate can be obtained by taking the minimum
of the two limits, in terms of absolute values, i.e.,
\begin{equation}
\hat{\lambda}_{0,\rm p,q} \coloneqq \max \left\{ \lambda_{0,\rm p,q}
(E_{\rm p} > E_{\rm c}),\,\hat{\lambda}_{0,\rm p,q}(E_{\rm p}\! \rightarrow - \infty) 
\right\}.
  \label{eq:lambda_rev}
\end{equation}
The results of this estimate are shown in Fig.~\ref{fig:Fig7}(b). In
slow remodeler binding rate $\pa$ limit, the estimate in
Eq.~\eqref{eq:lambda_rev} provides a good approximation to the
numerical results. In large $\pa$ regime, the most probable state on
$\Omega$ switches to the state on the boundary $\partial \Omega$.
Thus, $\lambda_0$ should be rate-limited by the DNA-histone unbinding
rate $\koff$ \deleted{$(/\kon)$} from the boundary state to
$\Omega^*$, while the first passage time $\mathbb{E} \big[T({\bf
1})\big]$ starting from the most interior state, ${\bf
1}=(1,1,1,1,1,(0,0))$, is approximately $N/\koff$.

%\section{Reversible multimeric histone detachment}
%\label{Appendix:multimeric}

\section{Quantifying contributions from the monomeric and multimeric pathways}
\label{Appendix:multimeric_pathway}
We adapt the facilitated, multimeric model to quantify the relative
contributions of the monomeric and multimeric disassembly pathways.
In the original model, the histone can leave the DNA either as an
intact octamer or by disassembling into dimers and tetramers.  For
example, in the high free histone concentration limit ($\Delta E_{\rm
s} \lesssim 0$) and low remodeler binding rate limit ($\pa \rightarrow
0$), as is shown in Fig.~\ref{fig:Fig7}(b), the histone is prevented
from breaking apart since any partial loss of histone modules will be
immediately replaced by free histone modules.  In this limit, the
histone can leave the DNA only as an intact octamer with the slow rate
$-\lambda_0 \approx N e^{N E_{\rm c}}$ associated with the
unfacilitated simple intact-histone model given by
Eq.~\eqref{EQUATION:PERTURBED}.

There are different ways to quantify the relative contributions of the
different disassembly pathways.  One possibility is to evaluate the
flux contribution of the monomeric pathway to the total flux
associated with the principal eigenvector of the transition matrix.
However, in the strong facilitation limit, the principal eigenvector
differs significantly from the fully attached state from which we wish
to quantify the probability flux. To overcome this discrepancy, we
adopt an alternative approach in the FPT
formalism. Suppose we start with the state ${\bf
1}=(1,1,1,1,1,(0,0))$, and split the target state $\Omega^*$ into two
parts: $\Omega^* = \Omega^*_1 \cup \Omega^*_2$.

We define the FPT to $\Omega^*_1$ and $\Omega^*_2$ as
$T_1$ and $T_2$, respectively.  The original first passage time to
$\Omega^*$ is thus $T = \min \{T_1, T_2\}$. Transitions into
$\Omega^*_1$ define histones that leave as an intact
octamer, \textit{i.e.}, $\Omega^*_1 = \{ (1,1,1,1,1,(n_1 + n_2 =
N)) \}$.  The relative contribution of the pathway that leads to
$\Omega^*_1$ can be quantified by the probability $\mathbb{P}[T_1 <
T_2]$ that $T_1 < T_2$. To compute $\mathbb{P}[T_1({\bf 1}) < T_2({\bf
1})]$, we employ the standard approach of first passage time
formalism. For completeness, we briefly describe the general method
below where the symbols used do not necessarily correspond to those
previously used.

Consider a continuous-time Markov chain with a discrete state space
$\Omega$ and the transition matrix $\Q$ defined by $\frac{d}{d t}
{\x}=\Q \x.$ Let $A, B \subseteq \Omega$ and $T$ be the
FPT to $A \cup B$. We then find $\mathbb{P}\left(x_T \in
A \,\vert\,x_0=x\right) \equiv P_A(x)$.

We first discretize the Markov chain $\left\{x_{t} : t \in
\mathbb{R}^+ \right\}$ into a sequence of states $\left\{x_{t_i} : t_i
\in \mathbb{R}^+ \right\}$, where $t_i$ is the $i$-th time point at which 
the $i$-th jump occurs.  The sequence $\left\{ x_{t_i} \equiv x_i:
i \geq 0 \right\}$ is a discrete-time Markov chain, with the
transition probability given by
\begin{equation*}
\mathbb{P}(x_{i+1}=y \mid x_i=x) = \frac{W_{y,x}}{-W_{x,x}}.
\end{equation*}
Conditioning on the first jump time $t_1$, we derive the recursion
relation for $P_A(x)$:
\begin{equation*}
\begin{aligned}
 \: & \hspace{-5mm}\mathbb{P}\left(x_T \in A \!\mid\! x_0=x\right) \\
\: \hspace{5mm} & = \sum_y \mathbb{P}\left(x_T \in A\,\vert\, x_0=x, x_1=y\right)
 \mathbb{P}\left(x_1=y\,\vert\,x_0=x\right) \\
\: \hspace{5mm} & =  \sum_y \mathbb{P}\left(x_T \in A\,\vert\, x_0=y\right)
 \mathbb{P}\left(x_1=y\,\vert\, x_0=x\right) \\
\: \hspace{5mm}& =  \sum_y P_A(y) \mathbb{P}(y\,\vert\, x).
\end{aligned}
\end{equation*}
Rearranging, we find
\begin{equation*}
  \begin{aligned}  
  W_{x, x} P_A(x)+\sum_{y \neq x} P_A(y) W_{y, x} & =0 . \\
 {\bf P}_{A}^{\intercal} \Q & =0,
  \end{aligned}
\end{equation*}
which can be solved with boundary condition

\begin{equation*}
  P_A(x)= \begin{cases}1 & x \in A, \\ 0 & x \in B.\end{cases}
\end{equation*}

%Otherwise, there can be infinitely many solutions to ${\bf P} W =
%0$.

In order to more efficiently solve the problem, it is helpful to
decompose $\Q$ and $P_A$ according to the decomposition of the state
space $\Omega = \Omega_* \cup A \cup B$, where $A$, $B$, and
$\Omega_*$ are disjoint.  We represent ${\bf W}$ by
\begin{equation*}
\Q=\left[\begin{array}{lll}
\Q_{\Omega_*, \Omega_*} & \Q_{\Omega_{ *}, A} & \Q_{\Omega_{ *}, B} \\
\Q_{A, \Omega_*} & \Q_{A, A} & \Q_{A, B} \\
\Q_{B, \Omega *} & \Q_{B, A} & \Q_{B, B}
\end{array}\right]
\end{equation*}
and ${\bf P}_A^{\intercal}$ by
\begin{equation*}
\P_A^{\intercal}=\left[\begin{array}{lll}
\P_{A \mid \Omega_*}^{\intercal} & \P_{A \mid A}^{\intercal} & \P_{A \mid B}^{\intercal}
\end{array}\right]=\left[\begin{array}{lll}
\P_{A \mid \Omega *}^{\intercal} & {\bf 1}^{\intercal} & 0^{\intercal}
\end{array}\right], 
\end{equation*}
where the second equality arises from the boundary condition.
Solving for $\P_{A \mid \Omega_*}^{\intercal}$, we find

\begin{equation*}
\begin{aligned}
& \P_{A \mid \Omega_*}^{\intercal} \Q_{\Omega_*, \Omega_*}+{\bf 1}^{\intercal} \Q_{A, \Omega_*}=\0 \\
& \P_{A \mid \Omega_*}=-\left(\Q_{\Omega_*, \Omega_*}\right)^{-\intercal}
\Q_{A, \Omega_*}^{\intercal} {\bf 1}
\end{aligned}
\label{PA}
\end{equation*}
The numerical solution is shown in Fig.~\ref{fig:Fig9}.  In the limit
$p_{\rm a} \rightarrow 0$, the multimeric pathway is typically faster
than the monomeric pathway, leading to a smaller probability of the
histone leaving as an intact octamer.  When $p_{\rm
a} \rightarrow \infty$, the multimeric pathway is rate-limited by the
histone-module dissociation rate $q_{\rm d}^*$, while both pathways
are also limited by the histone-DNA dissociation rate $\koff$.  When
$q_{\rm d}^* \ll \koff$, the monomeric pathway is faster than the
multimeric pathway, and the probability of the histone leaving as an
intact octamer is close to $1$. If $q_{\rm d}^* \gg \koff$ and $p_{\rm
a} \rightarrow \infty$, the multimeric pathway and the monomeric
pathway carry similar rates and the probability of the histone leaving
as an intact octamer is close to $1/2$, as shown in
Fig.~\ref{fig:Fig9}(b).

% \nocite{*}
% \bibliographystyle{plain}
\section*{References}
\bibliography{revision6_TC_clean2}

%%%%%%%%%%%%%%%%%%%%%%%%%%%%%%%%%%%%%%%%%%%%%%%%%%%
%%%%%%%%%%%%%%%%%%%%%%%%%%%%%%%%%%%%%%%%%%%%%%%%%%%
\end{document}